\documentclass{article}

\usepackage[utf8]{inputenc}
\usepackage[T1]{fontenc}
\usepackage{lmodern}
\usepackage{amsmath}
\usepackage{graphicx}
\usepackage{tikz}
\usetikzlibrary{shapes.geometric,arrows.meta, positioning, calc}
\usepackage{caption}
\usepackage{subcaption}
\usepackage[colorlinks=true,linkcolor=blue]{hyperref}
\usepackage{amssymb}
\usepackage{todonotes}
\usepackage{makecell}
\usepackage{enumitem}
\usepackage{pifont}
\definecolor{olivine}{rgb}{0.6, 0.73, 0.45}

\let\oldabstract\abstract
\let\oldendabstract\endabstract
\makeatletter
\renewenvironment{abstract}
{%
               {\list{}{\addtolength{\leftmargin}{-2.8em} 
                        \listparindent 1.5em%
                        \itemindent    \listparindent%
                        \rightmargin   \leftmargin%
                        \parsep        \z@ \@plus\p@}%
                \item\relax}%
               {\endlist}%
\oldabstract}
{\oldendabstract}
\makeatother

\begin{document}

\title{Learning Transient Convective Heat Transfer\\ with Geometry Aware World Models}

\author{Onur T.\ Doganay${}^1$, Alexander Klawonn${}^2$, \\Martin Eigel${}^{1,3}$ and Hanno Gottschalk${}^1$\\
~
\\
${}^1$Institute of Mathematics, TU Berlin, 
${}^2$Siemens Energy AG,\\ 
${}^3$Weierstrass Institute for Applied Analysis and Stochastics \\
\vspace{-.1cm}\\
\texttt{\footnotesize $\{$doganay,gottschalk$\}$@math.tu-berlin.de}, \texttt{\footnotesize martin.eigel@wias-berlin.de} \\ \texttt{ \footnotesize alexander.klawonn@siemens-energy.com}}

\maketitle

\begin{abstract}
\noindent Partial differential equation (PDE) simulations are fundamental to engineering and physics but are often computationally prohibitive for real-time applications. While generative AI offers a promising avenue for surrogate modeling, standard video generation architectures lack the specific control and data compatibility required for physical simulations. This paper introduces a geometry aware world model architecture, derived from a video generation architecture (LongVideoGAN), designed to learn transient physics. We introduce two key architecture elements: (1) a twofold conditioning mechanism incorporating global physical parameters and local geometric masks, and (2) an architectural adaptation to support arbitrary channel dimensions, moving beyond standard RGB constraints. We evaluate this approach on a 2D transient computational fluid dynamics (CFD) problem involving convective heat transfer from buoyancy-driven flow coupled to a heat flow in a solid structure. We demonstrate that the conditioned model successfully reproduces complex temporal dynamics and spatial correlations of the training data. Furthermore, we assess the model's generalization capabilities on unseen geometric configurations, highlighting both its potential for controlled simulation synthesis and current limitations in spatial precision for out-of-distribution samples.
\end{abstract}


\section{Introduction}

In engineering, design and operation of technical devices requires insight into underlying physical processes. Historically, this was first achieved through experiments and thereafter by numerical simulations calibrated on experiments. While numerical simulation in most simulation is considerably faster and more flexible than experimentation, it mostly does not yet permit real-time applications like extensive exploration studies, global optimization, interactive digital twins and online control. In recent years, generative learning models trained on data from numerical simulations promised to reach real-time capability.

Numerous methods have been suggested under the category of scientific machine learning (SciML) to solve physical equations with neural networks, with physics informed neural networks (PINN) \cite{raissi2019physics,cai2021physics}, operator learning (OL) \cite{lu2021learning,li2020fourier} and graph neural networks (GNN) \cite{sanchez2018graph,xmeshgraphnet2024,meshmask2025} as main trends. Many of these models allow control of initial conditions and physical parameters via model inputs and recently geometry aware methods thrived. While many of these methods show promising results with a considerably reduced compute time, some challenges remain. PINN suffer from the delicate task to balance various loss terms during training, which usually can be resolved, but requires skilled and time intensive numerical experimentation. Operator learning methods provide excellent solutions for dynamical systems on short time horizons, but struggle with long time stability. Graph neural networks generalize nicely over geometries, but require set up of simulation meshes, which often requires human intervention.

Generative learning promises to be an alternative to the previous methods. While generative models have excelled in text, image and video generation, simulation of physical systems with generative AI (GenAI) models is drawing more and more attention in the wake of recent developments towards  physical and embodied AI. In particular, the concept of world models emerged \cite{ha2018world} which represents dynamics in a lower dimensional latent space. Unlike model order reduction (MOR), the latent dynamics is not obtained from  a differential equation, but follows an auto-regressive approach derived from large language models (LLM). Under certain assumptions on the tokenizer, i.e. the map that computes the latent representation from the full physical state, this enables the full reconstruction of the dynamics for the full state using a super-resolution network applied to a history of several frames in the latent space. This approach has been shown to outperform baseline OL methods especially on longer time horizons \cite{ross2025worldmodelssuccessfullylearn}. 

Long time stability is particularly important for transient simulations like convective heat transfer. In buoyancy-driven heat transfer, chaotic fluid movement from ascending warm matter of lower density contributes to the effective distribution of the heat and thus to the cooling of the heat source that injects a certain amount of power into the system. technically, this is important in the operation of substations, which, in times of volatile energy supply due to renewable energy sources, in some cases might need to bear an overload during a limited time-frame. While instationary, turbulent fluid problems have been widely studied in SciML, transience adds another level of difficulty that so far has rarely been approached.

In this work, we develop a world model approach for the aforementioned transient heat transfer problem in a two dimensional setting. Our contributions are as follows:

    \begin{itemize}[label=\ding{227}]
        \item We modify a common wold model architecture (LongVideoGAN) such that it generalizes from three rgb channels to an arbitrary number of channels.  
        \item We introduce two conditioning mechanism which enable flexible conditioning on geometries by providing 2D masks and external setting of further physical parameters like the power of the heat source.    
        \item We publish a data set based on numerous numerical simulations representing a buoyancy-driven heat transient heat transfer inside of substations with active and passive heat sources which can serve as a benchmark problem for future developments in the field.
        \item We develop an extensive methodology based on metrics for the evaluation of the accuracy of the synthesized flow solutions in average, with respect to spatial and temporal fluctuations and with respect to the long term transient rise of the temperature.
        \item We present our results from training our architecture of the given data set and the evaluation of our set of metrics. This is done for in distribution and out of distribution data sets. We give a detailed account on the success and the remaining problems. 
        \item We also present details on training and inference and highlight the real-time capability of our methods dramatically reducing the simulation time compared to numerical simulation. 
    \end{itemize}

Our paper is structured as follows. Section \ref{sec:related_work} briefly discusses the state of the art of SciML for turbulent flow.  Section \ref{sec:governing_equations} introduces the underlying physics. The detailed description of our world model's architecture can be found in Section \ref{sec:architecture}. This section also introduces our dual conditioning strategy for geometry and parameters. Section \ref{sec:dataset_and_evaluation} deals with the metrics we develop for a comprehensive evaluation. We present our numerical results in Section \ref{sec:results} before we conclude and provide an outlook to future research in Section \ref{sec:conclusion_and_outlook}. Our data and code is publicly available under \href{https://github.com/otd11/TransientGeoAwareWorldModel}{github}.  See \href{https://www.youtube.com/watch?v=mo_mJpTB-Qs}{youtube} for a video visualizing our results.




\section{Related Work}
\label{sec:related_work}

There is a large variety of recent machine learning methods related to our approach. In the following, we only provide an overview and references to some of the most prominent DNN architectures.

\paragraph{Physics Informed Neural Networks.}  
Physics Informed Neural Networks (PINNs) \cite{raissi2019physics,cai2021physics} offer a data-free alternative by training networks to directly minimize PDE residuals, thereby bypassing data generation from numerical simulation. While early formulations were restricted to simple rectangular domains, recent work has significantly expanded their geometric capabilities. Notably, Laplace-Beltrami PINNs introduce a novel positional encoding mechanism based on manifold eigenfunctions, allowing the framework to solve heat transfer equations on complex, irregular topologies like heat sinks without explicit meshing \cite{costabal2024deltapinns}.

Furthermore, the stability of PINNs in stiff, coupled multi-physics regimes (such as the flow problem considered in this work) has been improved through Curriculum Learning strategies \cite{haddout2025deep,muenzer2022curriculum}. Recent applications to stratified forced convection demonstrate that adaptive training strategies together with residual-network architectures can overcome the divergence issues typically associated with high-stiffness regimes \cite{haddout2025deep,wang2024piratenets,wang2021gradient,maddu2022inverse}. Additionally, Transfer Learning techniques have been proposed to handle transient heat sources by initializing time-steps from previous solutions, thereby reducing the computational overhead of training \cite{kalyan2025movingheatsource}.
However, despite these advances, PINNs remain fundamentally optimization-based, requiring substantial compute time for each new case and reaching high accuracy only slowly.

\paragraph{Operator Learning.}
While foundational methods like the Fourier Neural Operator (FNO) demonstrated significant speed-ups for simple domains, their reliance on Fast Fourier Transforms initially restricted them to rectangular grids. Recent advancements have focused on extending operator learning to complex, arbitrary geometries. The Geometry-Informed Neural Operator (GINO) \cite{li2023geometry} addresses this by lifting irregular mesh data onto latent uniform grids using graph networks, thereby combining geometric flexibility with spectral efficiency. Similarly, the General Neural Operator Transformer (GNOT) \cite{hao2023gnot} utilizes heterogeneous attention mechanisms to handle arbitrary input shapes and scales, achieving state-of-the-art performance on diverse physical benchmarks including heat sinks.

In the context of transient dynamics, standard operator models often suffer from autoregressive error accumulation over long time horizons. To mitigate this, Time-Integrated DeepONet (TI-DeepONet) \cite{nayak2025tideeponet} proposes learning the temporal derivative operator rather than the state mapping, utilizing numerical integrators during inference to preserve stability. Sequential DeepONet (S-DeepONet) \cite{he2024sequential} alternatively embeds recurrent units to capture history-dependent loads in transient heat transfer problems. Unlike these methods, which often require complex hybrid architectures or numerical integration schemes, our approach leverages a unified generative architecture where geometry and temporal dynamics are handled via learned conditioning masks and 3D convolutions.

\paragraph{Graph Neural Networks.}
Graph Neural Networks (GNNs) explicitly model the physical domain as a mesh, treating fluid dynamics as message passing between spatial nodes. This allows for robust generalization across unseen geometries, as rigorously benchmarked for MeshGraphNets in flow prediction tasks around varying obstacles \cite{schmocker2024generalization, pfaff2021learning}. For specific applications in heat transfer, Multi-Stage GNNs \cite{ahangarkiasari2025multistage} have been recently developed to predict natural convection in enclosed cavities—the exact physical regime considered in this work. These architectures utilize pooling layers to capture the long-range global dependencies characteristic of buoyancy-driven flows.

However, GNNs introduce significant preprocessing overhead for mesh generation and higher inference latency compared to grid-based methods due to irregular memory access patterns during message passing. Furthermore, capturing fine-scale transient turbulence often requires very dense meshes, scaling the computational cost linearly or super-linearly with node count. Our world model avoids explicit meshing by operating in the pixel/latent space, employing binary masks to represent geometry, which decouples inference speed from geometric complexity.

\paragraph{Generative Models for State Snapshots.}
Generative learning has first been applied to state snapshots for turbulent flow. Variational autoencoders, diffusion models and generative adversarial neural networks (GAN) have all been applied for the generation task, see \cite{drygala2024comparisongenerativelearningmethods} for a comparison. Theoretical guarantees for accurate learning based on  Birkhoff's ergodic theorem are given in \cite{Drygala_2022}. \cite{drygala2023generalizationcapabilitiesconditionalgan} considers generalization over geometries. Snapshot models, however, do not represent the temporal features of unsteady flows. Transient long time dynamics can only be learned via conditioning, but not be synthesized directly.

\paragraph{Video Generation Models.}
Video generation with temporal consistency started with GAN based models \cite{vondrick2016generating,saito2017temporal}. Model architectures soon shifted to transformer and diffusion based video generation \cite{he2024latentvideodiffusion}. Commercial video generation models are available and produce video sequences of intriguing quality \cite{ho2022imagenvideo}. Our approach here is based on LongVideoGAN \cite{Brooks2022_lvg} by NVIDIA. We do not choose diffusion models as their inference time is considerably longer as multiple denoising steps are required, each coming with a forward pass through the network. As the key advantage of generative models over classical simulation is speed, we use GAN-based architectures which generate sequences in a single forward pass.


\paragraph{Fluid simulation using world models.}
Several authors have trained video generation models on physical data, see e.g.\ \cite{klemmer2022spate,skorokhodov2022stylegan}. Most of these works however only test for video similarity and doe not provide an extensive comparison with simulation based on physical quantities. In \cite{ross2025worldmodelssuccessfullylearn} the authors investigated the foundations of statistical learning of world models in the context of dynamical systems and provided detailed physics based evaluations. They show superiority of the world model approach compared to standard operator learning methods like DeepONet and FNO especially in terms of long time stability where OL methods often diverged. However, neither geometry awareness nor transient flows were considered, as we do in this article.   


\paragraph{Data Sets.}
Numerous data sets have been published to enable evaluation and training of neural network generated solutions. \cite{takamoto2022pdebench} offers data for benchmarking SciML models byond specific application domains. The Well \cite{ohana2024well} offers a broad range of instationary but mostly non transitive simulations. Unifoil \cite{kanchi2025unifoil} is a vast collection of stationary RANS flow simulations around airfoils. FlowBench \cite{tali2025flowbench} and CFDBench \cite{luo2023cfdbench} offer data from flows around various geometries in two and three dimensions.  \cite{Drygala_2022} published a high quality 2D data set of the Kármán vortex street and of a low pressure turbine with a circulating wake generator for geometric conditioning. See also the references in \cite{tali2025flowbench} for further data sets. While many of these data sets ar more general in scope and many of them contain more data in absolute, the data set published with this paper is special as it contains flow data that is instationary and transient, where transience is understood as change of the physical states over a longer timeframes as typical instationary effects like recurring turbulent flow features.     

\section{Governing Equations of Convective Heat Transfer}
	\label{sec:governing_equations}	
	To evaluate the proposed architecture, we utilize a dataset obtained from transient Computational Fluid Dynamics (CFD) simulations coupled to a heat flow through a solid, see also Section~\ref{sec:dataset_and_evaluation}. The corresponding governing equations, together with boundary and interface conditions are summarized in this section.\par 
    \paragraph{Equations.}
	The fluid dynamics are modeled using the compressible two-dimensional Navier-Stokes equations including gravitational body forces, coupled with an energy equation. The system of governing equations is closed by linking fluid density and temperature via the ideal gas law. No turbulence model was used. The governing momentum and continuity equations are
	\begin{subequations}
		\label{eq:navier-stokes}
		\begin{align}
			\rho\left(
			\frac{\partial u}{\partial t}
			+ u\frac{\partial u}{\partial x}
			+ v\frac{\partial u}{\partial y}
			\right)
			&=
			-\frac{\partial p}{\partial x}
			+ \mu\left(
			\frac{4}{3}\frac{\partial^2 u}{\partial x^2}
			+ \frac{\partial^2 u}{\partial y^2}
			+ \frac{1}{3}\frac{\partial^2 v}{\partial x\,\partial y}
			\right)
			&&\text{$x$-momentum},\\[0.5em]
			\rho\left(
			\frac{\partial v}{\partial t}
			+ u\frac{\partial v}{\partial x}
			+ v\frac{\partial v}{\partial y}
			\right)
			&=
			-\frac{\partial p}{\partial y}
			+ \mu\left(
			\frac{\partial^2 v}{\partial x^2}
			+ \frac{4}{3}\frac{\partial^2 v}{\partial y^2}
			+ \frac{1}{3}\frac{\partial^2 u}{\partial x\,\partial y}
			\right)
			- \rho g
			&&\text{$y$-momentum},\\[0.5em]
			\frac{\partial \rho}{\partial t}
			+ \frac{\partial (\rho u)}{\partial x}
			+ \frac{\partial (\rho v)}{\partial y}
			&= 0
			&&\text{continuity}.
		\end{align}
	\end{subequations}
	where $u$ and $v$ are the velocity components in the $x$- and $y$-directions, respectively, $\rho$ is the fluid density, $p$ is the pressure, $\mu$ is the dynamic viscosity, and $g=9.81\,\mathrm{m}/\mathrm{s}^2$ is the magnitude of gravitational acceleration (with $y$ directed upwards).\par 
	Since we want to model conjugate convective heat transfer between fluid and solid domains, we need to include an energy equation for both types of domain. Neglecting pressure work and viscous dissipation, while assuming constant specific heat capacities ($c_p$, $c_{p,s}$) and thermal conductivities ($k$, $k_s$), the energy equations read
	\begin{subequations}
		\begin{align}
			\label{eq:energy-equations}
			\rho c_p\left(
			\frac{\partial T}{\partial t}
			+ u\frac{\partial T}{\partial x}
			+ v\frac{\partial T}{\partial y}
			\right)
			&=
			k\left(
			\frac{\partial^2 T}{\partial x^2}
			+ \frac{\partial^2 T}{\partial y^2}
			\right)
			&&\text{energy, fluid},\\[0.5em]
			\rho_s c_{p,s}\,\frac{\partial T_s}{\partial t}
			&=
			k_s\left(
			\frac{\partial^2 T_s}{\partial x^2}
			+ \frac{\partial^2 T_s}{\partial y^2}
			\right)
			+ q
			&&\text{energy, solid}
		\end{align}
	\end{subequations}
	where the subscript $s$ indicates a material parameter or a field variable on the solid domain (if the same parameter or field is also defined on the fluid domain). The values for the different material parameters are given in Table~\ref{tab:mat-parameters}. Note that the volumetric heat source $q$ is only defined on the solid domain, and is constant in space and time.\par 
	As mentioned before, the equations are closed by linking the fluid density $\rho$ with the temperature field $T$ via the ideal gas state equation which is only defined on the fluid domain
	\begin{align}
		\label{eq:ideal-gas-law}
		\rho = \frac{pM}{R T}.
	\end{align}
	Here, $R$ is the universal gas constant and $M$ denotes the fluid's molecular weight for which we used a value of $M=29~\mathrm{kg}/\mathrm{kmol}$, a typical value for air.\par 
	\paragraph{Boundary conditions and coupling.} On the fluid-solid interface $\Gamma_\mathrm{int}$, see Figure~\ref{fig:dataset_parameters}, we impose continuity of temperature and normal conductive heat flux, i.e., 
	\begin{align}
		\label{eq:interface-condition}
		T\big|_{\Gamma_\mathrm{int}} = T_s\big|_{\Gamma_\mathrm{int}}, \qquad  
		-k \frac{\partial T}{\partial n}\bigg|_{\Gamma_\mathrm{int}}
		= -k_s \frac{\partial T_s}{\partial n}\bigg|_{\Gamma_\mathrm{int}},
	\end{align}
	with $n$ being the coordinate normal to the interface. The fluid velocity satisfies a no-slip condition at this interface.\par 
	On the external boundary of the fluid domain $\Gamma_b$, see Figure~\ref{fig:dataset_parameters}, we prescribe a fixed wall with no-slip for the velocity and an isothermal Dirichlet boundary condition for the temperature,
	\begin{align}
		\label{eq:dirichlet-condition}
		T\big|_{\Gamma_b} = T_\mathrm{amb},
	\end{align}
	which allows heat exchange between the fluid domain and an ambient environment but no mass exchange.

\section{Architecture Overview and Conditioning}
\label{sec:architecture}

Our approach is based on the architecture introduced in \cite{Brooks2022_lvg} which is a generative adversarial network designed to create temporally consistent videos spanning hundreds of frames. To this end, a hierarchical, two-stage architecture is employed. A low-resolution generator learns long term temporal dynamics over extended sequences using latent style-vectors, see, e.g., \cite{Karras2018A_SG1, Karras2019AnalyzingAI_SG2, Karras2021_SG3}, and a wide array of ranges of temporal context. While a separate super-resolution GAN refines spatial details using a sliding window approach. Furthermore, temporal coherence is achieved through multi-scale latent filtering, 3D convolutions, and temporal upsampling. This architecture is based on the {StyleGAN} architectures, see, e.g., \cite{Karras2018A_SG1, Karras2019AnalyzingAI_SG2, Karras2021_SG3}, with some task specific modifications. By training separately on long, low-resolution sequences and short, super-resolution videos, the  architecture balances computational efficiency with realism. It is capable of producing long, coherent videos that maintain both motion continuity and image quality. 

In \cite{Brooks2022_lvg} the GANs are used to generate realistic time coherent RGB videos in first-person perspective, while in \cite{ross2025worldmodelssuccessfullylearn} the architecture was used to learn CFD simulations of a K\'arm\'an vortex channel given as gray-scale images which were processed as RGB frames. In both cases, the frame data of videos was given as images, e.g., '.jpg' or '.png' files, which were converted to tensors $x^\textrm{DATA}\in\mathbb{R}^{3\times \mathcal{T}\times H \times W}$ for further processing. The generated output videos were given as $x^\textrm{OUT}\in\mathbb{R}^{3\times \mathcal{T}'\times H \times W}$, where $\mathcal{T}, \mathcal{T}'$ are the time dimensions, i.e., the sequence lengths of the videos, $H$ the height of the frames, and $W$ their width. Since we intend to learn from numerical simulations, we interpolate finite element simulation data of $C_{\textrm{SIM}}$ physical fields to a grid of height $H$ and width $W$, i.e., we require $x^\textrm{DATA}\in\mathbb{R}^{C_{\textrm{SIM}}\times \mathcal{T}\times H \times W}$ and $x^\textrm{OUT}\in\mathbb{R}^{C_{\textrm{SIM}}\times \mathcal{T}'\times H \times W}$. The architecture is already constructed to have configurable $C_{\textrm{SIM}}$ output channels. Nonetheless, in practice we encountered that even though $C_{\textrm{SIM}}$ as an input parameter exists, there is no argument pass-through of $C_{\textrm{SIM}}$ to the relevant parts of the architecture where instead  $C_{\textrm{SIM}}=3$ is forced. We adjusted the architecture accordingly enabling it now to handle arbitrary channel dimensions. Note that we also enabled the dataloader to directly load tensors with $C_{\textrm{SIM}}$ channels instead of reading RGB images to convert them to 3 channels. Hence, from here on we refer to the output channel dimension as $C_{\textrm{SIM}}$ instead of 3.

Furthermore, we incorporate static, i.e., constant in time, conditioning information to the LongVideoGAN architecture in two ways. We utilize a ``global'' conditioning vector $c^{\textrm{num}}\in\mathbb{R}^{n_{\textrm{num}}}$ that can be used to encode labels or features. In our case it contains boundary values and parameters of the PDEs. Furthermore, we also follow a more ``local'' approach, i.e., we incorporate (binary) masks $c^{\textrm{mask}}\in\mathbb{R}^{n_{\textrm{mask}}\times H \times W}$
, where $H$ and $W$ are the output dimensions of the generator the conditioning is applied to, to achieve a pixel-by-pixel conditioning. This approach allows maximum flexibility when defining geometries, they just have to be provided as a binary mask. In the following, a brief overview of the relevant components of the architecture as in \cite{Brooks2022_lvg} 
and our modifications to condition it on $c=(c^{\textrm{num}}, c^{\textrm{mask}})$ are given.

\paragraph{Low-resolution GAN: Generator $G^\textrm{lr}$.}

The low-resolution generator $G^\textrm{lr}$ learns to generate videos/frames from noise inputs, i.e., i.i.d.\ Gaussian noise 

\begin{equation*}
\epsilon_t \in \mathbb{R}^d,\quad \epsilon_t \overset{\mathrm{i.i.d.}}{\sim} \mathcal{H}(0, I_d).    
\end{equation*} 
To capture long-term patterns $K$ {temporal low-pass filters} $h^{(k)}:\mathbb{Z}\to \mathbb{R}$, i.e., {discrete-time finite impulse response (FIR) filters}, are applied

\begin{equation}    
\tilde{\epsilon}_t^{(k)} = \sum_{\tau=-\lfloor \mathcal{L}_k/2\rfloor}^{\lfloor \mathcal{L}_k/2\rfloor} h^{(k)}(\tau)\,\epsilon_{t+\tau}, \quad k=1,\dots, K.
\label{eq:fir}
\end{equation}
Here, $\mathcal{L}_k\in\mathbb{H}$ is the {temporal filter length} for the $k$-th filter $h^{(k)}$. For further details we refer to \cite{Brooks2022_lvg, kaiser1974nonrecursive}. Each $h^{(k)}$ captures temporal structure at a different scale, where smaller $\mathcal{L}_k$ correspond to sharp temporal transitions and larger values to smoother variations. Through this multiscale filtering a temporal context over a window of size $\mathcal{L}_k$ is provided to the latent code. In a next step, all $K$ filtered vectors are concatenated
\begin{equation}
    \tilde{\epsilon}_t := \left[ \tilde{\epsilon}_t^{(1)} \,\|\, \cdots \,\|\, \tilde{\epsilon}_t^{(K)} \right] \in \mathbb{R}^{Kd}.
    \label{eq:lres_gen_concat_eps}
\end{equation}

We incorporate our global conditioning information $c^{\textrm{num}}$ by concatenating it to the $K$ filtered vectors in \eqref{eq:lres_gen_concat_eps} and obtain
\begin{equation}
    \tilde{\epsilon}_t :=\tilde{\epsilon}_{t, c^{\textrm{num}}_t} := \left[ \tilde{\epsilon}_t^{(1)} \,\|\, \cdots \,\|\, \tilde{\epsilon}_t^{(K)} \,\|\, c^{\textrm{num}}_t \right] \in \mathbb{R}^{Kd + n_{\textrm{num}}},
    \label{eq:lres_gen_concat_eps_cond}
\end{equation}

where in our case $c^{\textrm{num}}_t = c^{\textrm{num}}$ since our conditioning is static.

Each of these concatenated $\tilde{\epsilon}_t$ are then transformed to {style vectors} $w_t$ that are passed to the layers of the generator via a shared mapping network given as an MLP
\begin{equation}
    w_t := M(\tilde{\epsilon}_t) \in \mathbb{R}^d,\quad
M = \phi \circ A_n \circ \cdots \circ A_1,
\label{eq:lres_gen_styleMLP}
\end{equation}
where all layers $A_i$ are affine and $\phi$ is {Leaky ReLU}. Note that, corresponding to our modifications we adjust the input dimension of \eqref{eq:lres_gen_styleMLP}, in particular of layer $A_1$, to accept  \eqref{eq:lres_gen_concat_eps_cond} instead of \eqref{eq:lres_gen_concat_eps} as an input, i.e., the input dimension has to be increased by $n_{\textrm{num}}$ while the output dimension stays unchanged.

The generator $G^{\textrm{lr}}$ initializes with a learned constant tensor $x^{(0)} \in \mathbb{R}^{C_0 \times \mathcal{T}_0 \times H_0 \times W_0}$ and passes it through $L$ sequential blocks $\mathcal{B}_\ell$. In $\mathcal{B}_\ell$, a modulated 3D convolution via the style vectors $w_t$ and optionally up-scaling on the dimensions $(\mathcal{T},H,W)$ or $(H,W)$ is applied before passing through {Leaky ReLU}. During the forward pass we then have

\begin{equation}
    x^{(\ell)} = \mathcal{B}_\ell(x^{(\ell-1)}; w)\in \mathbb{R}^{C_\ell \times \mathcal{T}_\ell \times H^{\textrm{lr}}_\ell \times W^{\textrm{lr}}_\ell}, \quad \ell=1,\dots,L.
    \label{eq:lres_gen_fwdPass}
\end{equation}

To incorporate the spatial conditioning information $c^{\textrm{mask}}\in\mathbb{R}^{n_{\textrm{mask}}\times H^{\textrm{lr}} \times W^{\textrm{lr}}}$ we (down-)scale $c^{\textrm{mask}}$ spatially to $(H_\ell, W_\ell)$ with a nearest neighbor approach for each $\ell$. We then obtain 
\begin{equation}
    c^{\textrm{mask},(\ell)}\in\mathbb{R}^{n_{\textrm{mask}}\times \mathcal{T}_\ell \times H_\ell \times W_\ell}
    \label{eq:scaled_mask_cond}
\end{equation}
by duplicating in dimension $\mathcal{T}_\ell$. Note that we use this duplication approach since the conditioning information is static in our case. Further, there exists a $\hat{\ell}$ such that $H^{\textrm{lr}}= H_\ell$ and $W^{\textrm{lr}}= W_\ell$ for all $ \ell \geq \hat{\ell}$. We concatenate these spatially scaled $c^{\textrm{mask},(\ell)}$ with the inputs $x^{(\ell)}$ of \eqref{eq:lres_gen_fwdPass} and increase the input dimensions of $\mathcal{B}_\ell$ by $n_\textrm{mask}$, i.e., $\mathcal{B}_\ell$ now expects the input dimension $C_{\ell-1} + n_\textrm{mask}$ instead of $C_{\ell-1}$  for $\ell=1,\dots,L$. We then obtain for the conditioned forward pass
\begin{align}
    \begin{split}
        \tilde{x}^{(\ell)} &= \left[ x^{(\ell)} \,\|\, c^{\textrm{mask},(\ell)}  \right], \quad \ell=0,\dots,L\,,\\
        x^{(\ell)} &= \mathcal{B}_\ell(\tilde{x}^{(\ell-1)}; w, c^{\textrm{mask}}), \quad \ell=1,\dots,L.
    \end{split}
    \label{eq:lres_gen_fwdPass_cond}
\end{align}

\begin{figure}[h!]
    \centering
    \begin{subfigure}[t]{0.68\textwidth}
    \scalebox{0.6}{
    \begin{tikzpicture}[
        rect/.style={draw, rectangle, text width=4.5cm, minimum height=1cm, align=center, fill=olivine!50},
        smallrect/.style={draw, rectangle, text width=3.4cm, minimum height=0.8cm, align=center, fill=olivine!50},
        imgrect/.style={draw, rectangle, minimum width=2.4cm, minimum height=2.4cm, align=center},
        squares/.style={draw, rectangle, minimum width=0.8cm, minimum height=0.8cm, align=center, fill=olivine!50},
        pluscircle/.style={ draw, circle,minimum size=6mm,  inner sep=0pt,align=center,
        font=\small},
        arrow/.style={->, thick},
        darrow/.style={->, thick, dotted},
        node distance=6mm,
        highlight/.style={fill=blue!20}, 
        cond/.style={fill=orange!20}, 
        noise/.style={rectangle, text width=4.5cm, minimum height=1cm, align=center} 
    ]

    \def\colsep{2.6cm}
    
    \node[] (C11) {Temporal noise};
    \node[rect, highlight, below=3mm of C11] (C12) {Temporal lowpass filters};

    \node[squares, cond, draw, dashed, left=1mm of C11] (Cnum11)
    {$c^{num}$};
    
    \node[rect, below=3mm of C12] (C14) {Mapping network -- $T\times1\times1$};
    
    \node[below=10mm of C14] (C14b) {Sequence of $w_t \in \mathcal{W}$, $0\le t \le 1$};
    
    \node[squares,highlight, below=60mm of Cnum11] (C15) {};
    \node[right=1mm of C15] (T15) {Fixed};
    \node[squares, below=3mm of C15] (C16) {};
    \node[right=1mm of C16] (T16) {Learned};
    \node[squares,highlight, below=3mm of C16] (C17) {D};
    \node[right=1mm of C17] (T17) {Temporal downsample};
    \node[squares, below=3mm of C17] (C18) {A};
    \node[right=1mm of C18] (T18) {Affine};
    \node[squares, cond, below=3mm of C18, dashed] (C19) {};
    \node[right=1mm of C19] (T19) {Condition};

    \draw[arrow] (C11) -- (C12);
    \draw[arrow] (C12) -- (C14);
    \draw[darrow] (Cnum11) -- (C12.north);
    \node[smallrect, right=60mm of Cnum11] (C21) {Spatial input - $1 \times 4 \times 4$}; 
    \node[pluscircle, below=3mm of C21] (C22) {+};
    \node[squares, left=7mm of C22] (A22) {A};
    \node[squares, highlight, left=2mm of A22] (D22) {D};

    \node[smallrect, below=3mm of C22] (C23) {$\textrm{\textbf{ST}}_0$ - $(T/16) \times 4 \times 4$};
    \node[smallrect, below=3mm of C23] (C24) {$\textrm{\textbf{ST}}_1$ - $(T/8) \times 8 \times 8$};
    \node[smallrect, below=3mm of C24] (C25) {$\textrm{\textbf{ST}}_2$ - $(T/4) \times 8 \times 8$};
    \node[smallrect, below=3mm of C25] (C26) {$\textrm{\textbf{ST}}_3$ - $(T/2) \times 16 \times 16$};
    \node[smallrect, below=3mm of C26] (C27) {$\textrm{\textbf{ST}}_4$ - $T \times 16 \times 16$};
    \node[smallrect, below=3mm of C27] (C28) {$\textrm{\textbf{ST}}_5$ - $T \times 16 \times 16$};
    \node[smallrect, below=3mm of C28] (C29) {$\textrm{\textbf{S}}_0$ - $32 \times 32$};
    \node[smallrect, below=3mm of C29] (C210) {$\textrm{\textbf{S}}_1$ - $32 \times 32$};
    \node[smallrect, below=3mm of C210] (C211) {$\textrm{\textbf{S}}_2$ - $64 \times 64$};
    \node[smallrect, below=3mm of C211] (C212) {$\textrm{\textbf{S}}_3$ - $64 \times 64$};
    \node[smallrect, highlight, below=3mm of C212] (C213) {EMA norm};
    \node[smallrect, below=3mm of C213] (C214) {$1\times 1$ ModConv};
    \node[squares, left=2mm of C214] (A214) {A};
    \node[below=3mm of C214] (C215) {};
    \foreach \i [evaluate=\i as \j using int(\i+1)] in {1,...,14}{
        \draw[arrow] (C2\i) -- (C2\j);
    }
    
    \def\imggroupsep{2.8cm}  
    \def\imginnerspacing{0.6cm} 
    
    \node[smallrect, cond, draw, dashed, align=center, right= of C21] (C3desc) {$c^{mask}$};
    
    \node[imgrect, cond, dashed, below=of C3desc] (C31) {\includegraphics[width=2.2cm]{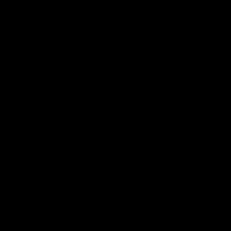}};
    \node[below=\imginnerspacing of C31] (C32) [imgrect, cond, dashed] {\includegraphics[width=2.2cm]{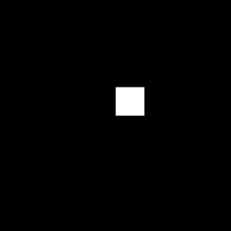}};
    
    \node[below=\imginnerspacing of C32] (C33) [imgrect, cond, dashed] {\includegraphics[width=2.2cm]{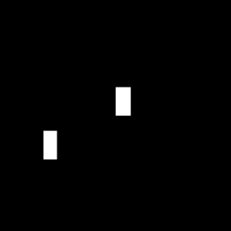}};

    \node[imgrect, cond, dashed, below=\imginnerspacing of C33] (C34) {\includegraphics[width=2.2cm]{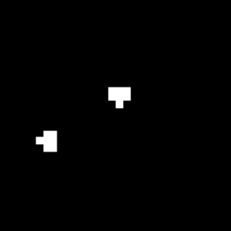}};
    \node[below=\imginnerspacing of C34] (C35) [imgrect, cond, dashed] {\includegraphics[width=2.2cm]{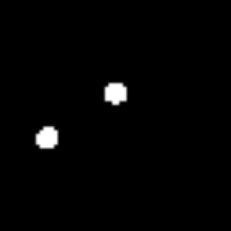}};

    \draw[darrow] (C31.west) -- (C23.east);
    \draw[darrow] (C32.west) -- (C24.east);
    \draw[darrow] (C32.west) -- (C25.east);
    \draw[darrow] (C33.west) -- (C26.east);
    \draw[darrow] (C33.west) -- (C27.east);
    \draw[darrow] (C33.west) -- (C28.east);
    \draw[darrow] (C34.west) -- (C29.east);
    \draw[darrow] (C34.west) -- (C210.east);
    \draw[darrow] (C35.west) -- (C211.east);
    \draw[darrow] (C35.west) -- (C212.east);
    
    \coordinate (busX)   at ($(D22.west)+(-0.4cm,0)$);
    \coordinate (busTop) at (busX |- D22);
    \coordinate (busBot) at (busX |- C214);
    
    \draw[thick] (C14.east) -- (busX |- C14);
    \draw[thick] (busTop) -- (busBot);

    \foreach \i in {3,...,12}{
        \pgfmathtruncatemacro{\c}{\i}
        \draw[arrow] (busX |- C2\c.west) -- (C2\c.west);
    }
    
    \draw[arrow] (busX |- D22.west) -- (D22.west);
    \draw[arrow] (D22.east |- A22.west) -- (A22.west);
    \draw[arrow] (A22.east |- C22.west) -- (C22.west);
    
    \draw[arrow] (busX |- A214.west) -- (A214.west);
    \draw[arrow] (A214.east |- C214.west) -- (C214.west);
    
    \end{tikzpicture}
}
        
        \caption{Generator $G^\textrm{lr}$\label{fig:arch_lr_gen}}
    \end{subfigure}
    \hfill
    \begin{subfigure}[t]{0.30\textwidth}
        \centering

    \scalebox{0.6}{
    \begin{tikzpicture}[
        rect/.style={draw, rectangle, text width=4.5cm, minimum height=1cm, align=center, fill=olivine!50},
        smallrect/.style={draw, rectangle, text width=3.4cm, minimum height=0.8cm, align=center, fill=olivine!50},
        imgrect/.style={draw, rectangle, minimum width=2.4cm, minimum height=2.4cm, align=center},
        squares/.style={draw, rectangle, minimum width=0.8cm, minimum height=0.8cm, align=center, fill=olivine!50},
        pluscircle/.style={ draw, circle,minimum size=6mm,  inner sep=0pt,align=center,
        font=\small},
        arrow/.style={->, thick},
        darrow/.style={->, thick, dotted},
        node distance=6mm,
        highlight/.style={fill=blue!20}, 
        cond/.style={fill=orange!20}, 
        noise/.style={rectangle, text width=4.5cm, minimum height=1cm, align=center} 
    ]
    
    \node[smallrect] (Video) {Video - $128 \times 64 \times 64$};
    \node[smallrect, below=3mm of Video] (1x1_conv) {$1 \times 1$ Conv};
    \node[smallrect, below=3mm of 1x1_conv] (B1) {\textbf{B} - $128 \times 32 \times 32$};
    \node[smallrect, below=3mm of B1] (B2) {\textbf{B} - $64 \times 16 \times 16$};
    \node[smallrect, below=3mm of B2] (B3) {\textbf{B} - $32 \times 8 \times 8$};
    \node[smallrect, below=3mm of B3] (B4) {\textbf{B} - $16 \times 4 \times 4$};
    \node[smallrect, highlight, below=3mm of B4] (R1) {Reshape};
    \node[smallrect, below=3mm of R1] (D1) {1D Conv};
    \node[smallrect, below=3mm of D1] (D2) {1D Conv};
    \node[smallrect, below=3mm of D2] (D3) {1D Conv};
    \node[smallrect, below=3mm of D3] (D4) {1D Conv};
    \node[smallrect, highlight, below=3mm of D4] (R2) {Reshape};
    \node[smallrect, below=3mm of R2] (L1) {Linear};
    \node[smallrect, below=3mm of L1] (L2) {Linear};
    \node[below=3mm of L2] (L3) {};
    
    \node[smallrect, cond, draw, dashed, align=center, right=of Video] (C3desc) {$c^{mask}$};

    \node[imgrect, cond, dashed, below=of C3desc] (C31) {\includegraphics[width=2.2cm]{fig/masks/mask_64x64.pdf}};
    
    \node[smallrect, cond, draw, dashed, align=center, right=of R2] (Cnum) {$c^{num}$};
    \draw[darrow] (C31.west) -- (Video.east);
    \draw[darrow] (Cnum.west) -- (R2.east);
    
    \foreach \from/\to in {Video/1x1_conv, 1x1_conv/B1, B1/B2, B2/B3, B3/B4, B4/R1, R1/D1, D1/D2, D2/D3, D3/D4, D4/R2, R2/L1, L1/L2, L2/L3}resolution
        \draw[arrow] (\from) -- (\to);
    \end{tikzpicture}
    }

        \caption{Discriminator $D^\textrm{lr}$\label{fig:arch_lr_disc}}
    \end{subfigure}
    \label{fig:arch_lr}
    \caption{Architectures of the conditioned low-resolution generator (left, compare with Figure~3 in \cite{Brooks2022_lvg}) and discriminator (right, compare with Figure~16 \cite{Brooks2022_lvg}).}
\end{figure}

Finally, we modify the last layer, i.e., a 3D convolution, of $G^\textrm{lr}$ that produces the output to accept $\tilde{x}^{(L)}$ instead of ${x}^{(L)}$. Hence, we have 

\begin{equation}
    x^{\mathrm{OUT}}_\textrm{lr} := \tanh\left( \mathrm{Conv3D}(\tilde{x}^{(L)}) \right) \in [-1,1]^{C_{\textrm{SIM}} \times \mathcal{T} \times H^\textrm{lr} \times W^\textrm{lr}}.
    \label{eq:lres_gen_outputLayer_cond}
\end{equation}
Thus, concluding our modifications to the low-resolution generator. In Figure~\ref{fig:arch_lr_gen}, the architecture of our conditioned low-resolution generator is depicted.

\paragraph{Low-resolution GAN: Discriminator $D^\textrm{lr}$.}

In \cite{Brooks2022_lvg}, the initial convolution layer of the discriminator $D^\textrm{lr}$ applies a channel expansion on a given input video $x'\in\mathbb{R}^{C_{\textrm{SIM}}\times \mathcal{T}\times H^\textrm{lr} \times W^\textrm{lr}}$. 
To incorporate $c^{\textrm{mask}}$ we first follow \eqref{eq:scaled_mask_cond} to expand it to $c^{\textrm{mask}, (L)}$ matching the $(\mathcal{T}, H, W)$ dimensions of $x'$. In a next step, we concatenate $c^{\textrm{mask}, (L)}$ and $x'$ to obtain
\begin{equation}
    \bar{x}=\left[x \| c^{\textrm{mask}, (L)} \right],
    \label{eq:lres_disc_mask_concat_cond}
\end{equation}
 as the modified conditioned input to the first convolutional layer of $D^\textrm{lr}$, for which we adjust the input dimension to $C_{\textrm{SIM}} + n_\textrm{mask}$ instead of $C_{\textrm{SIM}}$ channels. Hence, we have
\begin{equation}
    x^{(0)} = \phi\left( \mathrm{Conv3D}_{1 \times 1 \times 1}(\bar{x}) \right), \quad x^{(0)} \in \mathbb{R}^{128 \times \mathcal{T} \times H^\textrm{lr} \times W^\textrm{lr}}.
    \label{eq:lres_discriminator_initial_cond}
\end{equation}
Note that the dimension of $x^{(0)}$ stays unchanged after modifications, i.e., it can be passed to subsequent blocks without further modifications. Next, by combining spatial and channel dimensions, features are then reshaped followed by 4 temporal 1D convolutional layers. Their output is then flattened to 
\begin{equation}
    h^\textrm{flat}\in\mathbb{R}^{\mathcal{T}\cdot C_{\textrm{SIM}} \cdot H^\textrm{lr} \cdot W^\textrm{lr}},
    \label{eq:lres_disc_flatten}
\end{equation}
and further processed by 2 linear layers $V_1\in \mathbb{R}^{m\times \mathcal{T}\cdot C_{\textrm{SIM}} \cdot H^\textrm{lr} \cdot W^\textrm{lr}}$ and $V_2\in \mathbb{R}^{1\times m}$, where $m$ is the hidden width of the first fully connected layer, to produce output logits.

To incorporate $c^{\textrm{num}}$ we concatenate it to the flattened extracted features $h^\textrm{flat}$ and obtain
\begin{equation}
    \bar{h}^\textrm{flat}=\left[h^\textrm{flat} \| c^\textrm{num}\right].
    \label{eq:lres_disc_flatten_cond}
\end{equation}

We also adjust $V_1$ such that $V_1\in \mathbb{R}^{m\times \mathcal{T}\cdot C_{\textrm{SIM}} \cdot H^\textrm{lr} \cdot W^\textrm{lr} + n_\textrm{num}}$ instead of $V_1\in \mathbb{R}^{m\times \mathcal{T}\cdot C_{\textrm{SIM}} \cdot H^\textrm{lr} \cdot W^\textrm{lr}}$ yielding 
\begin{align}
\begin{split}
    z &= \phi(V_1 \bar{h}^\textrm{flat} + b^{V_1}),\quad \phi = \mathrm{Leaky ReLU}\\
    \hat{y} &= V_2 z + b^{V_2} \in \mathbb{R},
\end{split}
\label{eq:lres_disc_linLayers_cond}
\end{align}
where $\hat{y}$ is the scalar logit for the discriminator $D^\textrm{lr}$. The architecture of the conditioned low-resolution discriminator is illustrated in Figure~\ref{fig:arch_lr_disc}.

\paragraph{Super-resolution GAN: Generator $G^\textrm{sr}$.}
The super-resolution GAN in \cite{Brooks2022_lvg} with generator $G^\textrm{sr}$ and discriminator $D^\textrm{sr}$ incorporates a conditioning on temporal context for a temporal window size of $2Q$, i.e., to generate a super-resolution frame $x^\textrm{sr}_t\in \mathbb{R}^{ C_{\textrm{SIM}} \times H^\textrm{sr} \times W^\textrm{sr}}$ a temporal conditioning window
\begin{equation}
    \{ x_{t-Q}^{\mathrm{lr}}, \dots, x_t^{\mathrm{lr}}, \dots, x_{t+Q}^{\mathrm{lr}} \} \in \mathbb{R}^{(2Q+1) \times C_{\textrm{SIM}} \times H^\textrm{sr} \times W^\textrm{sr}}
    \label{eq:sres_temp_window}
\end{equation}
is used. Here, $H^\textrm{sr}$ and $W^\textrm{sr}$ are the height and width of super-resolution frames. Note that no latent code is used for super-resolution synthesis. The generator $G^\textrm{sr}$ is entirely conditioned on low-resolution frames $x_i^\textrm{lr}$. Each super-resolution frame $x_t^\textrm{sr}$ is generated independently per frame, i.e., there is no temporal modeling in $G^\textrm{sr}$. Temporal coherence is incorporated indirectly through the temporal window \eqref{eq:sres_temp_window}.

Since the super-resolution GAN already incorporates conditioning information, we utilize the same framework to incorporate $c^\textrm{mask}\in \mathbb{R}^{ n_\textrm{mask} \times H^\textrm{sr} \times W^\textrm{sr}}$ into the architecture. We omit $c^\textrm{num}$ since, similar to temporal coherence, it is already implicitly represented in the low-resolution frames. This choice allows us to modify the architecture with only minimal changes. In the following, we give an overview of our modifications to the super-resolution generator  $G^\textrm{sr}$.

For the super-resolution generator each of the frames $\{ x_{t-Q}^{\mathrm{lr}}, \dots, x_t^{\mathrm{lr}}, \dots, x_{t+Q}^{\mathrm{lr}} \}$ are up-sampled using a fixed operator \( \mathcal{U} \)
\begin{equation}    
\tilde{x}_{t-k}^{\mathrm{lr}} := \mathcal{U}(x_{t-k}^{\mathrm{lr}}) \in \mathbb{R}^{C_{\textrm{SIM}} \times H^\textrm{sr} \times W^\textrm{sr} } \quad \text{for } k = -Q,-Q+1, \dots, Q
\label{eq:super_res_upsampled_context}
\end{equation}
and then concatenated along the channel dimension
\begin{equation}
c^G_t := \left[ \tilde{x}_{t-Q}^{\mathrm{lr}}\| \dots \| \tilde{x}_{t+Q}^{\mathrm{lr}} \right] \in \mathbb{R}^{C_{\textrm{SIM}}(2K+1) \times H^\mathrm{sr} \times W^\mathrm{sr}}.
    \label{eq:super_res_gen_concat}
\end{equation}
This vector $c^G_t$ serves as the conditioning input to the super-resolution generator \( G^{\mathrm{sr}} \) which is a StyleGAN2/3-style \cite{Karras2019AnalyzingAI_SG2, Karras2021_SG3} convolutional network to produce the super-resolution output:
\[
x_t^{\mathrm{sr}} = G^{\mathrm{sr}}(c^G_t) \in \mathbb{R}^{ C_{\textrm{SIM}} \times H^{\mathrm{sr}} \times W^{\mathrm{sr}}}
\]
To introduce our conditioning information, we use the same fixed operator $\mathcal{U}$ to ensure dimensional compatibility, i.e.,
\begin{equation}
    \tilde{c}^\textrm{mask} := \mathcal{U}(c^\textrm{mask}) \in \mathbb{R}^{n_\textrm{mask} \times H^\mathrm{sr} \times W^\mathrm{sr}}.
\end{equation}
In a next step, we concatenate $\tilde{c}^\textrm{mask}$ with  \eqref{eq:super_res_gen_concat}, and obtain
\begin{equation}
    c^G_t := c^G_{t, \textrm{mask}} :=\left[ \tilde{x}_{t-Q}^{\mathrm{lr}}\| \dots \| \tilde{x}_{t+Q}^{\mathrm{lr}} \| \tilde{c}^\textrm{mask}\right] \in \mathbb{R}^{C_{\textrm{SIM}}(2K+1) + n_\textrm{mask} \times H^\mathrm{sr} \times W^\mathrm{sr}}.
    \label{eq:super_res_gen_concat_cond}
\end{equation}

The generator \( G^{\mathrm{sr}} \) in \cite{Brooks2022_lvg} consists of $L'$  convolutional blocks \( \mathcal{B}_\ell \), for $\ell=1,\dots, L'$, where 2D convolutions and optionally up-sampling is applied before passing through {Leaky ReLU}. Let the initial input be
\begin{equation}
x^{(0)} := \text{Conv}(c^G_t),
\label{eq:super_res_gen_initial}
\end{equation}
where we adjusted the input dimensions w.r.t.\ \eqref{eq:super_res_gen_concat_cond}. 
Then for \( \ell = 1, \dots, L' \), it is
\[
x^{(\ell)} = \mathcal{B}_\ell(x^{(\ell-1)}).
\]
Finally, a 2D convolution followed by a {tanh} nonlinearity yields 
\[
x_t^{\mathrm{sr}} = \tanh\left( \text{Conv}_{\text{out}}(x^{(L')}) \right).
\]

\paragraph{Super-resolution GAN: Discriminator $D^\textrm{sr}$.}
The super-resolution discriminator $D^\textrm{sr}$ is a StyleGAN2-style discriminator, see \cite{Karras2019AnalyzingAI_SG2}, that concatenates the up-sampled temporal context \eqref{eq:super_res_upsampled_context} with a super-resolution video frame $x^\textrm{sr}_t\in\mathbb{R}^{C_{\textrm{SIM}} \times H^\mathrm{sr} \times W^\mathrm{sr}}$ to obtain the conditioning input of the discriminator

\begin{equation}
    c^D_t  :=\left[ \tilde{x}_{t-Q}^{\mathrm{lr}}\| \dots \| \tilde{x}_{t+Q}^{\mathrm{lr}} \| x^\textrm{sr}_t\right] \in \mathbb{R}^{C_{\textrm{SIM}}(2K+2) \times H^\mathrm{sr} \times W^\mathrm{sr}}.
    \label{eq:super_res_disc_concat}
\end{equation}
For our modifications we further concatenate $c^\textrm{mask}$ with \eqref{eq:super_res_disc_concat} and obtain the conditioning input
\begin{equation}
c^D_t  := c^D_{t, \textrm{mask}} :=\left[ \tilde{x}_{t-Q}^{\mathrm{lr}}\| \dots \| \tilde{x}_{t+Q}^{\mathrm{lr}} \| x^\textrm{sr}_t \| c^\textrm{mask}\right] \in \mathbb{R}^{C_{\textrm{SIM}}(2K+2) + n_\textrm{mask} \times H^\mathrm{sr} \times W^\mathrm{sr}}.
    \label{eq:super_res_disc_concat_cond}
\end{equation}
This input is then passed through a stack of $\hat{L}$ 2D residual convolutional blocks $\mathcal{D}_\ell$, $\ell=1,\dots,\hat{L}$, where down-sampling is applied followed by {Leaky ReLU}. Note that corresponding to our modification \eqref{eq:lres_disc_mask_concat_cond} we increase the input channel dimension of the first block $\mathcal{D}_1$ from $C_{\textrm{SIM}}(2K+2)$ to $C_{\textrm{SIM}}(2K+2) + n_\textrm{mask}$. Thus, we obtain

\begin{equation}
f^{(0)} := c_t, \quad f^{(\ell)} := \mathcal{D}_\ell(f^{(\ell-1)}) \in \mathbb{R}^{ C_{\ell} \times H^{\mathrm{sr}}_{\ell} \times W^{\mathrm{sr}}_{\ell}}, \quad \ell = 1, \dots, \hat{L}.
    \label{eq:super_res_disc_blocks}
\end{equation}
Which is then flattened in a next step

\[
f^\textrm{flat} := \text{Flatten}(f^{(\hat{L})}) \in \mathbb{R}^{C_{\ell} \cdot H^{\mathrm{sr}}_{\ell} \cdot W^{\mathrm{sr}}_{\ell}}
\]
and passed to two dense layers $W_1 $ and $W_2 $

\begin{align}
\begin{split}
    z &= \phi(W_1 f^\textrm{flat} + b^{W_1}),\quad \phi = \mathrm{LeakyReLU},\\
    \hat{y} &= W_2 z + b^{W_2} \in \mathbb{R},
\end{split}
\label{eq:super_res_disc_linLayers}
\end{align}
The final logit score is then given by
\[
D^{\mathrm{sr}}(c^D_t ) := \hat{y} \in \mathbb{R}.
\]

\section{Data Set and Evaluation Metrics}
\label{sec:dataset_and_evaluation}
\paragraph{Data set.} The data for model training and testing was provided by a total of 10 transient CFD-simulations using Ansys Fluent (2024R1) as the CFD-software. Nine of these simulations could be mirrored with respect to the $y$-axis resulting in effectively 19 simulations for training purposes. Each simulation was describing 1 hour of flow-time exported in 3600 frames (1-second steps with 50 iterations/second, \textit{Coupled} scheme for pressure-velocity coupling, \textit{PRESTO!} pressure discretization and otherwise default solver settings). We prepared a simple 2D setup describing two solid circles positioned in a square fluid domain as visualized in Figure \ref*{fig:dataset_parameters}. This configuration was supposed to describe a cross section of two electrical conductors immersed in some kind of insulating fluid, for example a pressurized gas.\par
\begin{figure}[h!]
    \centering
    \includegraphics[width=\textwidth]{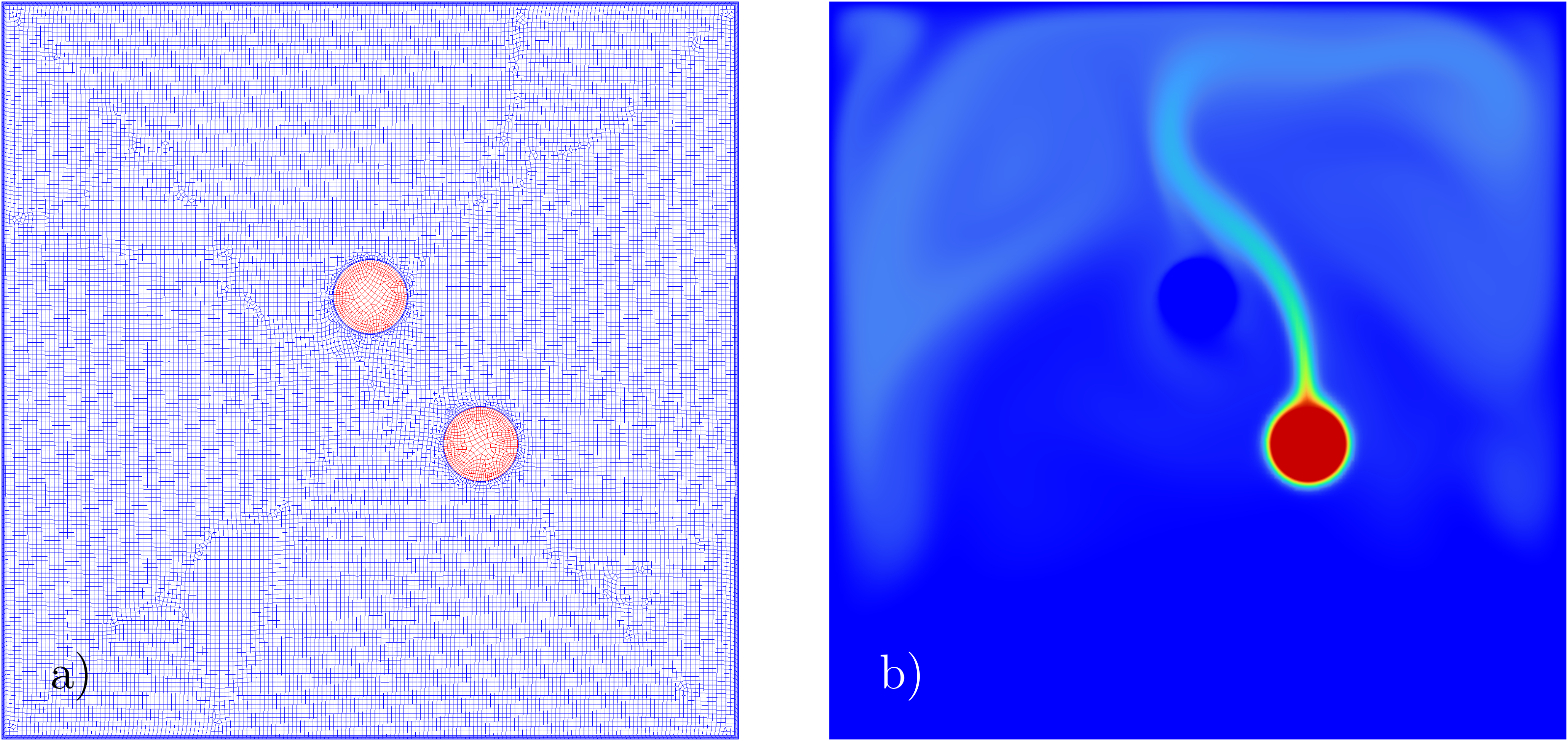}
    \caption{Mesh (a) and solution snapshot (b) of a CFD-simulation from the dataset. In this case, only on the bottom circle a positive heat source was defined. The upper circle is passive.}
    \label{fig:mesh} 
\end{figure}

Starting from a spatially constant temperature field $T\equiv T_\mathrm{amb}$, the circles are heating up due to heat sources defined on them, and will be cooled by the surrounding fluid predominantly via convection and to some lower extend by diffusion (no radiation model). More specifically, the circles are heating up the fluid around them, lowering its density and letting it rise up due to the emerging buoyancy forces (see also Figure \ref{fig:mesh}b). Note that gravity is considered in this model ($g=9.81\,\mathrm{m/s^2}$), acting in negative $y$-direction. On the boundaries of the box, Dirichlet-type boundary conditions with $T=T_\mathrm{amb}$ are imposed, allowing energy to exit the system.\par
For the fluid model we used the compressible 2D Navier-Stokes equations (including volume forces) coupled with the energy equation and closed by linking density and temperature via the ideal gas law, see Section~\ref{sec:governing_equations}.\par
Note that due to the 2D-setup no turbulence model was used. Also note that in all following plots and tables $T$, $u$, $v$, and $p$ are expressed in SI units, i.e., $T$ is in Kelvin, $u$ and $v$ are in m/s, and $p$ is in Pascals. The pressure $p$ represents a fluctuation about the operating pressure $p_{\mathrm{op}} = 101325~\mathrm{Pa}$. The material parameters for the two different domains are listed in Table~\ref{tab:mat-parameters}. Additionally, a molecular weight of $M=29\,\mathrm{kg/kmol}$ was used for the fluid in the ideal gas law.\par
\renewcommand{\arraystretch}{1.2}
\begin{table}[h]  
    \centering  
    \begin{tabular}{lccccc}  
        \hline  
        & $\rho\;\left[\mathrm{kg/m^3}\right]$ & $c_p\;\left[\mathrm{J/(kg\,K)}\right]$ & $k\;\left[\mathrm{W/(m\,K)}\right]$ & $\mu\;\left[\mathrm{kg/(m\,s)}\right]$ \\  
        \hline  
        \textbf{Solid} & $2720$ & $870$ & $200$ & --- \\  
        \textbf{Fluid} & ideal gas & $1000$ & $0.024$ & $1.8\times10^{-5}$ \\  
        \hline  
    \end{tabular}  
    \caption{Material properties for solid and fluid domains.}  
    \label{tab:mat-parameters}
\end{table} 
\renewcommand{\arraystretch}{1.0}
We parameterized the CFD-model with nine parameters, see Table~\ref{tab:model-parameters}. Six of these nine parameters ($x_A, y_A, r_A, x_B, y_B, r_B$) describe the size and position of the two circles. Two parameters ($\dot{Q}_A, \dot{Q}_B$) describe the heat sources on the two circles. And finally, the last parameter ($T_\mathrm{amb}$) describes the ambient temperature which is applied to the outer boundary of the fluid domain as a Dirichlet boundary condition. In the dataset considered in this paper, only one of those parameters was varied - the horizontal position of the lower circle $x_A$. The parameter values and the range for $x_A$ are given in Table~\ref{tab:model-parameters}\par 
To create the setups for our 10 CFD-simulations we used a script that automatically generated the CFD-mesh for the given $x_A$ value and then ran the simulation. The used meshes in all simulations were of medium fidelity and had cell counts between 30 and 40 thousand, see also Figure \ref{fig:mesh}.
\newpage
\begin{figure}[h!]  
    \centering  
    \hspace{0.025\textwidth}
    \begin{minipage}{0.47\textwidth}  
        \centering  
        \includegraphics[width=\textwidth]{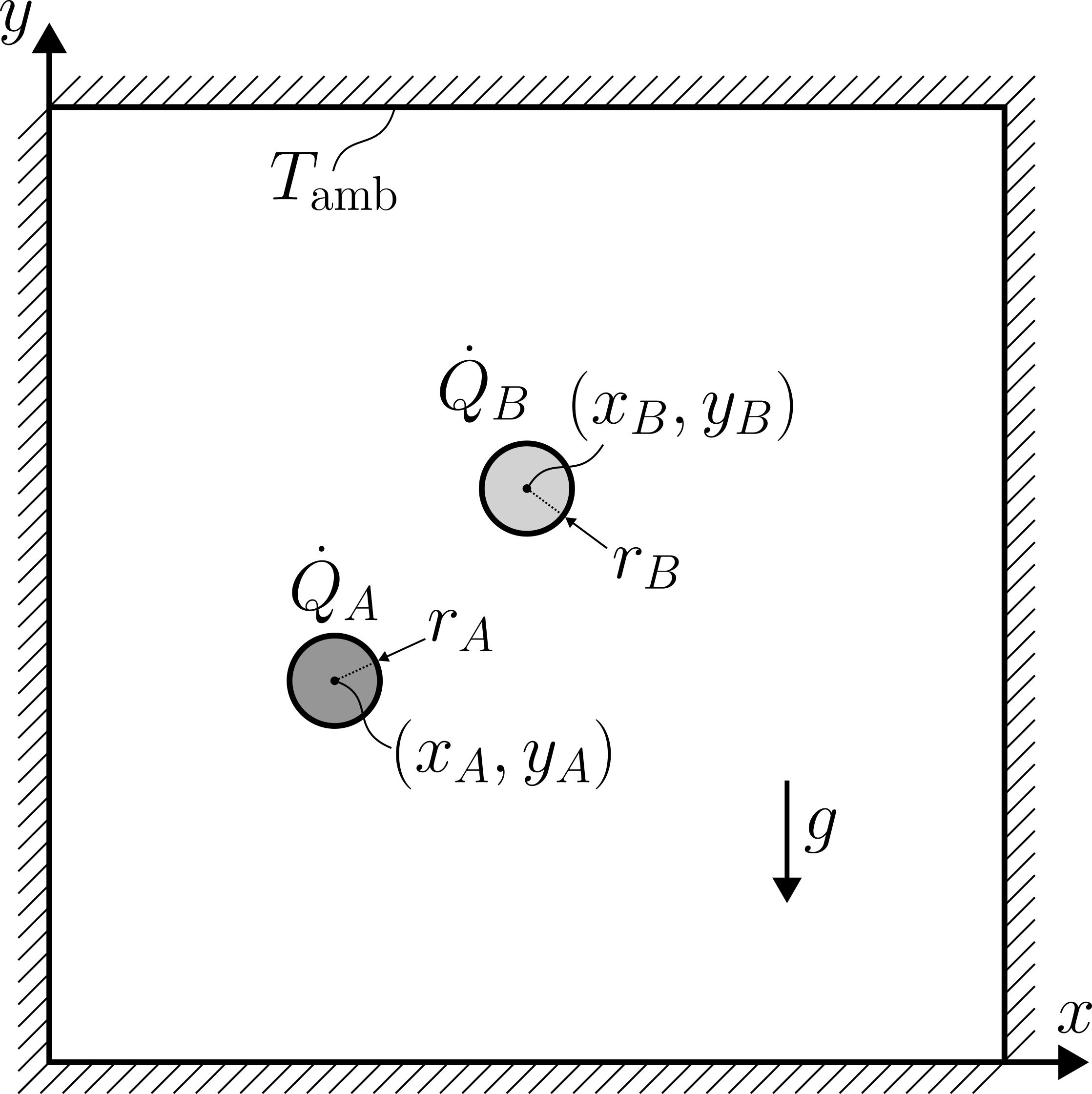}  
        \caption{Parameterization of the reference problem}  
        \label{fig:dataset_parameters} 
    \end{minipage}%
    \hspace{0.025\textwidth} 
    \begin{minipage}{0.45\textwidth}  
        \centering
        \small
        \renewcommand{\arraystretch}{1.2}  
            \begin{tabular}{lcc}  
            \hline  
            \textbf{Name} & \textbf{Unit} & \textbf{Range} \\  
            \hline  
            $T_\mathrm{amb}$ & $\mathrm{K}$ & 300 \\  
            $\dot{Q}_A$      & $\mathrm{W}$ & 100 \\  
            $\dot{Q}_B$      & $\mathrm{W}$ & 0 \\  
            $r_A$            & $\mathrm{m}$ & 0.05 \\  
            $r_B$            & $\mathrm{m}$ & 0.05 \\  
            $x_A$            & $\mathrm{m}$ & $\lbrack 0.2, 0.85 \rbrack$ \\  
            $y_A$            & $\mathrm{m}$ & 0.4 \\  
            $x_B$            & $\mathrm{m}$ & 0.5 \\  
            $y_B$            & $\mathrm{m}$ & 0.6 \\  
            \hline  
        \end{tabular}   
        \label{tab:param-ranges}  
        \renewcommand{\arraystretch}{1.0}
        \captionof{table}{Parameter values and ranges}  
        \label{tab:model-parameters}  
    \end{minipage}  
\end{figure}

\paragraph{Evaluation metrics.}
Following \cite{drygala2023generalizationcapabilitiesconditionalgan, drygala2024comparisongenerativelearningmethods, Drygala_2022, ross2025worldmodelssuccessfullylearn}, we compute {Pearson correlation curves}, see, e.g. \cite{Drygala_2022}, to evaluate the performance of our architecture relative to simulation data. We consider two types of correlations. We compute {spatial correlations} within each frame using cropped boxes $B$ and {temporal correlations} at fixed spatial points, computed over a sequence of frames. In Figure~\ref{fig:corr_setup020}, for one experiment the choice of $B$ (gray box) and fixed points (green dots) are illustrated. The placement and size of $B$, as well as the location of the fixed points, are chosen to capture the most relevant regions within a frame given the experiment. For every channel $c \in \{T,u,v,p\}$ we compute the correlations separately for simulation and generated data  $X^{(\nu, c)}$ with $\nu \in \{\text{sim}, \text{gen}\}$ and include confidence intervals using the Fisher $z$-transformation. For generated frames we also average over random seeds $s\in\{s_1,\dots,s_P\}$. Furthermore, we investigate the {mean values} of each frame of generated and simulated frames, i.e., the temporal behavior of the mean values, and the {variance} of frames computed  over a sequence of frames.


To compute spatial correlation, let $x^{(\nu, c)}_{t,i,j}$ denote the value at time $t\in\{0,\dots, \mathcal{T}^{(\nu)}\}$, row $i\in\{1,\dots, H_B\}$, and column $j\in\{1,\dots, W_B\}$, in a cropped box $B$ for 
$\nu \in \{\text{sim}, \text{gen}\}$  and $c \in \{T,u,v,p\}$. For two sequences extracted from $X^{(\nu, c)}$ along rows
\[
\{x^{(\nu, c)}_{t,i_0,j}\}_{j=1}^{W_B} \quad \text{and} \quad
\{x^{(\nu, c)}_{t,i_0+d,j}\}_{j=1}^{W_B},
\]
the Pearson correlation coefficient is defined as
\begin{align}
    \begin{split}
        \rho^{(\nu, c)}_{i_0, d, t} & := \rho^{(\nu, c)}(\{x^{(\nu, c)}_{t,i_0,j}\}_{j=1}^{W_B} , \{x^{(\nu, c)}_{t,i_0+d,j}\}_{j=1}^{W_B}) \\ &=
\frac{\sum_{j=1}^{W_B} 
\big(x^{(\nu, c)}_{t,i_0,j} - \mu^{(\nu, c)}_{i_0}\big) 
\big(x^{(\nu, c)}_{t,i_0+d,j} - \mu^{(\nu, c)}_{i_0+d}\big)}
{\sqrt{\sum_{j=1}^{W_B} \big(x^{(\nu, c)}_{t,i_0,j} - \mu^{(\nu, c)}_{i_0}\big)^2} \;
 \sqrt{\sum_{j=1}^{W_B} \big(x^{(\nu, c)}_{t,i_0+d,j} - \mu^{(\nu, c)}_{i_0+d}\big)^2}} ,
    \label{eq:pearson_row}
    \end{split}
\end{align}

where
\begin{equation*}
\mu^{(\nu, c)}_{i_0} = \frac{1}{W_B}\sum_{j=1}^{W_B} x^{(\nu, c)}_{t,i_0,j}, 
\quad
\mu^{(\nu, c)}_{i_0+d} = \frac{1}{W_B}\sum_{j=1}^{W_B} x^{(\nu, c)}_{t,i_0+d,j}.
\end{equation*}

Note that here $d$ is the relative offset from the reference row $i_0$ in a cropped box $B$. Furthermore, we average over all time steps, i.e., frames,
\begin{equation}
    \bar{\rho}^{(\nu, c)}_{i_0, d}=\frac{1}{\mathcal{T}^{(\nu)}}\sum_{t=0}^{\mathcal{T}^{(\nu)}}\rho^{(\nu, c)}_{i_0, d, t}.
    \label{eq:pearson_row_mean}
\end{equation}
For generated data, \eqref{eq:pearson_row_mean} is also averaged over the seeds $s\in\{s_1,\dots,s_P\}$. The spatial correlation w.r.t.\ the columns $\rho^{(\nu, c)}_{j_0, d, t}$ and $\bar{\rho}^{(\nu, c)}_{j_0, d}$ are computed analogous to \eqref{eq:pearson_row} and \eqref{eq:pearson_row_mean}, i.e., in $B$ the relative offset $d$ is applied to the column index $j$ for an initial $j_0$.

For temporal auto-correlations at a fixed spatial point $(i,j)$ and a temporal
lag $\tau \in \mathbb{Z}$, we fix a window length $M \in \mathbb{N}$.
Given the total number of frames $\mathcal{T}^{(\nu)}$, the admissible window
starting indices are
\[
t_0 \in \{0,  \dots, \mathcal{T}^{(\nu)} - M - |\tau|\},
\]
ensuring that both sequences below have equal length $M$. For any such $t_0$, we consider the sequences
\begin{equation*}
\{x^{(\nu, c)}_{t_0+t,\, i,j}\}_{t=0}^{M-1}
\quad\text{and}\quad
\{x^{(\nu, c)}_{t_0+t+\tau,\, i,j}\}_{t=0}^{M-1}.   
\end{equation*}
The corresponding temporal Pearson correlation for a window of length $M$ is defined as
\begin{align*}
    \begin{split}
        \rho^{(\nu, c)}_{t_0,\, \tau,\, i,j}
        &:=
        \rho^{(\nu, c)}\!\left(
        \{x^{(\nu, c)}_{t_0+t,\, i,j}\}_{t=0}^{M-1},
        \{x^{(\nu, c)}_{t_0+t+\tau,\, i,j}\}_{t=0}^{M-1}
        \right)
        \\[4pt]
        &=
        \frac{
        \sum_{t=0}^{M-1}
        \big(x^{(\nu, c)}_{t_0+t,\, i,j} - \mu^{(\nu, c)}_{t_0,i,j}\big)
        \big(x^{(\nu, c)}_{t_0+t+\tau,\, i,j} - \mu^{(\nu, c)}_{t_0+\tau,i,j}\big)
        }
        {
        \sqrt{
        \sum_{t=0}^{M-1}
        \big(x^{(\nu, c)}_{t_0+t,\, i,j}
        - \mu^{(\nu, c)}_{t_0,i,j}\big)^2}
        \;
        \sqrt{
        \sum_{t=0}^{M-1}
        \big(x^{(\nu, c)}_{t_0+t+\tau,\, i,j}
        - \mu^{(\nu, c)}_{t_0+\tau,i,j}\big)^2}
        } ,
    \end{split}
\end{align*}
with 
\begin{equation*}
\mu^{(\nu, c)}_{t_0,i,j}
= \frac{1}{M}\sum_{t=0}^{M-1} x^{(\nu, c)}_{t_0+t,\, i,j},
\qquad
\mu^{(\nu, c)}_{t_0+\tau,i,j}
= \frac{1}{M}\sum_{t=0}^{M-1} x^{(\nu, c)}_{t_0+t+\tau,\, i,j}.
\end{equation*}
There are
\begin{equation*}
N^{(\nu)}_{\tau}
= \mathcal{T}^{(\nu)} - M - |\tau| + 1
\end{equation*}
valid windows.  By averaging over all $N^{(\nu)}_{\tau}$ windows we obtain
\begin{equation}
\bar{\rho}^{(\nu, c)}_{\tau,i,j}
=
\frac{1}{N^{(\nu)}_{\tau}}
\sum_{t_0=0}^{N^{(\nu)}_{\tau}-1}
\rho^{(\nu, c)}_{t_0,\, \tau,\, i,j}.
\label{eq:pearson_temporal_average}
\end{equation}
Further, for the generated frames we average \eqref{eq:pearson_temporal_average} over all seeds $s\in\{s_1,\dots,s_P\}$.

We also compute confidence intervals for both types of correlations via the Fisher $z$-transformation \cite{fisher_z_1915}. For a correlation value $\rho$, the Fisher $z$-transform is given by

\begin{equation*}    
z = \frac{1}{2} \ln \frac{1+\rho}{1-\rho},
\end{equation*}
with standard error $\mathrm{SE} = 1/\sqrt{N-3}$ for $N$ samples.  Confidence intervals in $z$-space are computed as
\begin{equation*} 
z_{\textrm{lower}}= z - z_{\alpha/2}\,\mathrm{SE},\qquad z_{\textrm{upper}} = z + z_{\alpha/2}\,\mathrm{SE},
\end{equation*}
and then transformed back to correlation space via
\begin{equation*} 
\rho_{\textrm{lower}} = \frac{e^{2 z_{\textrm{lower}}} - 1}{e^{2 z_{\textrm{lower}}} + 1},\qquad \rho_{\textrm{upper}} = \frac{e^{2 z_{\textrm{upper}}} - 1}{e^{2 z_{\textrm{upper}}} + 1}.
\end{equation*}
Note that for simulation data we consider $|\{0,\dots,\mathcal{T}^{(\textrm{sim})}\}|$ samples, i.e., all frames, for spatial and $N^{(\textrm{sim})}_{\tau}$ samples, i.e., all windows, for temporal correlations. All confidence intervals corresponding to the generated frames are computed w.r.t.\ the $P$ seeds $\{s_1,\dots,s_P\}$.

In addition to correlation-based metrics, we also investigate the  consistency of the transient behavior for generated and simulated data by comparing {mean curves}.  
For each channel $c \in \{T,u,v,p\}$, we compute the field mean value of each frame
\begin{equation}
\bar{x}^{(\nu, c)}_{t} = \frac{1}{H\,W} 
\sum_{i=1}^{H}\sum_{j=1}^{W} x^{(\nu, c)}_{t,i,j},
\quad t = 1,\dots,\mathcal{T}^{(\nu)}.
\label{eq:mean_frame_sim}
\end{equation}
For generated frames we additionally average over all $P$ seeds. Hence, we obtain sequences 
$\{\bar{x}^{(\nu, c)}_{t}\}_{t=1}^{\mathcal{T}^{(\nu)}}$ of mean values over time.

We furthermore  consider lag-dependent temporal variance using the same windowing as in the temporal correlation analysis. For a fixed window length $M\in\mathbb{N}$ and a temporal lag $\tau\in\mathbb{Z}$, and for any admissible window starting index $t_0$, we consider the shortened sequence
\begin{equation*}
\{x^{(\nu,c)}_{t_0+t,\,i,j}\}_{t=0}^{M-|\tau|-1}.
\end{equation*}
The corresponding temporal variance is defined as
\begin{equation*}
\mathrm{Var}^{(\nu,c)}_{t_0,\tau,i,j}
=
\frac{1}{M-|\tau|}
\sum_{t=0}^{M-|\tau|-1}
\bigl(x^{(\nu,c)}_{t_0+t,\,i,j}
-
\mu^{(\nu,c)}_{t_0,\tau,i,j}\bigr)^2,
\end{equation*}
with
\begin{equation*}
\mu^{(\nu,c)}_{t_0,\tau,i,j}
=
\frac{1}{M-|\tau|}
\sum_{t=0}^{M-|\tau|-1}
x^{(\nu,c)}_{t_0+t,\,i,j}.   
\end{equation*}
Averaging over all $N^{(\nu)}_{\tau}=\mathcal{T}^{(\nu)}-M-|\tau|+1$ admissible windows yields the mean variance
\begin{equation}
\overline{\mathrm{Var}}^{(\nu,c)}_{\tau,i,j}
=
\frac{1}{N^{(\nu)}_{\tau}}
\sum_{t_0=0}^{N^{(\nu)}_{\tau}-1}
\mathrm{Var}^{(\nu,c)}_{t_0,\tau,i,j}.
    \label{eq:mean_variance}
\end{equation}

Finally, this quantity is averaged over all spatial locations $(i,j)$ and, for generated data, over all $P$ seeds.

\section{Results}
\label{sec:results}
\paragraph{Training and generation.} To investigate the performance of our approach, for given conditioning information $c=(c^\textrm{num}, c^\textrm{mask})$ from the dataset we generate sequences of $256$ super-resolution frames via the low-resolution and super-resolution networks for $P=100$ different seeds. The low-resolution GAN was trained for 700000 steps while the super-resolution GAN was trained for 300000 steps; both were trained using two A100 GPUs. Furthermore, for inference we utilize one A100 GPU. To generate $256$ super-resolution frames the inference time for the low-resolution generator is $\approx 0.017$ seconds and for the super-resolution frame $\approx 0.00001$ seconds. Including the loading time of the generators and some data handling operations, the generation of a sequence takes $\approx 1.93$ seconds. Since we apply only static conditioning information without explicit control of the time steps, at the moment the networks may produce sequences where a re-initialization happens, i.e., the sequence jumps back in time to mimic an earlier part of the simulation. Hence, we omit sequences w.r.t\ to seeds $s_i$ in which a drop in the mean temperature, i.e., a re-initialization, occurs and generate a new sequence for $s_i+1$. Towards this end, we omit seeds for which the temperature drops by $0.01$ over $50$ frames. On average, we evaluate $170$ seeds to filter out $P=100$ seeds, resulting in a manageable computational cost given the inference time.

Recall that the networks were trained on $17$ simulations with different $x$ positions of the lower circle and that $2$ positions were omitted from training to observe the generalization capabilities of the approach. In the following, we present the results for reproduction of the $17$ simulation frames that were used during training, and the generalization capabilities for the $2$ omitted simulations.

\begin{figure}[htbp]
    \centering
    \begin{subfigure}[b]{0.48\textwidth}
        \centering
        \includegraphics[width=\textwidth]{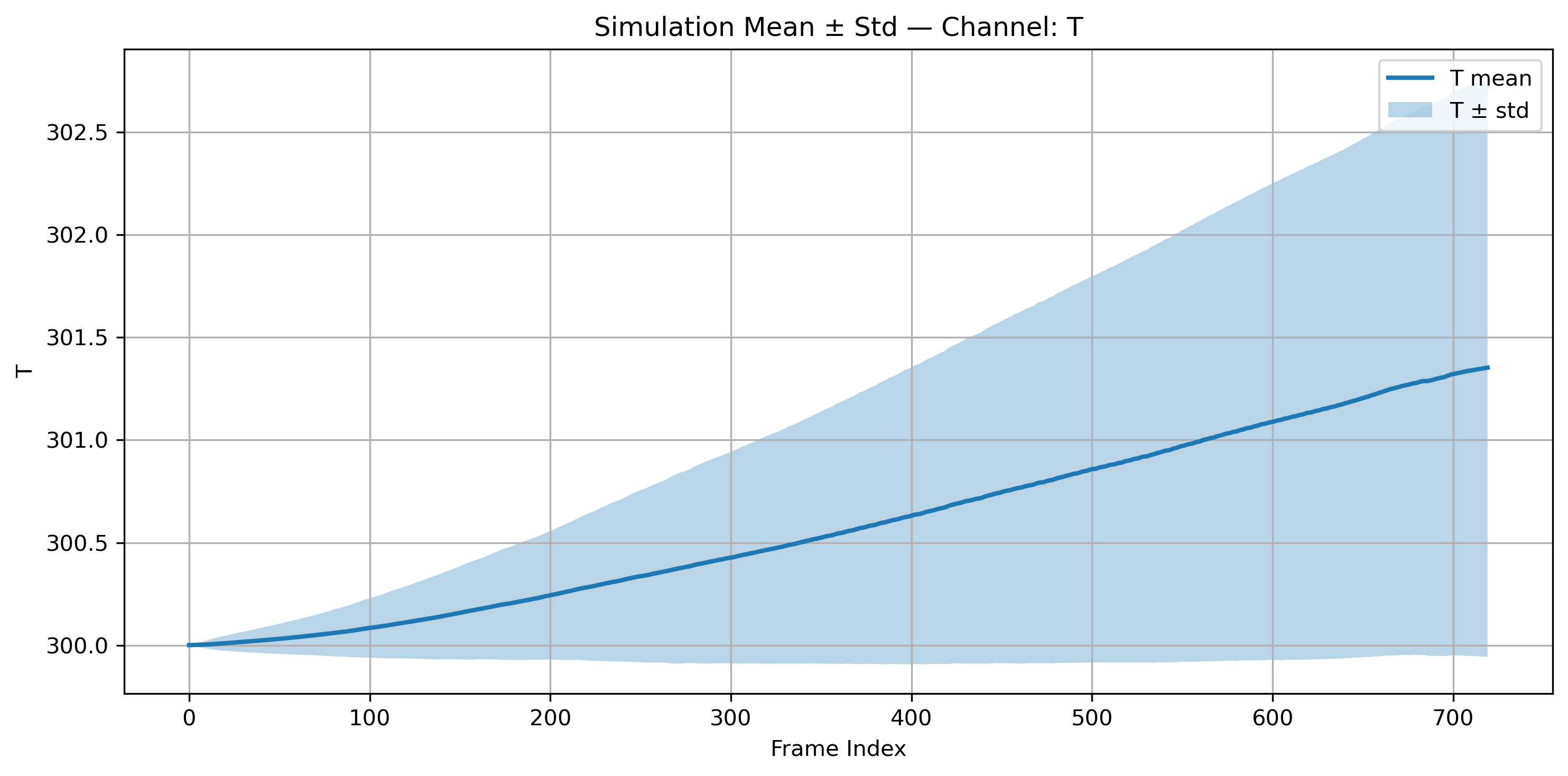}
        \caption{Mean and standard deviation of $T$.}
    \end{subfigure}
    \hfill
    \begin{subfigure}[b]{0.48\textwidth}
        \centering
        \includegraphics[width=\textwidth]{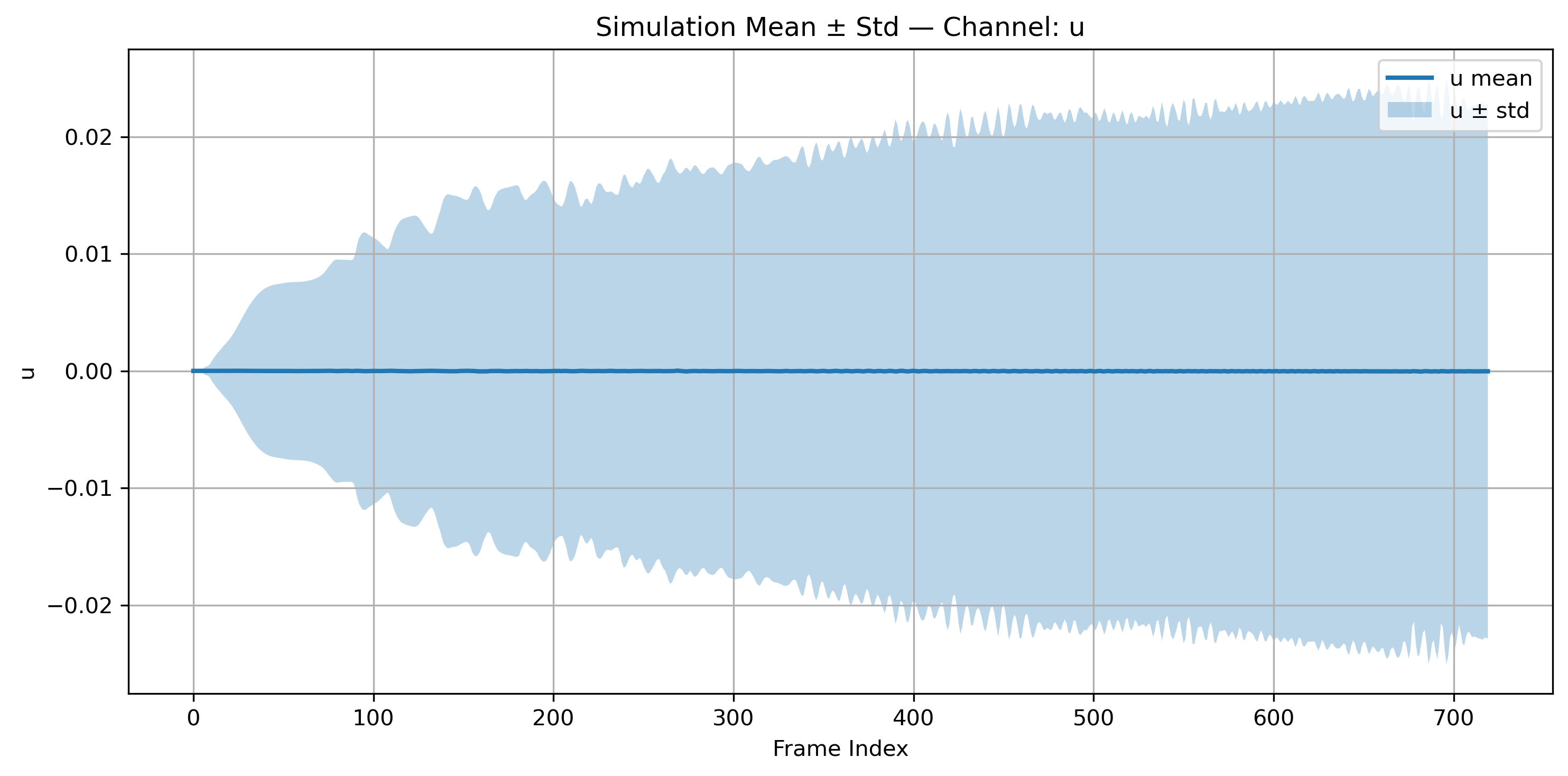}
        \caption{Mean and standard deviation of $u$.}
    \end{subfigure}
    \hfill
    \begin{subfigure}[b]{0.48\textwidth}
        \centering
        \includegraphics[width=\textwidth]{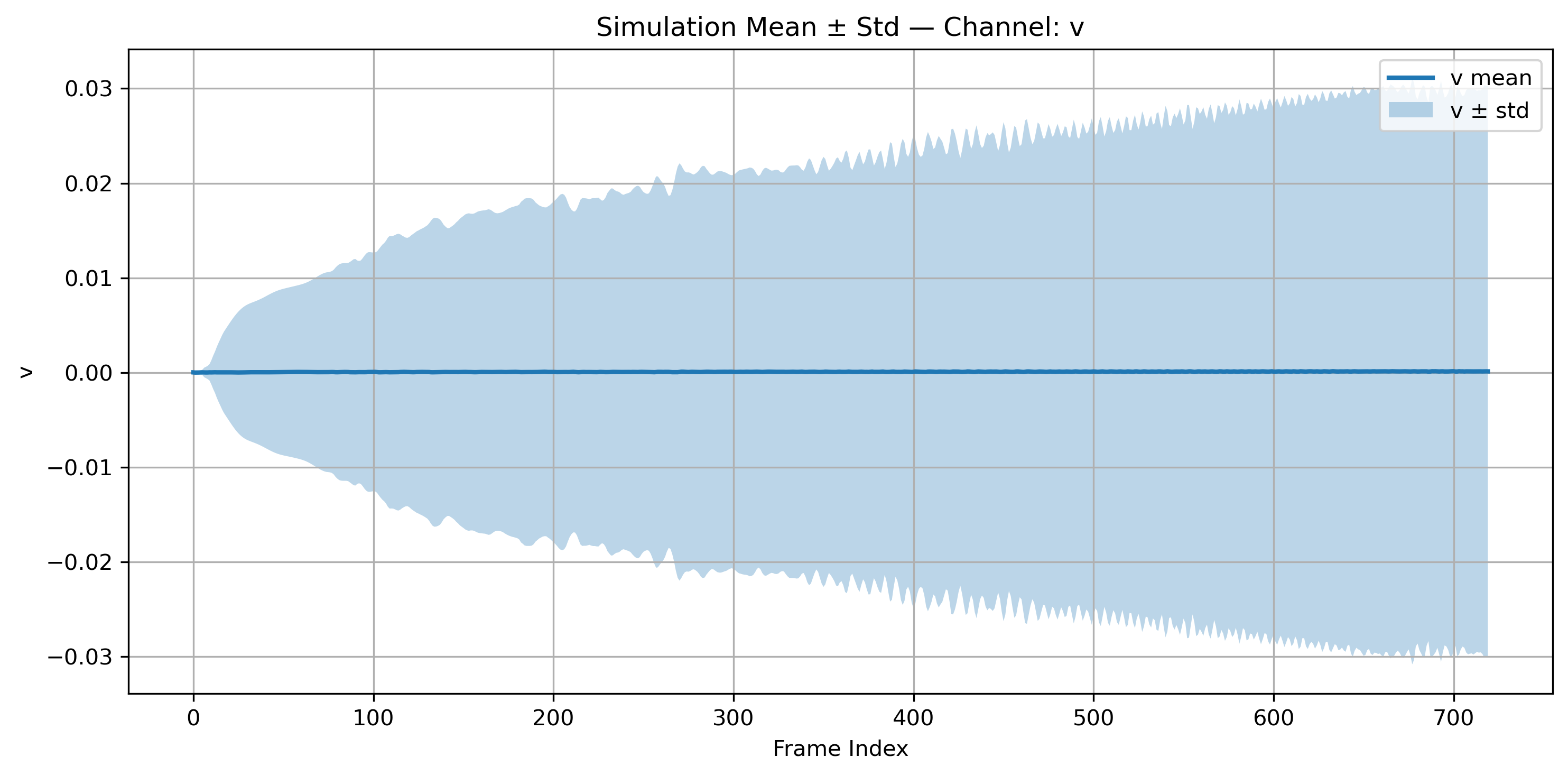}
        \caption{Mean and standard deviation of $v$.}
    \end{subfigure}
    \hfill
    \begin{subfigure}[b]{0.48\textwidth}
        \centering
        \includegraphics[width=\textwidth]{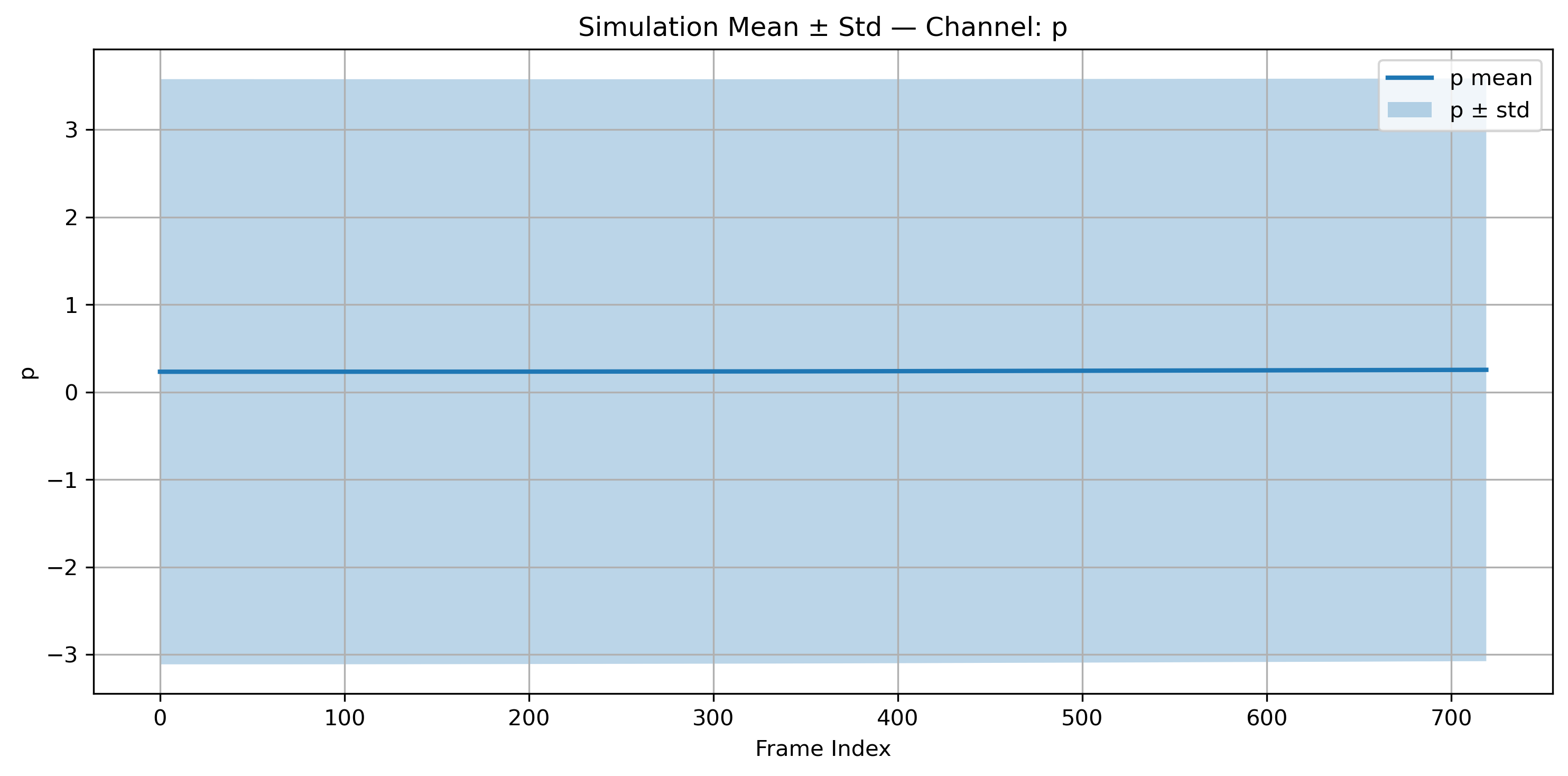}
        \caption{Mean and standard deviation of $p$.}
    \end{subfigure}

    \caption{Frame-wise simulation mean values (blue) and standard deviation (blue, shaded) of the fields $T,u,v,p$ of experiment $x_a=0.20$. }
    \label{fig:sim_mean_std}
\end{figure}

\begin{figure}[htbp]
\centering
    \begin{subfigure}[b]{0.48\textwidth}
        \centering
        \includegraphics[width=\textwidth]{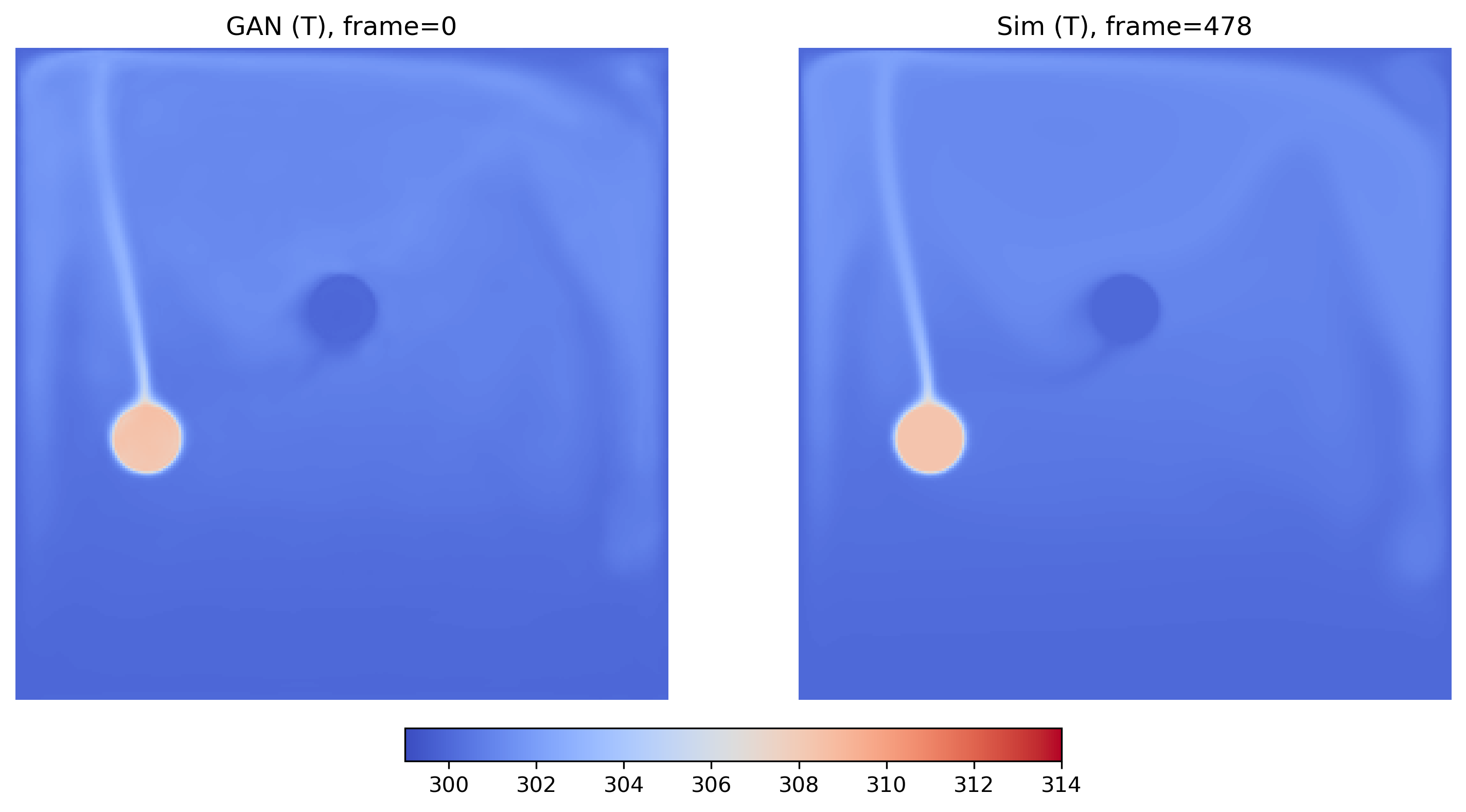}
        \caption{Comparison of $T$.}
    \end{subfigure} 
    \hfill
    \begin{subfigure}[b]{0.48\textwidth}
        \centering
        \includegraphics[width=\textwidth]{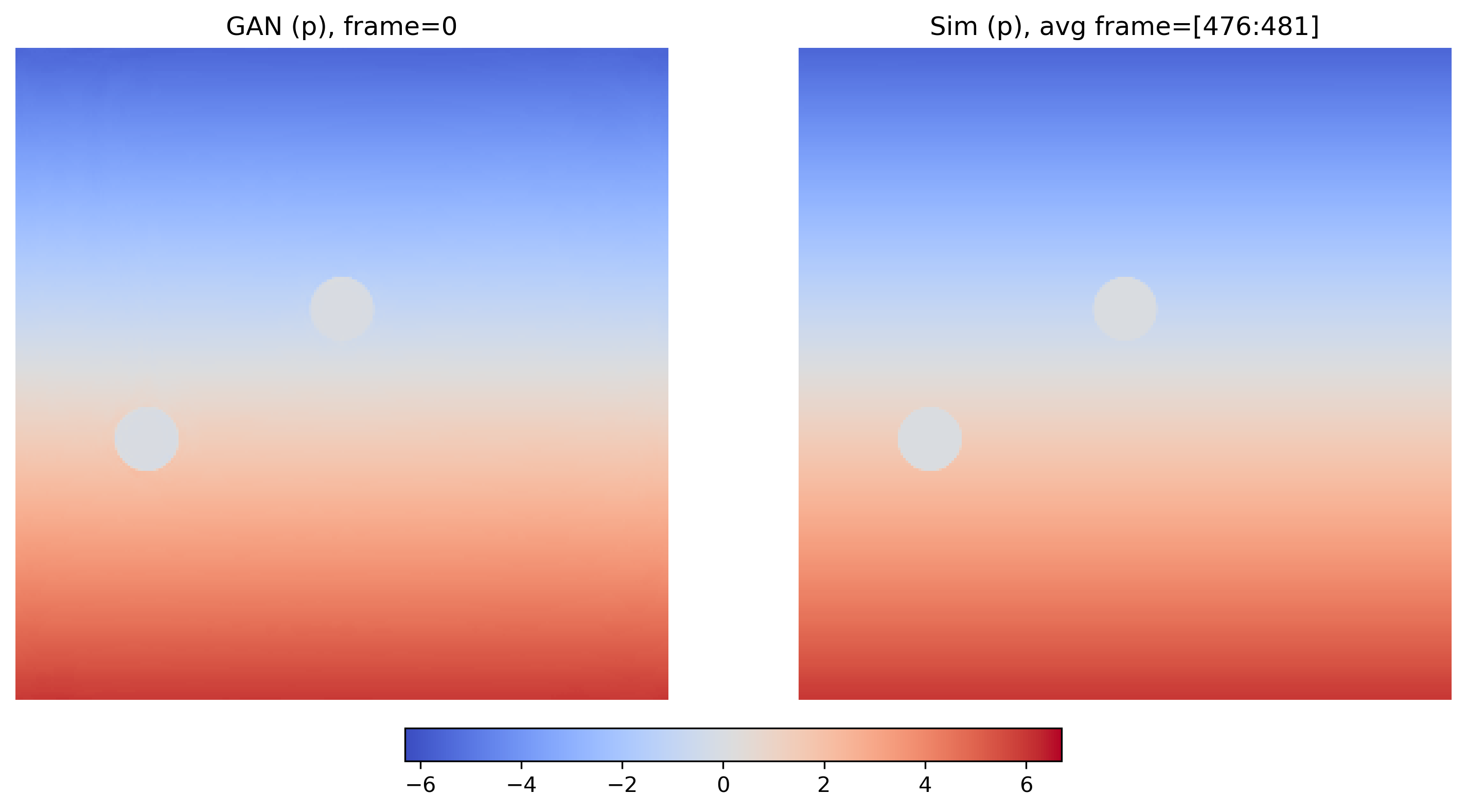}
        \caption{Comparison of $p$.}
    \end{subfigure}   
    \hfill
    \begin{subfigure}[b]{0.48\textwidth}
        \centering
        \includegraphics[width=\textwidth]{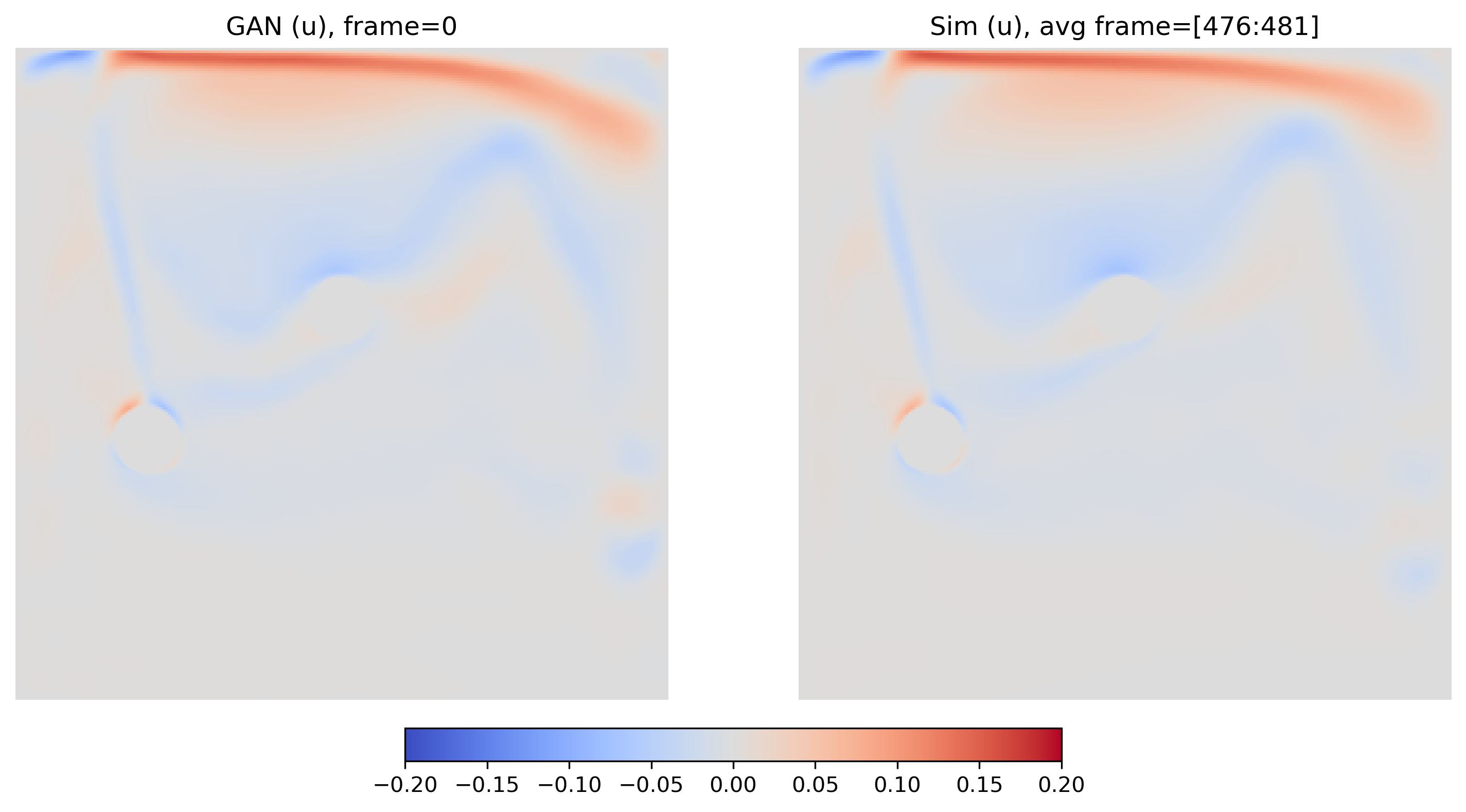}
        \caption{Comparison of $u$.}
    \end{subfigure}
    \hfill
    \begin{subfigure}[b]{0.48\textwidth}
        \centering
        \includegraphics[width=\textwidth]{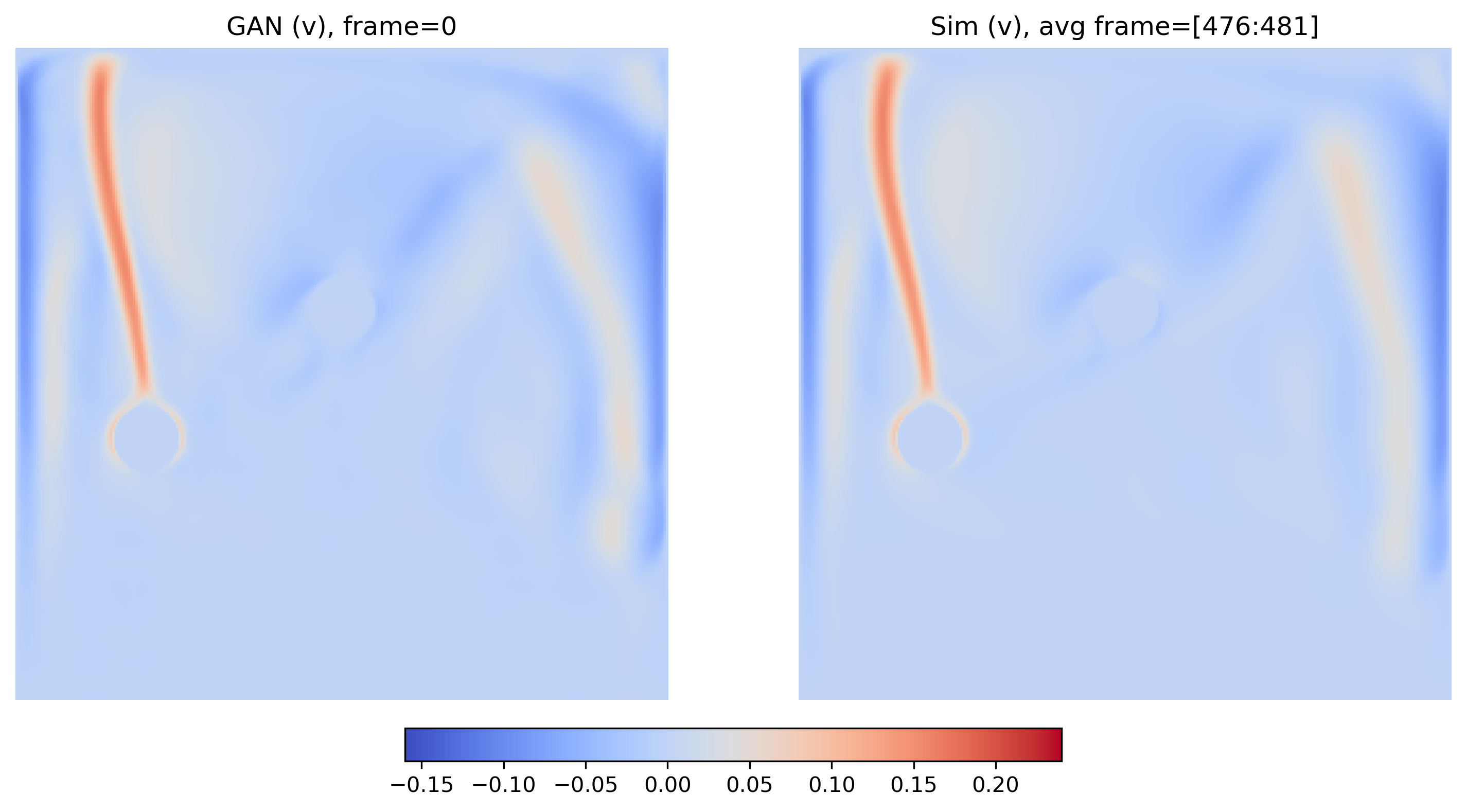}
        \caption{Comparison of $v$.}
    \end{subfigure}

    \caption{Experiment $x_a=0.20$: Comparison of GAN generated frames and simulation frames. The fields $u$ and $v$ are compared with an average over $5$ simulation frames.}
    \label{fig:comp_frames020}
\end{figure}

In Figure~\ref{fig:sim_mean_std}, for the simulation data of one experiment, i.e., $x_a=0.20$, the per frame means (blue) and the standard deviations (blue, shaded) of each channel are shown. Note that for other experiments the curves behave similar. Since the means of channels $u$ and $v$ are $0$ over the whole time horizon, we only report the maximum absolute mean value of the generated tensors to assess the quality of the generated frames. For channel $p$, in each frame the lower part of the domain takes values of $\approx 6$ and the upper part values of $\approx -6$ (see also Figure~\ref{fig:comp_frames020}) resulting in a mean that is close to zero and a substantial standard deviation. However, the pixel-wise values per frame do only minimally change over time, i.e., the average pixel-wise temporal standard deviation is smaller than $10^{-2}$. Under these circumstances, the correlation and variance computations become numerically unstable and do not reflect meaningful temporal dynamics. For this reason, correlation analysis is not considered for this channel, while mean curves are still evaluated. For channel $T$, both, mean curves and correlations are considered.

Furthermore, note that the number of simulation frames is greater than of generated frames, i.e., $\mathcal{T}^{(\text{gen})} < \mathcal{T}^{(\text{sim})}$.  
Consequently, the mean curves of the generated data cover a smaller time window. Since, we at the moment cannot control the initialization state of the generated frames we are interested in how good the means computed via \eqref{eq:mean_frame_sim}, $\bar{x}^{(\text{gen}, c)}_t$ and $\bar{x}^{(\text{gen}, c, s)}_t$, for seeds $s \in \{s_1,\dots,s_P\}$, fit a subsequence of $\bar{x}^{(\text{sim}, c)}_t$.  
Thus, for the first channel ($c = T$) each generated seed curve 
$\bar{x}^{(\text{gen}, T, s)}_{t}$ and $\bar{x}^{(\text{gen}, T)}_{t}$
are horizontally aligned to the simulation curve such that its initial mean value matches the closest simulation temperature mean value, i.e.,
\begin{equation}
\Delta t_s = 
\arg\min_{t}
\big|
\bar{x}^{(\text{sim}, T)}_{t} -
\bar{x}^{(\text{gen}, T, s)}_{1}
\big|,
\label{eq:shift_alignment}
\end{equation}
and the plot is translated by $\Delta t_s$. For the GAN generated pressure $p$ mean curves we apply the same shift w.r.t.\ $\Delta t_s$ to investigate if the values of $p$ correspond to the temperature values when compared to simulation data.

\begin{figure}[t]
\centering
    \begin{subfigure}[b]{0.48\textwidth}
        \centering
        \includegraphics[width=\textwidth]{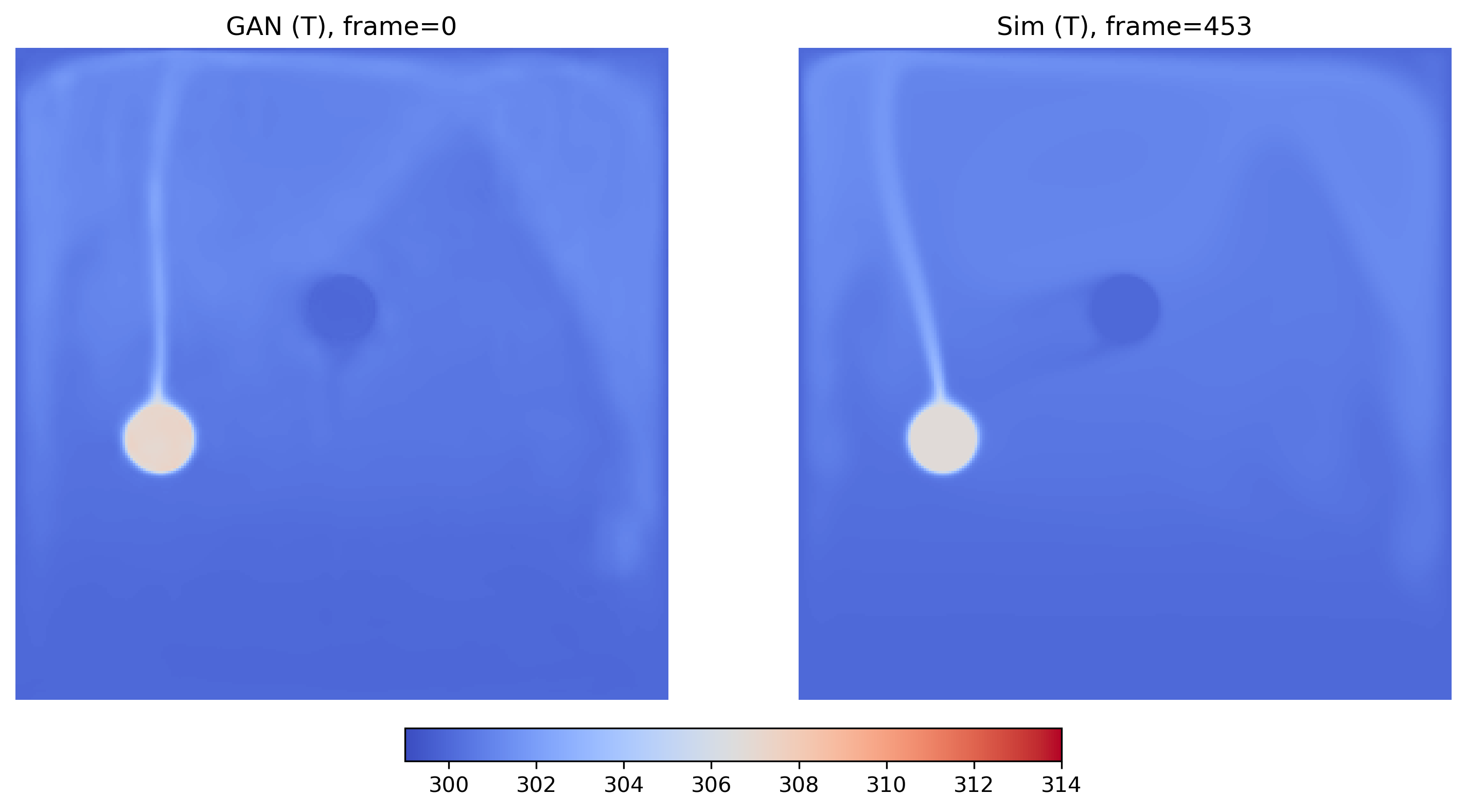}
        \caption{Comparison of $T$.}
    \end{subfigure} 
    \hfill
    \begin{subfigure}[b]{0.48\textwidth}
        \centering
        \includegraphics[width=\textwidth]{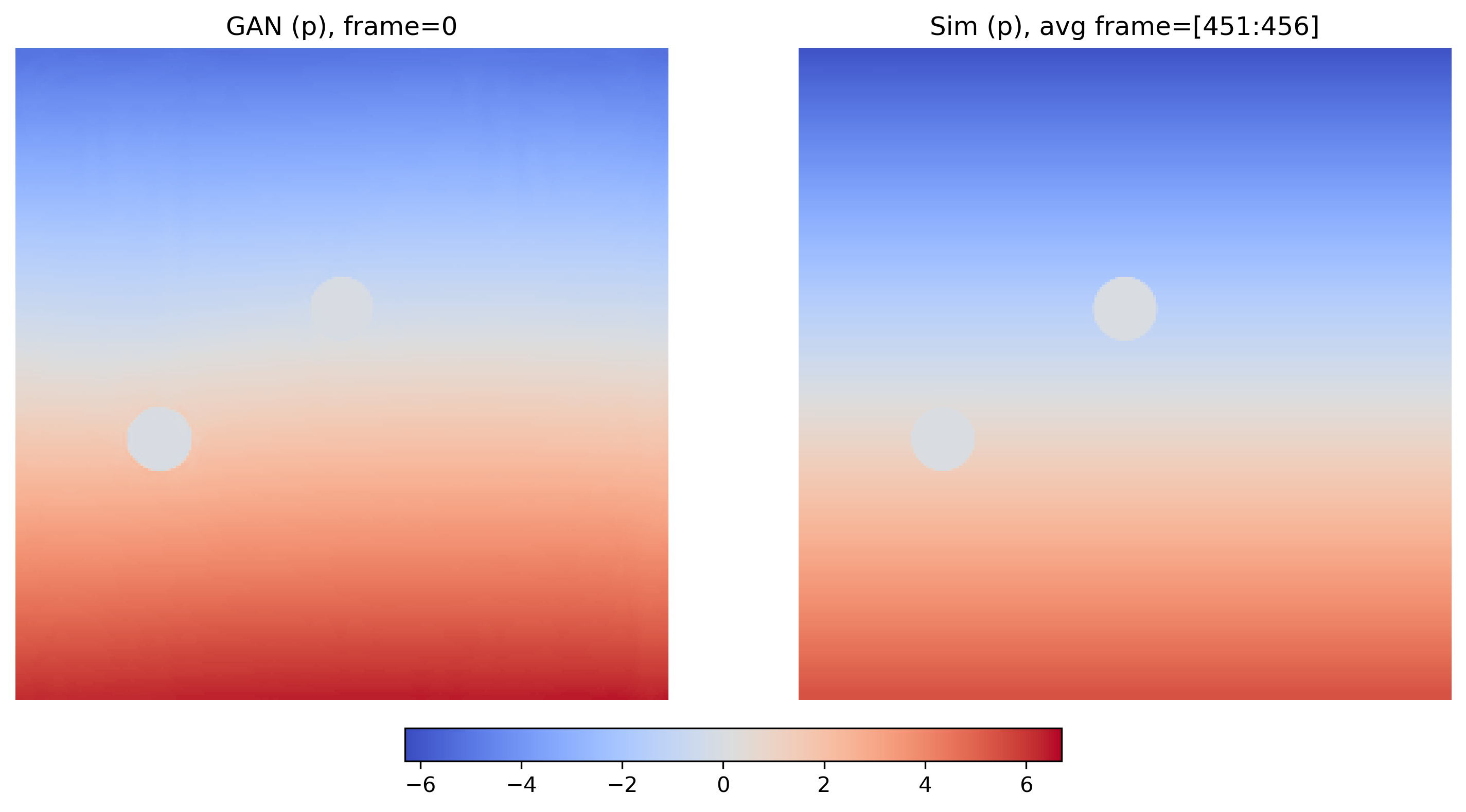}
        \caption{Comparison of $p$.}
    \end{subfigure}   
    \hfill
    \begin{subfigure}[b]{0.48\textwidth}
        \centering
        \includegraphics[width=\textwidth]{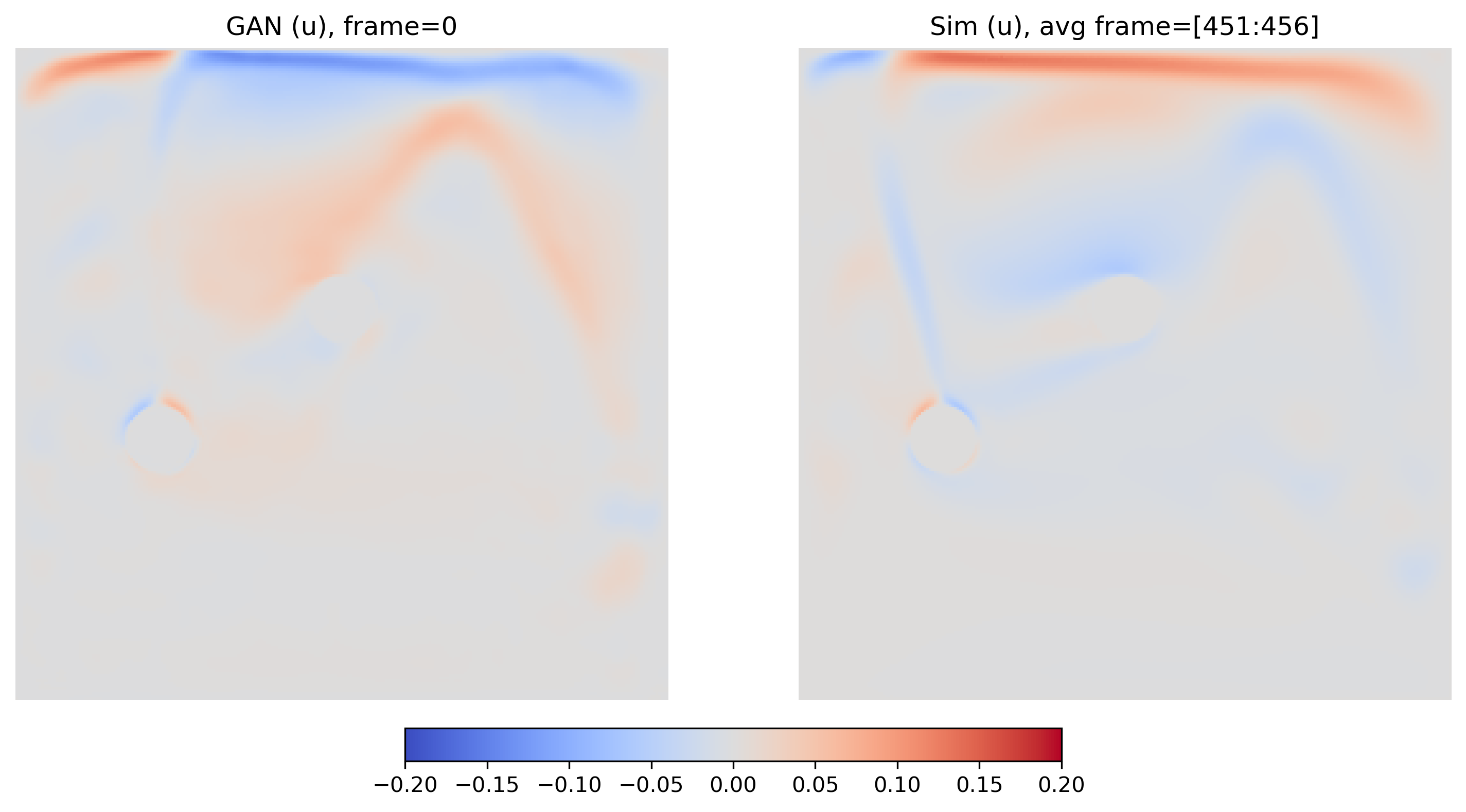}
        \caption{Comparison of $u$.}
    \end{subfigure}
    \hfill
    \begin{subfigure}[b]{0.48\textwidth}
        \centering
        \includegraphics[width=\textwidth]{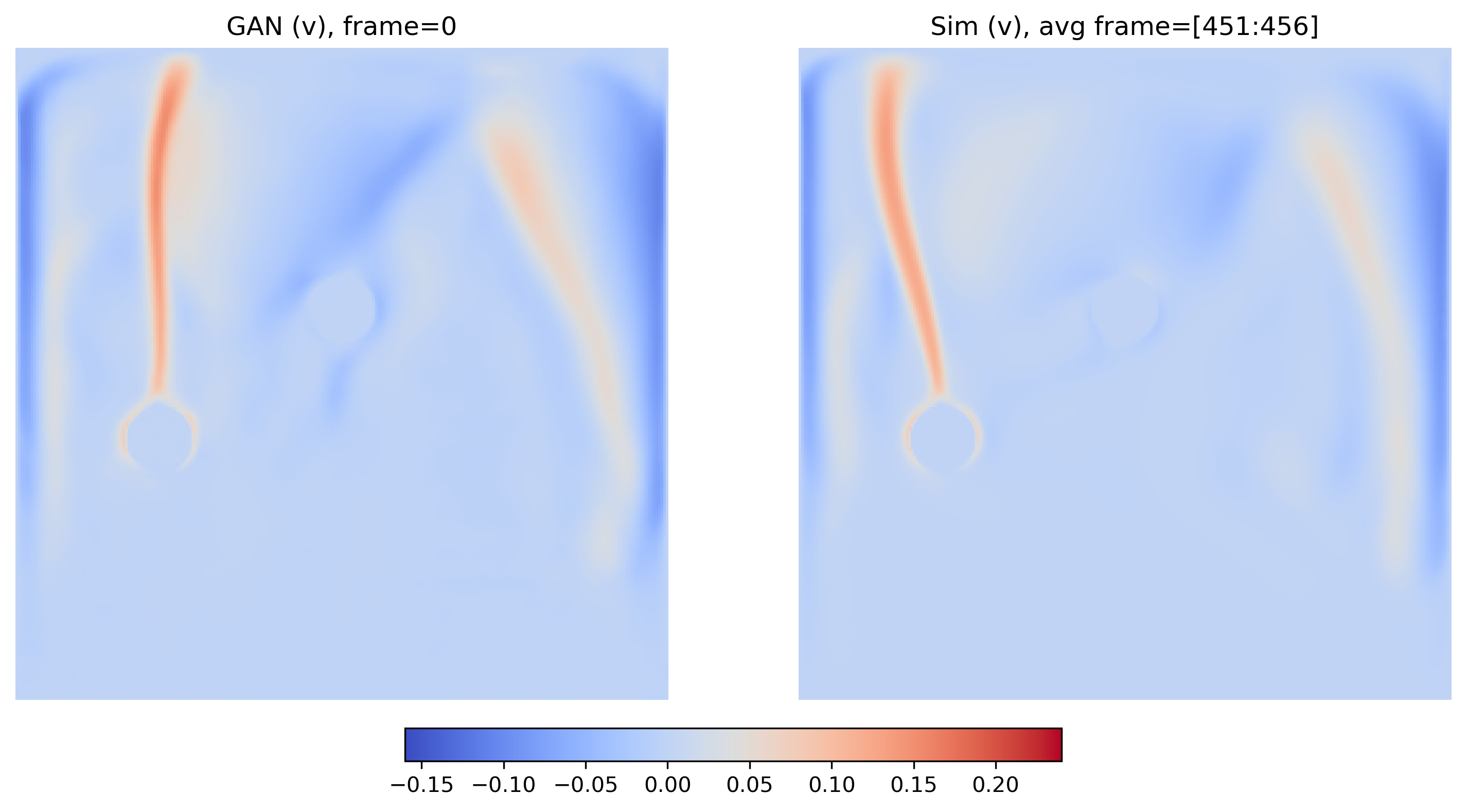}
        \caption{Comparison of $v$.}
    \end{subfigure}

    \caption{Experiment $x_a=0.22$: Comparison of GAN generated frames and simulation frames. The fields $u$ and $v$ are compared with an average over $5$ simulation frames.}
    \label{fig:comp_frames022}
\end{figure}

\paragraph{Reproduction.}
We observe that for conditioning information $c=(c^\textrm{num}, c^\textrm{mask})$ that were encountered during training the networks generate good approximations of the corresponding simulation frames. Through the spatial conditioning $c^\textrm{mask}$ we have full control of the placement of the circles in the domain, e.g., see Figure~\ref{fig:comp_frames020} and \ref{fig:comp_frames022}. Furthermore, the minimum and maximum values, and the ranges of the fields $T,u,v,p$ of the dataset are given in Table~\ref{tab:field-ranges}.

\begin{table}[h!]
\centering
\caption{Minimum, maximum, and range values for each field.}
\label{tab:field-ranges}
\[
\begin{array}{lccc}
\hline
\textbf{Field} & \textbf{Min} & \textbf{Max} & \textbf{Range} \\ 
\hline
T~[\mathrm{K}] & 299 & 314 & 15 \\[4pt]
u~[\mathrm{m/s}] & -0.2 & 0.2 & 0.4 \\[4pt]
v~[\mathrm{m/s}] & -0.16 & 0.24 & 0.40 \\[4pt]
p~[\mathrm{Pa}] & -6.3 & 6.7 & 13.0 \\ 
\hline
\end{array}
\]
\end{table}

\begin{figure}[htbp]
    \centering

    \begin{subfigure}[b]{0.48\textwidth}
        \centering
        \includegraphics[width=\textwidth]{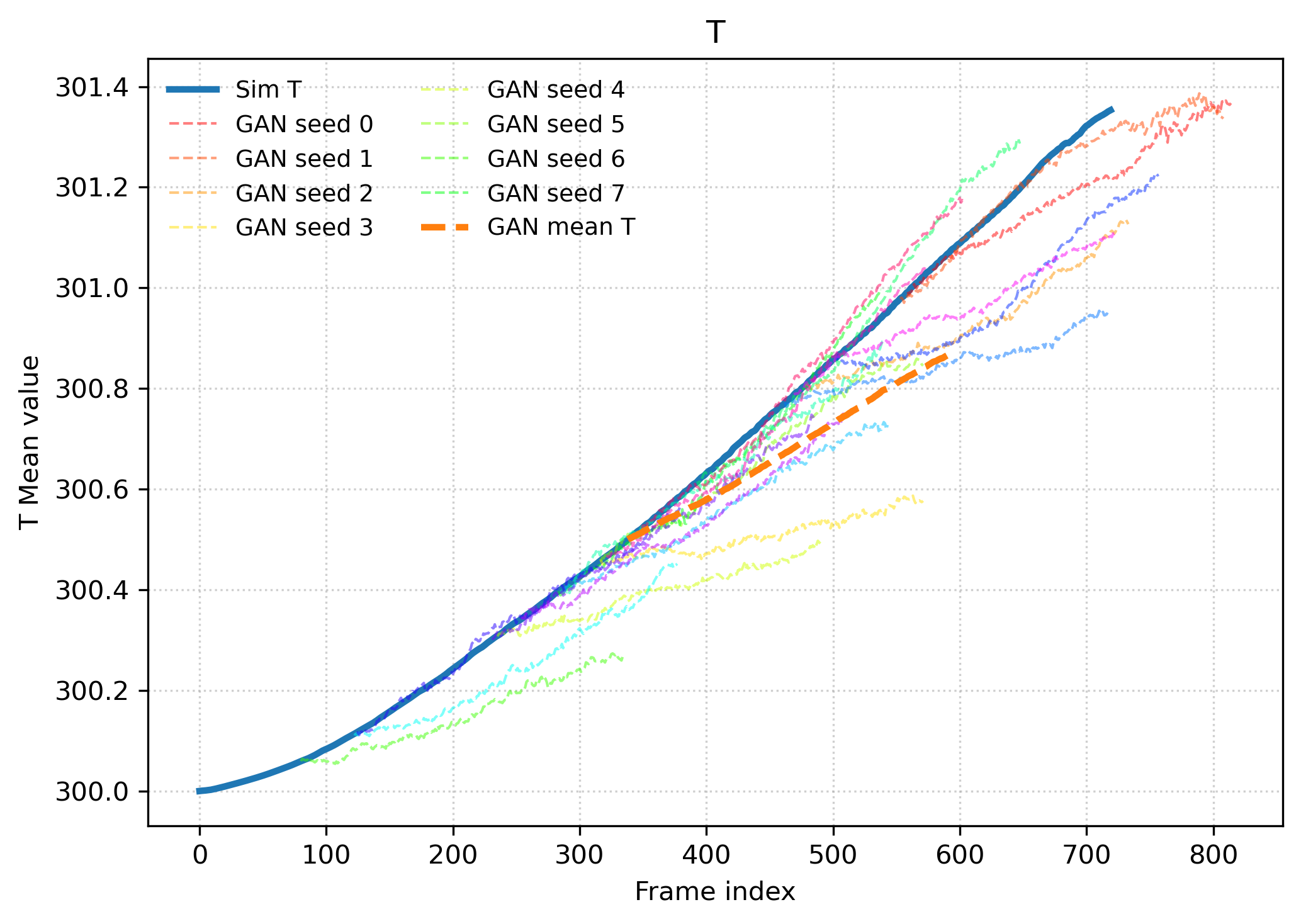}
        \caption{Experiments with $x_a = 0.20$}
    \end{subfigure}
    \hfill
    \begin{subfigure}[b]{0.48\textwidth}
        \centering
        \includegraphics[width=\textwidth]{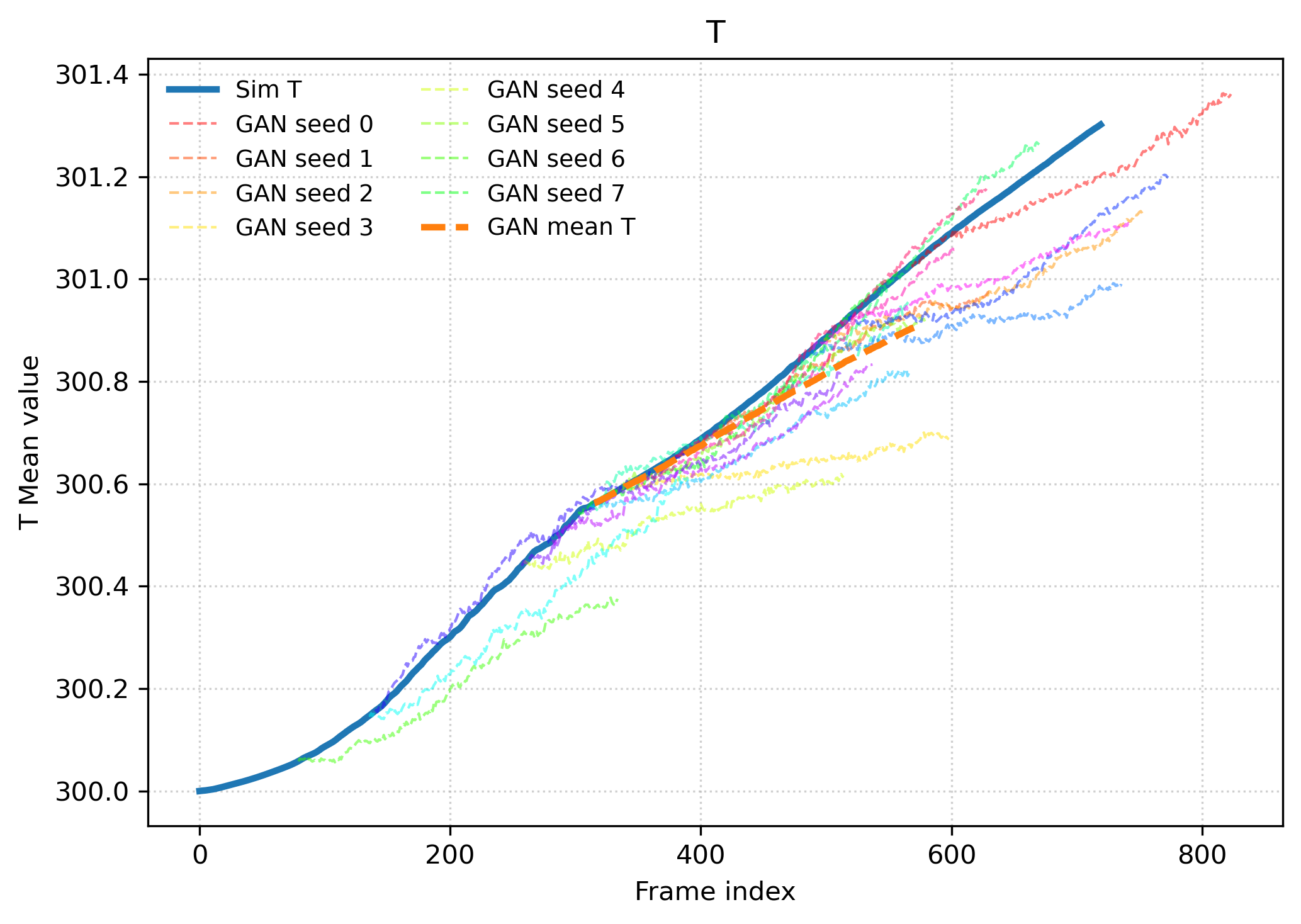}
        \caption{Experiments with $x_a = 0.35$}
    \end{subfigure}

    \medskip

    \begin{subfigure}[b]{0.48\textwidth}
        \centering
        \includegraphics[width=\textwidth]{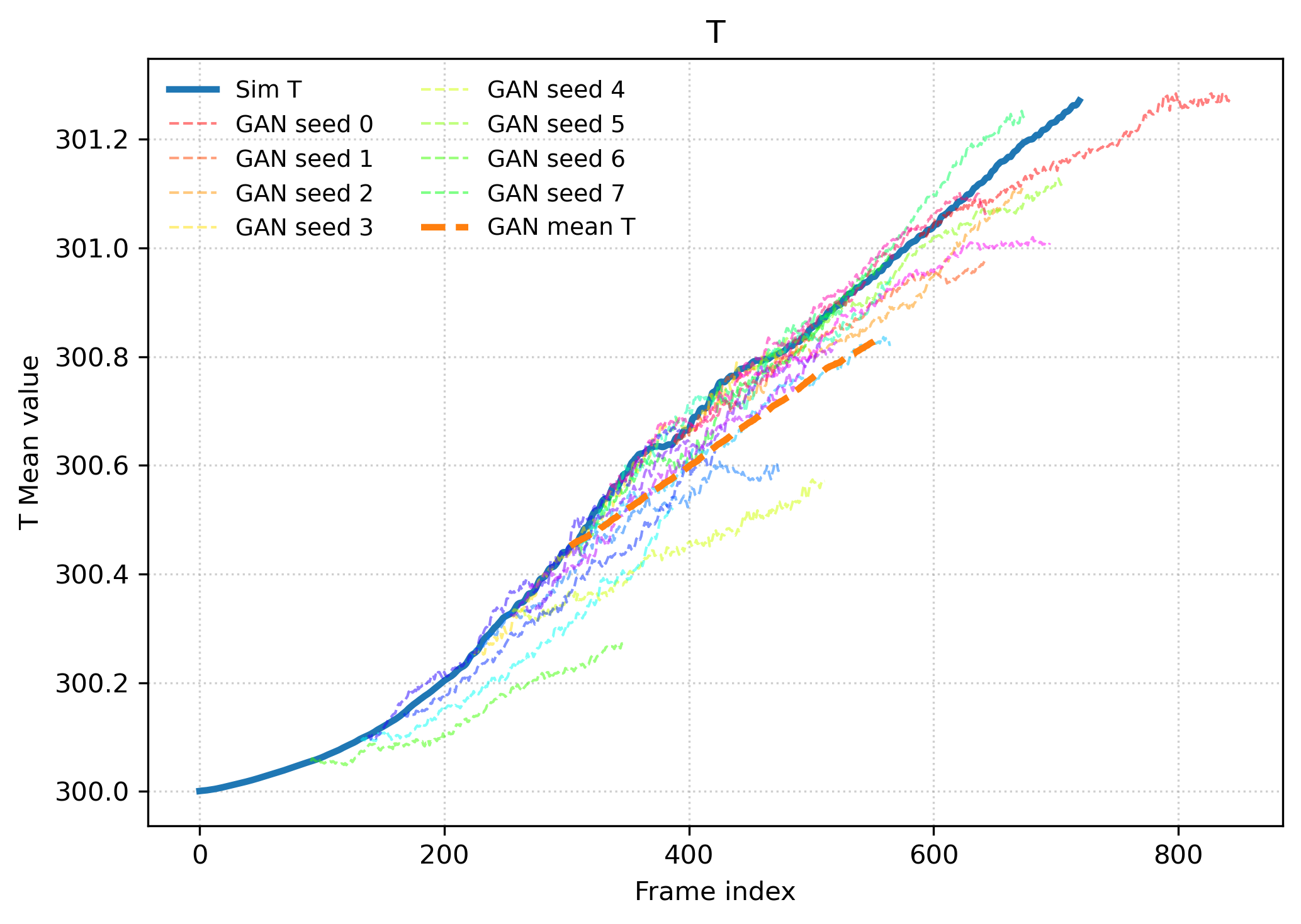}
        \caption{Experiments with $x_a = 0.48$}
    \end{subfigure}
    \hfill
    \begin{subfigure}[b]{0.48\textwidth}
        \centering
        \includegraphics[width=\textwidth]{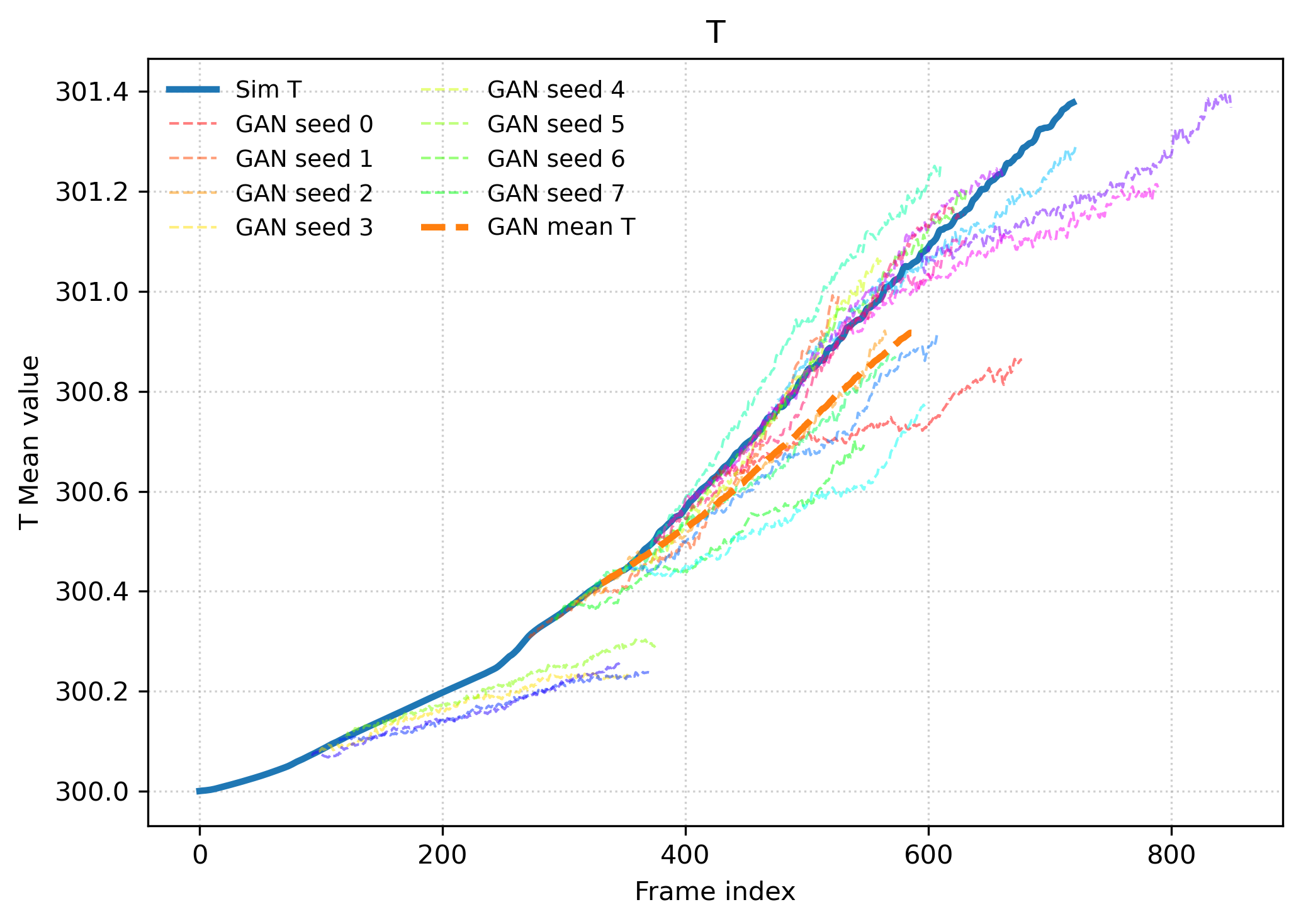}
        \caption{Experiments with $x_a = 0.85$}
    \end{subfigure}

    \caption{Experiments $x_a\in\{0.20, 0.35, 0.48, 0.85\}$: Comparison of the generated (orange, dashed) and simulation (blue) mean values of the temperature $T$. Some of the $P=100$ GAN realizations are included and the curves are shifted via \eqref{eq:shift_alignment}. }
    \label{fig:training_mean_values}
\end{figure}

In Figure~\ref{fig:training_mean_values}, for four different experiments, i.e., $x_a\in\{0.20, 0.35, 0.48, 0.85\}$, the transient behavior of mean values of the temperature $T$ of simulation frames (blue) and generated frames (orange) is illustrated. Recall that the GAN sequences are shorter than the simulation sequence. We also depict $10$ GAN realizations for different seeds and apply the translation described in \eqref{eq:shift_alignment}. For the temperature field $T$ the network manages for different seeds to approximate most of the simulation curve, underlining the transient capabilities of our approach. Furthermore, most GAN realizations and consequently the mean underestimate the simulation curve, where an underestimation of up to $0.4$ can occur which is around $3\%$ of the range of $T$. For the pressure $p$ we encountered absolute deviations w.r.t.\ mean simulation curve of up to $0.7$ which amounts to $5\%$ of the range of $p$. This deviation is negligible, since $p$ represents a fluctuation about the operating pressure $p_{op}=101325$ Pa.  Furthermore, for the channels $u$ and $v$ whose mean values stay around zero we are interested in the absolute fluctuations from the simulation mean values. For fields $u$ and $v$ the absolute deviation for all experiments is smaller than $2\times 10^{-3}$, which amounts to less than $1\%$ of the ranges of $u$  and $v$, respectively. Note that the model struggles to reproduce the fields for experiment $x_a=0.52$, see the video material \href{https://www.youtube.com/watch?v=mo_mJpTB-Qs}{youtube}, while it captures the behavior of its mirrored experiment $x_a=0.48$. In these two experiments, the lower circles are placed under the upper circle and shifted slightly to the left ($x_a=0.48$) and right ($x_a=0.52$) of the center ($x_a=0.50$). At $x_a=0.50$ the passive conductor causing the upward thermal flow to split into two streams. When moving $x_a$ to the left or right, the splitted flow transits to an unilateral flow on the respective side. The generative world model is struggling to capture this bifurcation \cite{kielhofer2012bifurcation} for $x_a=0.52$, while for the mirrored $x_a=0.48$ the correct split of flows is learned.

In the following, spatial and temporal correlations, and comparisons of GAN and simulation frames for an exemplary experiment $x_a=0.20$ are shown. Section~\ref{app:additional_results} in the appendix provides the same analysis for a second experiment $x_a=0.48$.

In a first experiment with $x_a=0.20$, the lower circle is near the left boundary of the domain. For seed $s\in\{s_1,\dots,s_P\}$ we compare the first GAN generated frame with the frame $j'$ of the simulation sequence whose mean temperature is closest to the mean temperature of the GAN frame. For the temperature field $T$ and pressure $p$ we compare the two frames directly. For the other two fields $u$ and $v$ we compare the GAN frame with an average of the simulation frames $\{j'-2,\dots,j',\dots,j'+2\}$ since these two fields move more chaotically than $T$, i.e., with a higher rate of change. In Figure~\ref{fig:comp_frames020}, the GAN and simulation frames and averages are shown. The GAN frames on the left are quite good approximations of the simulation frames and averages.

\begin{figure}[htbp]
    \centering
    \begin{subfigure}[b]{0.48\textwidth}
        \centering
        \includegraphics[width=\textwidth]{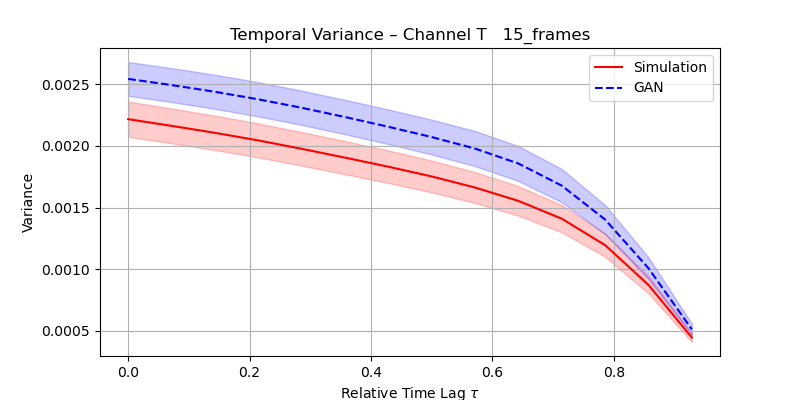}
        \caption{Comparison of $T$.}
    \end{subfigure}
    \hfill
    \begin{subfigure}[b]{0.48\textwidth}
        \centering
        \includegraphics[width=\textwidth]{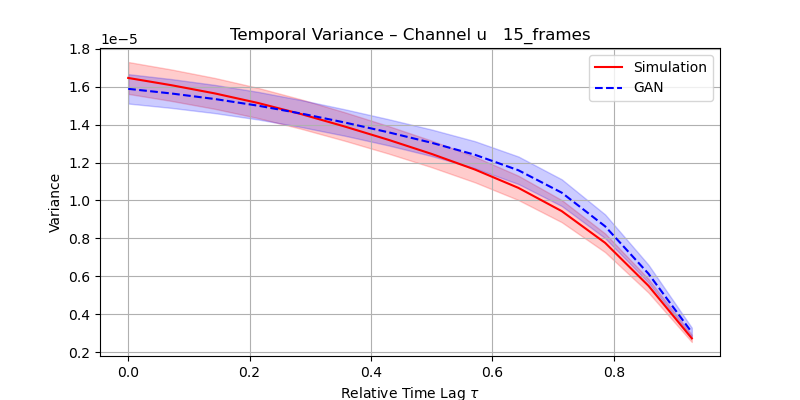}
        \caption{Comparison of $u$.}
    \end{subfigure}
    \hfill
    \begin{subfigure}[b]{0.48\textwidth}
        \centering
        \includegraphics[width=\textwidth]{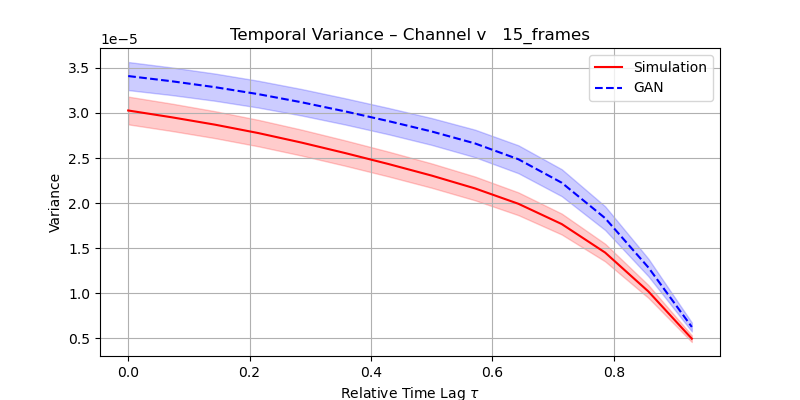}
        \caption{Comparison of $v$.}
    \end{subfigure}

    \caption{Experiment $x_a=0.20$: comparison of mean variance values computed via \eqref{eq:mean_variance} of generated (blue) and simulation frames (red). }
    \label{fig:mean_variance020}
\end{figure}

The windowed variance computation for $T$, $u$ and $v$ described in \eqref{eq:mean_variance} is illustrated in Figure~\ref{fig:mean_variance020}. The GAN generated frames satisfactorily approximate the variance behavior observed in the simulation data.

\begin{figure}[htbp]
    \centering
    \includegraphics[width=\linewidth]{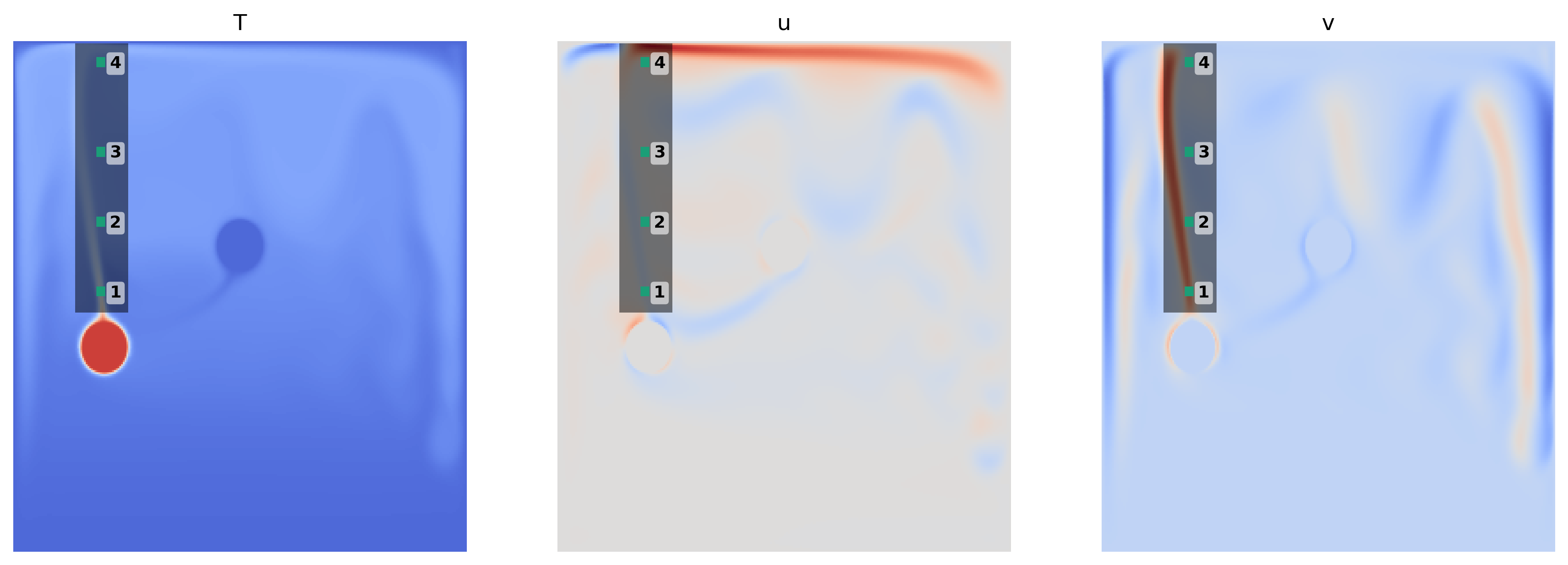}
    \caption{Experiment $x_a=0.20$: Placement of a cropped box $B$ (gray) and fixed spatial points (green) to compute spatial and temporal correlations.}
    \label{fig:corr_setup020}
\end{figure}

\begin{figure}[htbp]
    \centering
    \begin{subfigure}{\textwidth}
        \centering
        \includegraphics[width=0.32\textwidth]{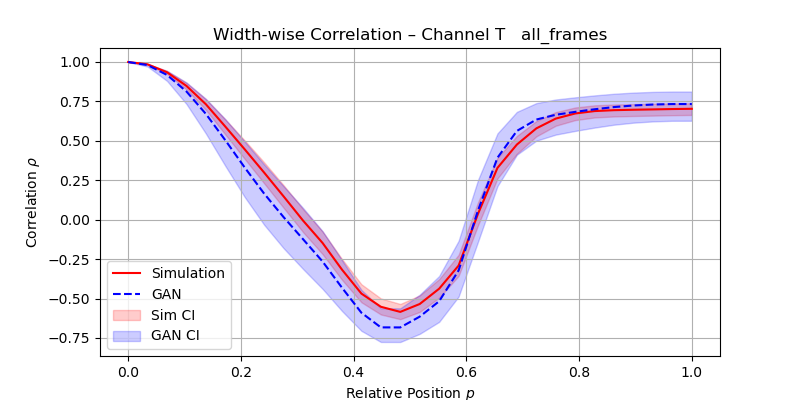}\hfill
        \includegraphics[width=0.32\textwidth]{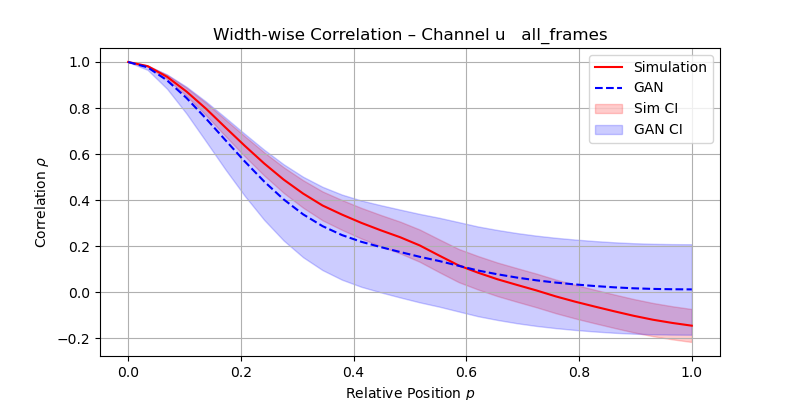}\hfill
        \includegraphics[width=0.32\textwidth]{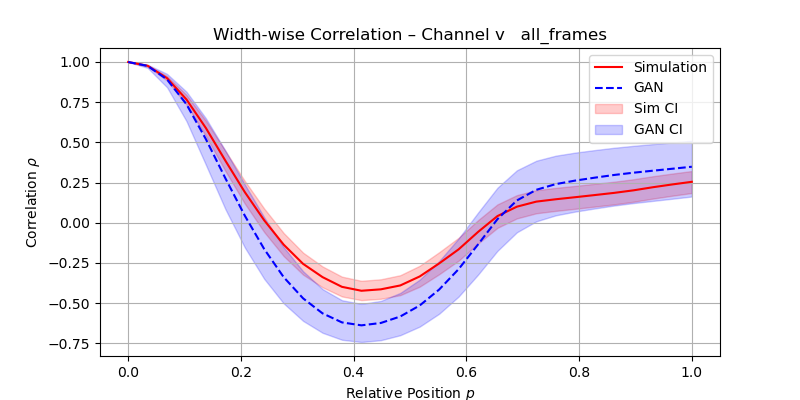}
        \caption{Column wise spatial correlations of $T$, $u$, and $v$.}
    \end{subfigure}

    \begin{subfigure}{\textwidth}
        \centering
        \includegraphics[width=0.32\textwidth]{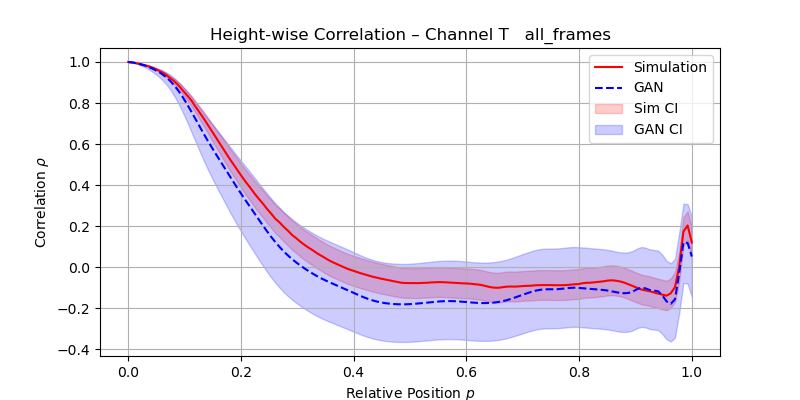}\hfill
        \includegraphics[width=0.32\textwidth]{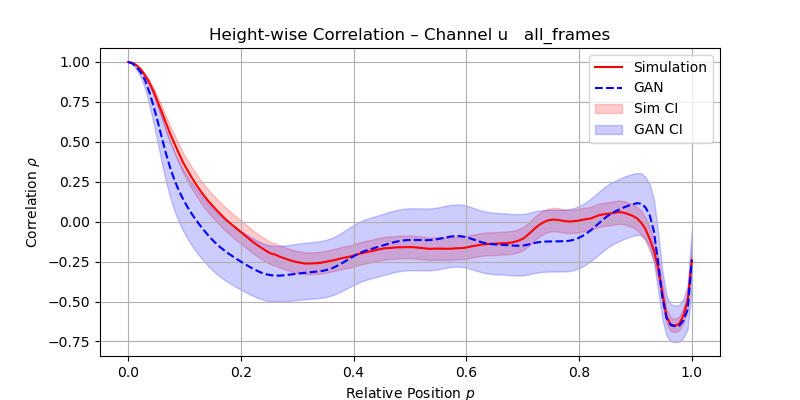}\hfill
        \includegraphics[width=0.32\textwidth]{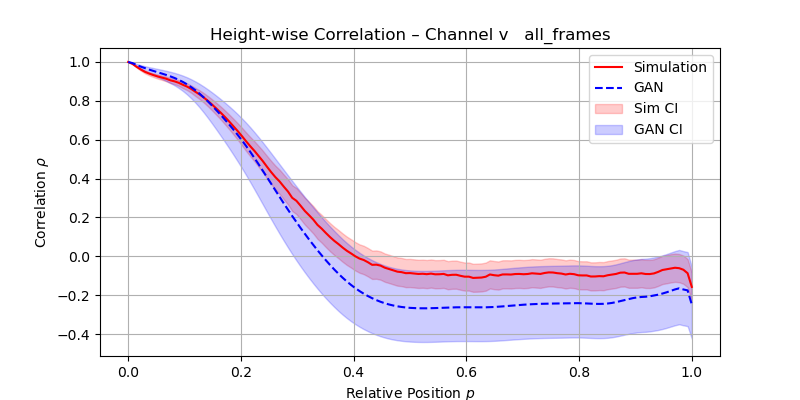}
        \caption{Row wise spatial correlations of $T$, $u$, and $v$.}
    \end{subfigure}
    \begin{subfigure}{\textwidth}
        \centering
        \includegraphics[width=0.32\textwidth]{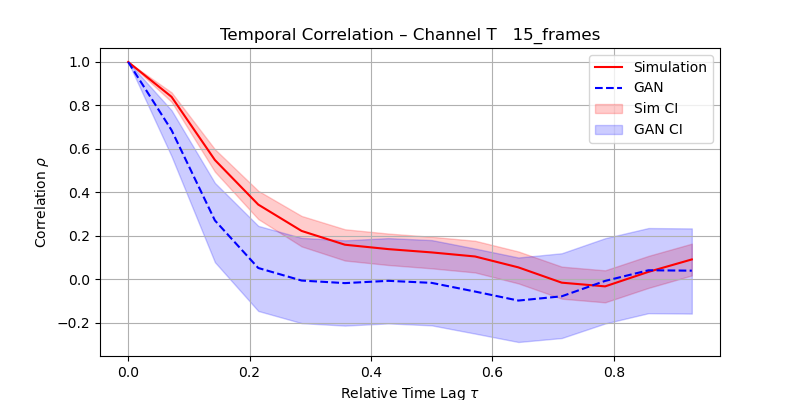}\hfill
        \includegraphics[width=0.32\textwidth]{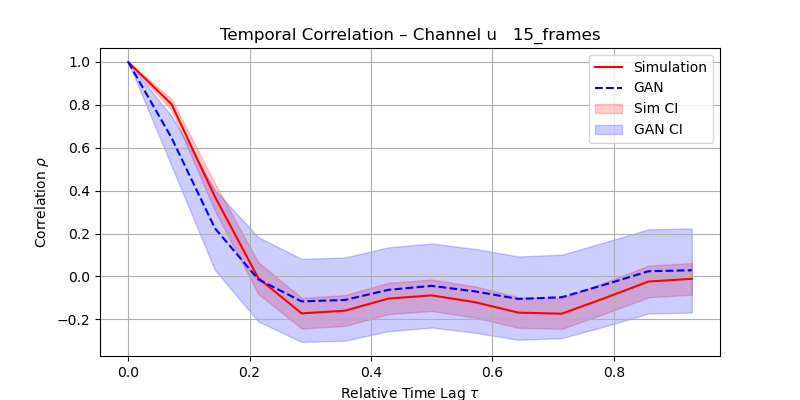}\hfill
        \includegraphics[width=0.32\textwidth]{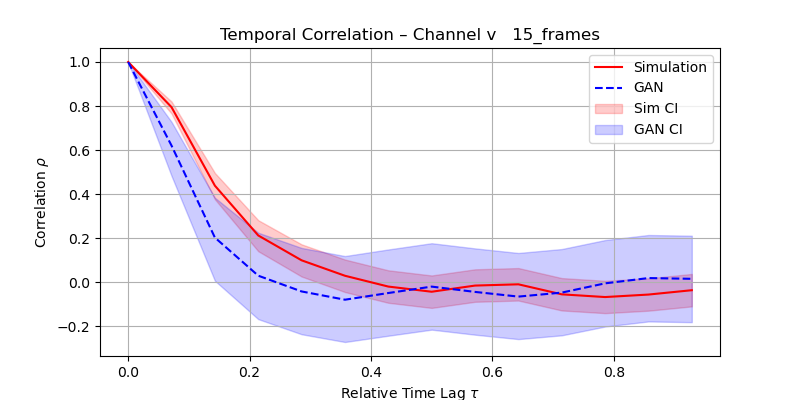}
        \caption{Temporal correlations of $T$, $u$, and $v$ in fixed spatial point 1.}
    \end{subfigure}

    \begin{subfigure}{\textwidth}
        \centering
        \includegraphics[width=0.32\textwidth]{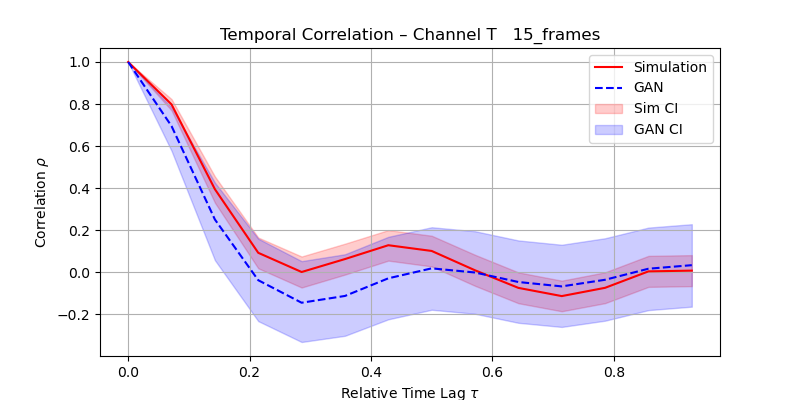}\hfill
        \includegraphics[width=0.32\textwidth]{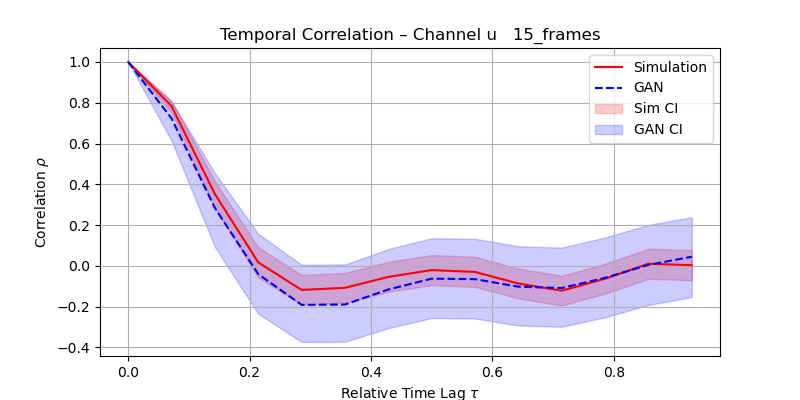}\hfill
        \includegraphics[width=0.32\textwidth]{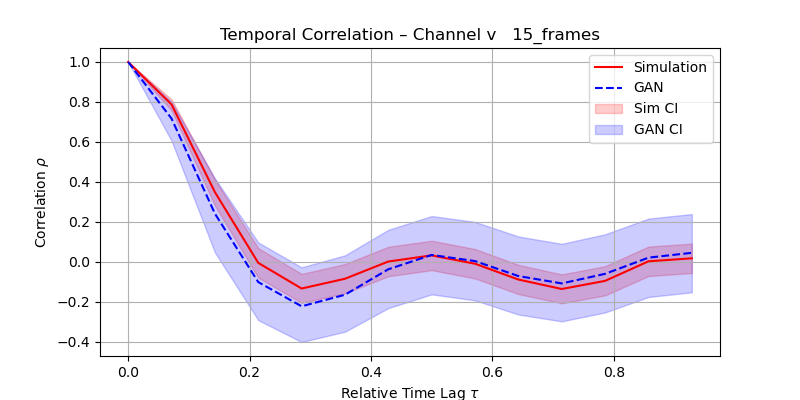}
        \caption{Temporal correlations of $T$, $u$, and $v$ in fixed spatial point 2.}
    \end{subfigure}

    \begin{subfigure}{\textwidth}
        \centering
        \includegraphics[width=0.32\textwidth]{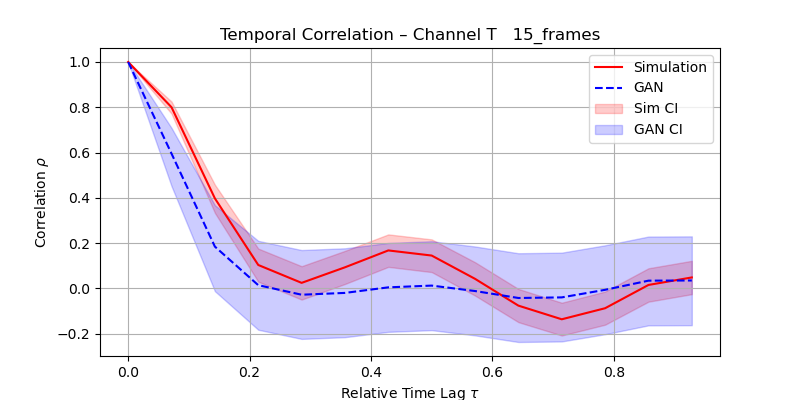}\hfill
        \includegraphics[width=0.32\textwidth]{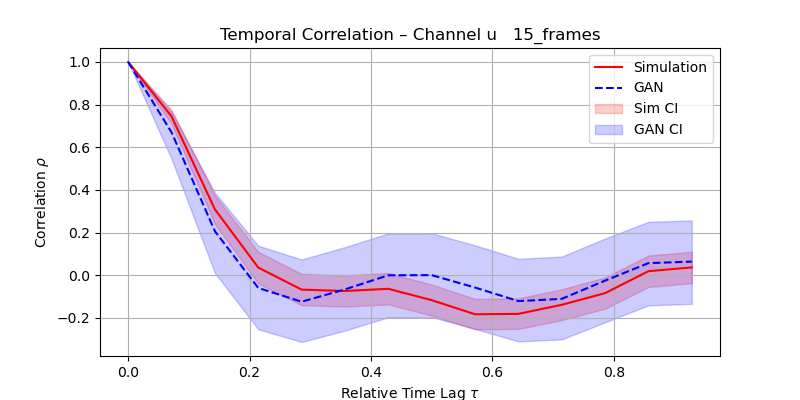}\hfill
        \includegraphics[width=0.32\textwidth]{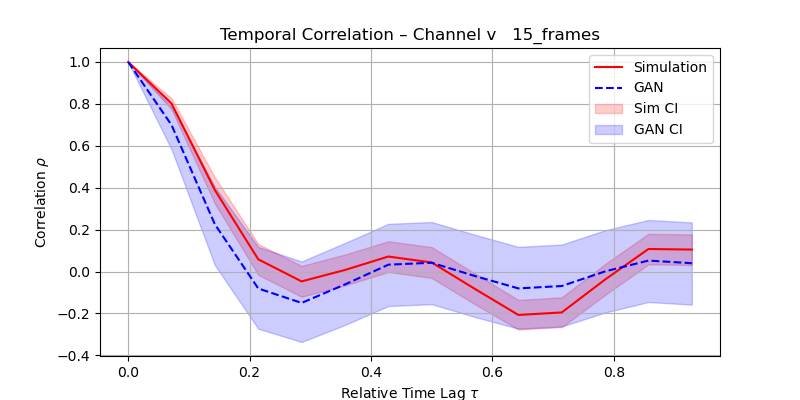}
        \caption{Temporal correlations of $T$, $u$, and $v$ in fixed spatial point 3.}
    \end{subfigure}

    \begin{subfigure}{\textwidth}
        \centering
        \includegraphics[width=0.32\textwidth]{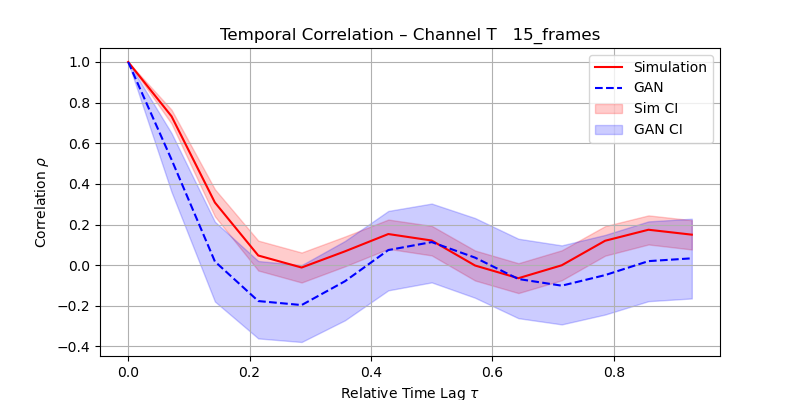}\hfill
        \includegraphics[width=0.32\textwidth]{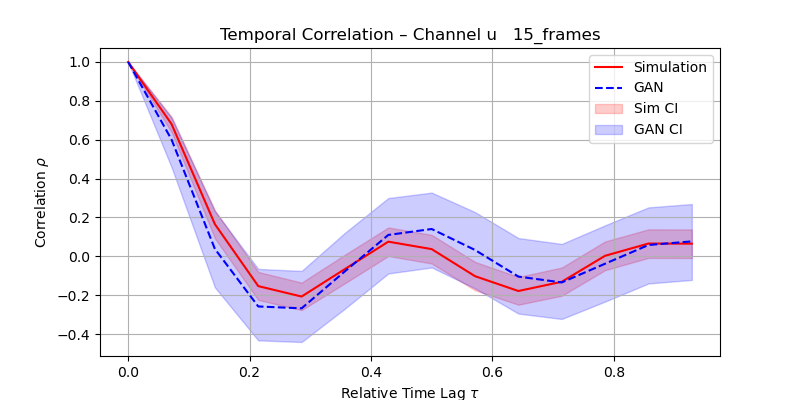}\hfill
        \includegraphics[width=0.32\textwidth]{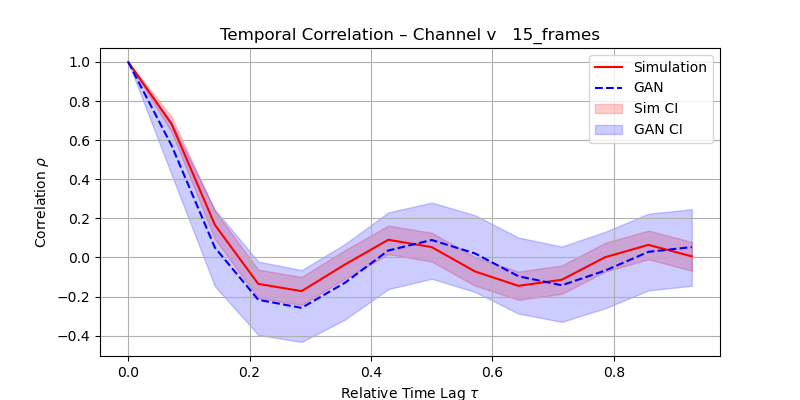}
        \caption{Temporal correlations of $T$, $u$, and $v$ in fixed spatial point 4.}
        
    \end{subfigure}

    \caption{Experiment $x_a=0.20$: Comparison of spatial and temporal correlations of simulation data (red) and GAN generated data (blue), including confidence intervals (shaded), for the fields $T$ (left), $u$ (middle), and $v$ (right).}
    \label{fig:correlations_xc020}
\end{figure}

Furthermore, the cropped box $B$ and the spatial points we chose for the spatial and temporal correlation analysis are illustrated in Figure~\ref{fig:corr_setup020}.
In Figure~\ref{fig:correlations_xc020}, the spatial and temporal correlations with their confidence intervals as described in Section~\ref{sec:dataset_and_evaluation} are shown. We observe that the GAN frames perform reasonable well for all measured entities.

\begin{figure}[htbp]
    \centering

    \begin{subfigure}[b]{0.48\textwidth}
        \centering
            \includegraphics[width=\textwidth]{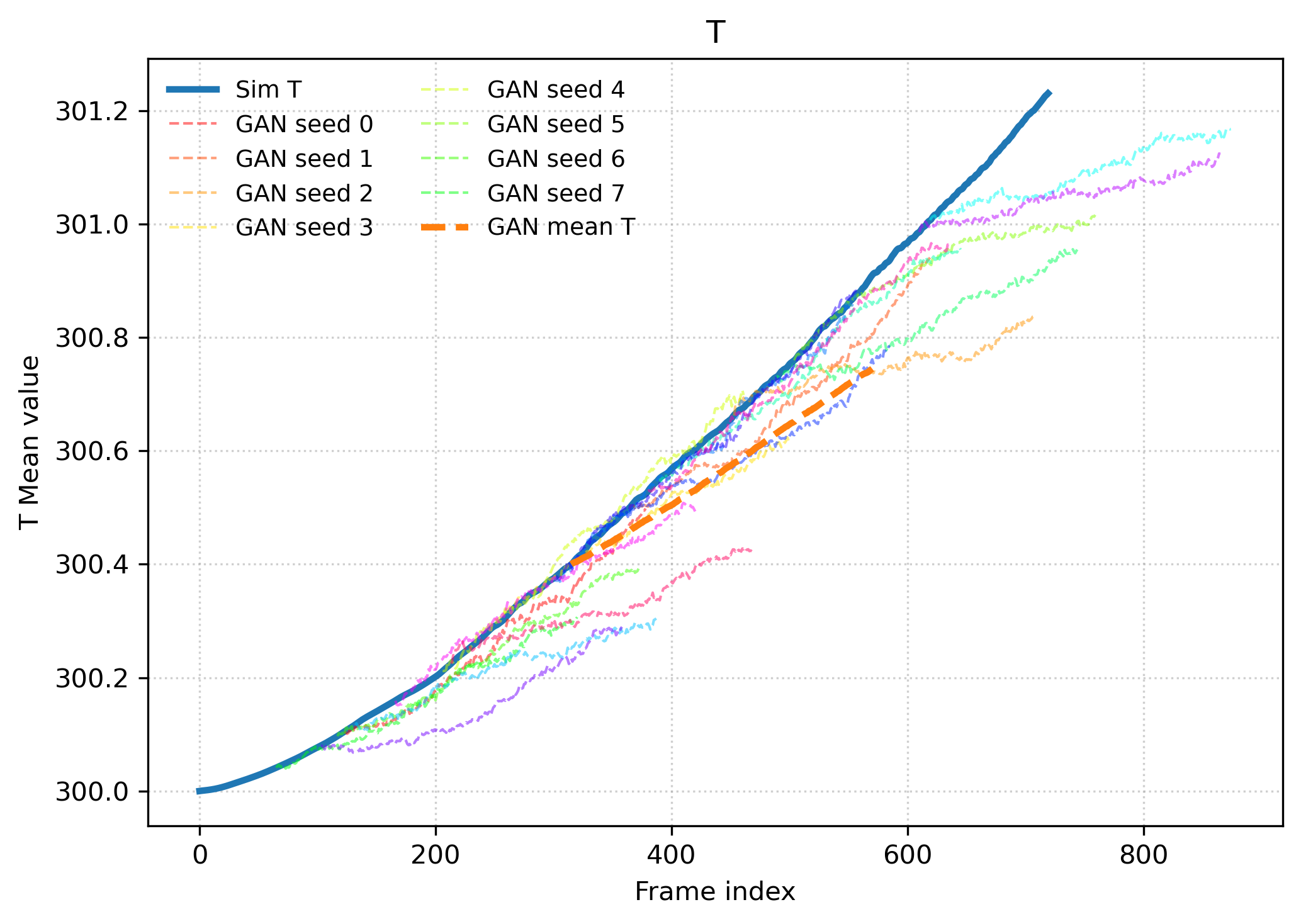}        
           \caption{Experiments with $x_a=0.22$}
    \end{subfigure}
    \hfill
    \begin{subfigure}[b]{0.48\textwidth}
        \centering
            \includegraphics[width=\textwidth]{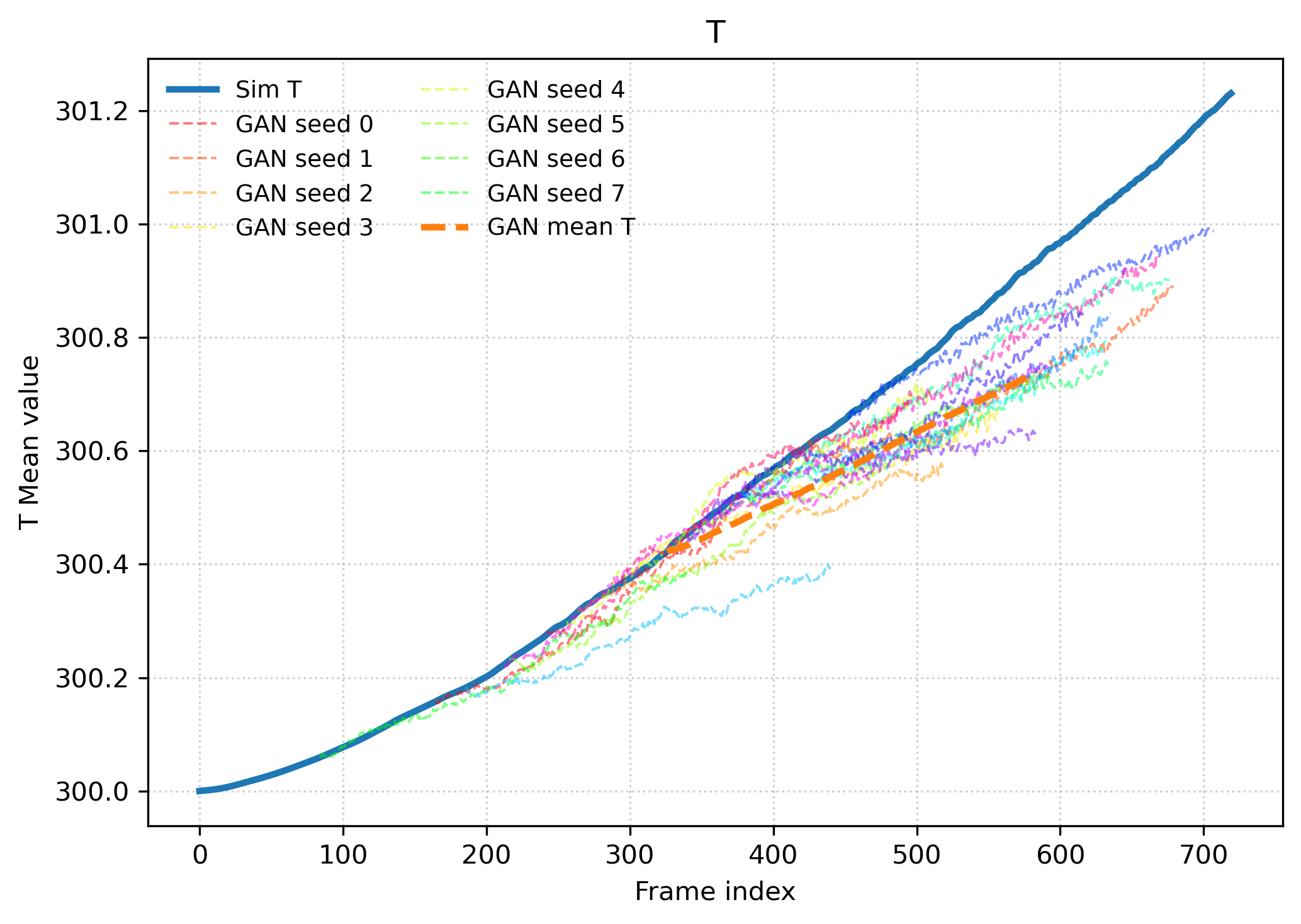}
        \caption{Experiments with $x_a=0.78$}
    \end{subfigure}

    \caption{Comparison of the generated (orange, dashed) and simulation (blue) mean values of the temperature $T$ for the experiments $x_a \in \{0.22, 0.78\}$ that were excluded from the training dataset. We also include some of the $P=100$ GAN realizations and shift the curves via \eqref{eq:shift_alignment}. }
    \label{fig:testing_mean_values}
\end{figure}

\paragraph{Generalization.}
To investigate the generalization capabilities of the networks, we excluded the experiments with $x_a \in \{0.22, 0.78\}$ from the training dataset and generate GAN sequences and evaluate them analogous to the reproduction results. These two configurations are symmetric with respect to the vertical midline of the domain at $x = 0.5$, i.e., the simulation fields for $x_a = 0.78$ correspond to the mirrored fields of $x_a = 0.22$. In Figure~\ref{fig:testing_mean_values}, a comparison of the field mean values of $T$ per frame of generated and simulation frames are shown. The deviations for $T$, $u$, $v$, and $p$ have the same magnitude as in the reproduction experiment.
We assume that further training may mitigate this behavior. In the following, we assess the performance of the GAN networks w.r.t.\ the unseen experiment $x_a=0.22$. The results for $x_a=0.78$ are provided in Section~\ref{app:additional_results}.

\begin{figure}[htbp]
    \centering
    \begin{subfigure}[b]{0.48\textwidth}
        \centering
        \includegraphics[width=\textwidth]{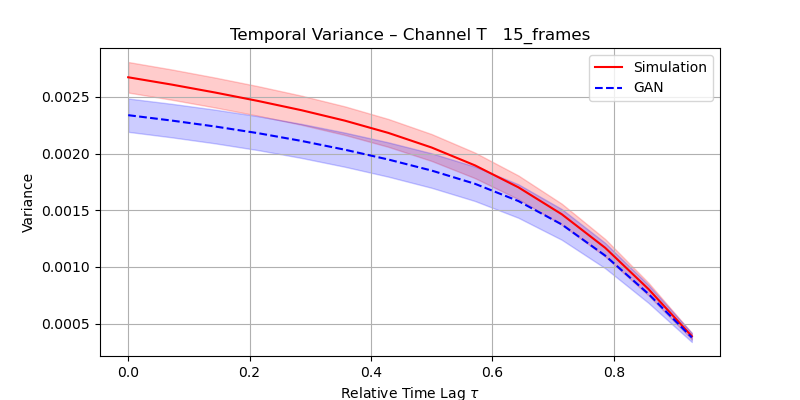}
        \caption{Comparison of $T$.}
    \end{subfigure}
    \hfill
    \begin{subfigure}[b]{0.48\textwidth}
        \centering
        \includegraphics[width=\textwidth]{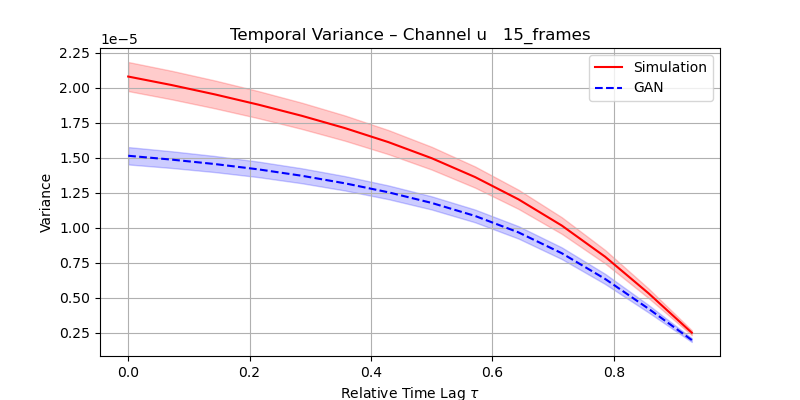}
        \caption{Comparison of $u$.}
    \end{subfigure}
    \hfill
    \begin{subfigure}[b]{0.48\textwidth}
        \centering
        \includegraphics[width=\textwidth]{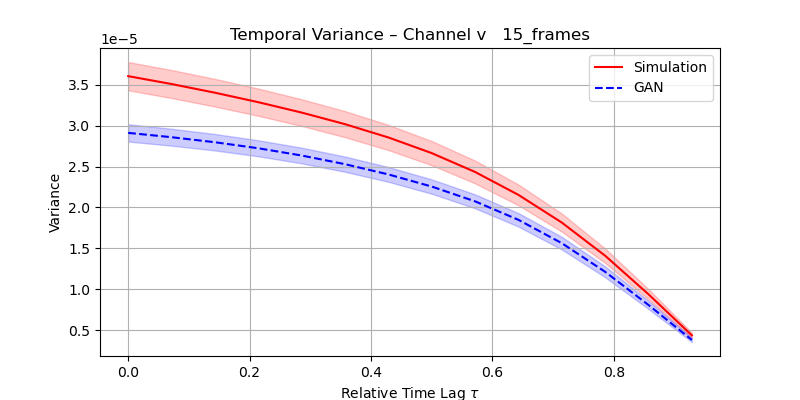}
        \caption{Comparison of $v$.}
    \end{subfigure}

    \caption{Experiment $x_a=0.22$: comparison of mean variance values computed via \eqref{eq:mean_variance} of generated (blue) and simulation frames (red). }
    \label{fig:mean_variance022}
\end{figure}

\begin{figure}[htbp]
    \centering
    \includegraphics[width=0.9\linewidth]{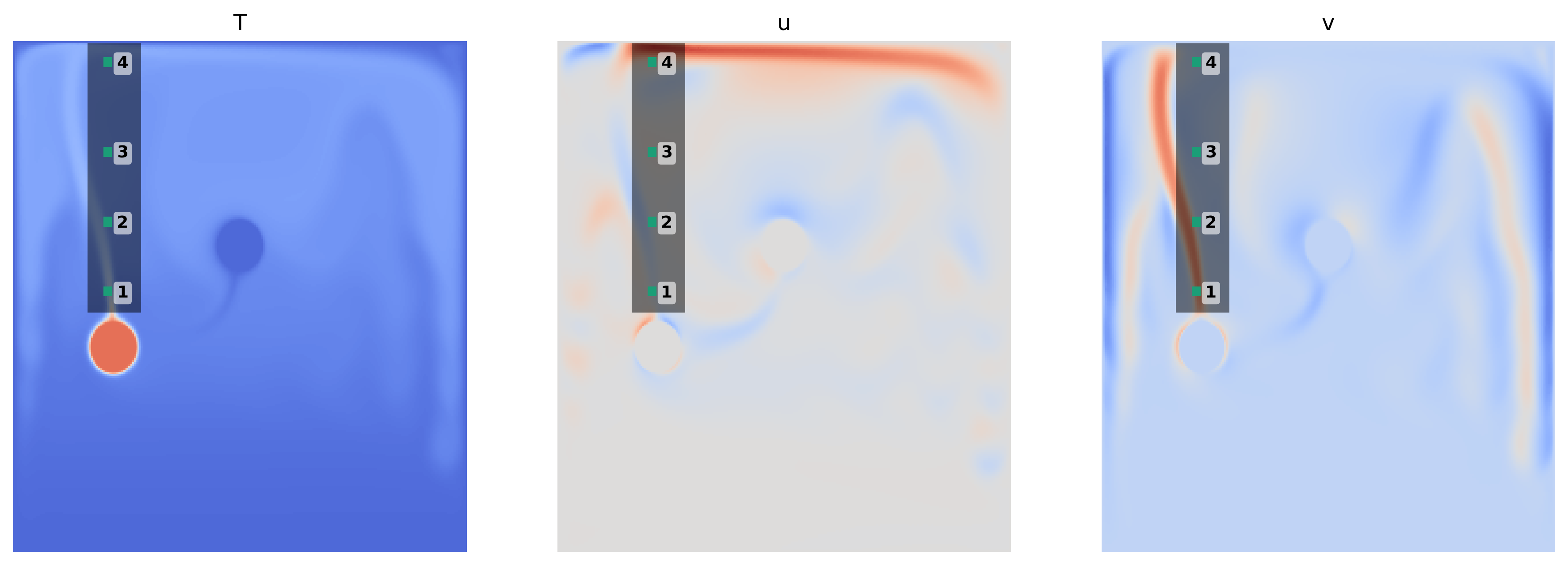}
    \caption{Experiment $x_a=0.22$: Placement of a cropped box $B$ (gray) and fixed spatial points (green) to compute spatial and temporal correlations.}
    \label{fig:corr_setup022}
\end{figure}

\begin{figure}[htbp]
    \centering
    \begin{subfigure}{\textwidth}
        \centering
        \includegraphics[width=0.32\textwidth]{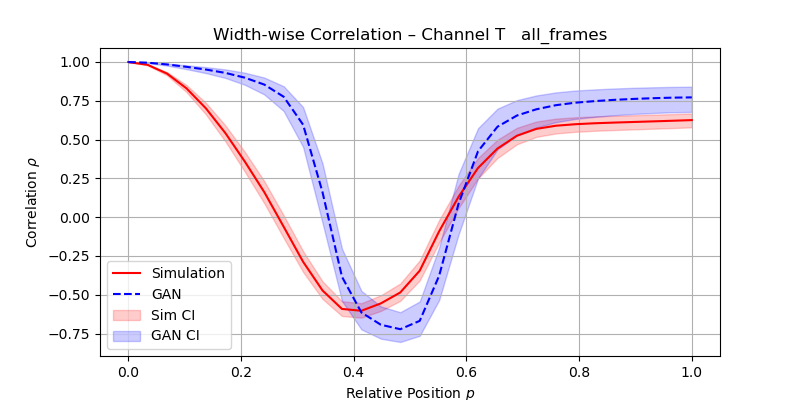}\hfill
        \includegraphics[width=0.32\textwidth]{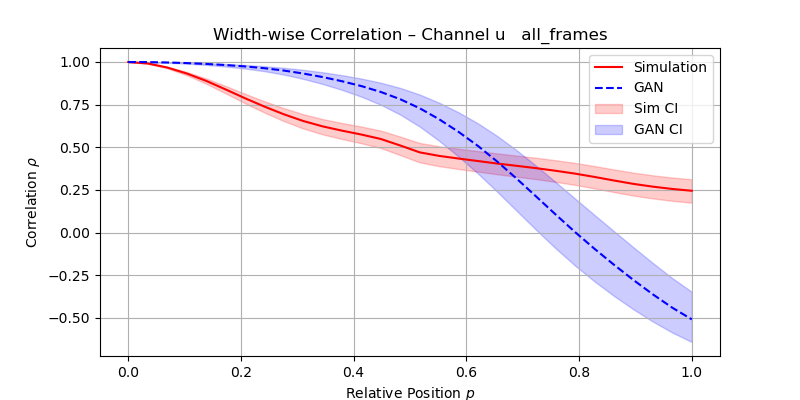}\hfill
        \includegraphics[width=0.32\textwidth]{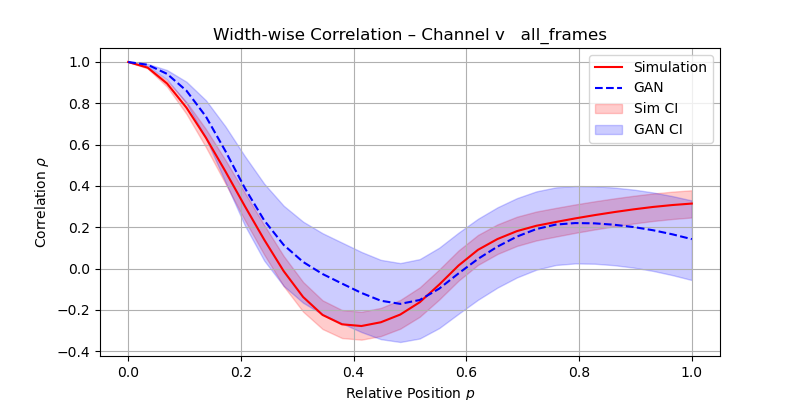}
        \caption{Column wise spatial correlations of $T$, $u$, and $v$.}
    \end{subfigure}

    \begin{subfigure}{\textwidth}
        \centering
        \includegraphics[width=0.32\textwidth]{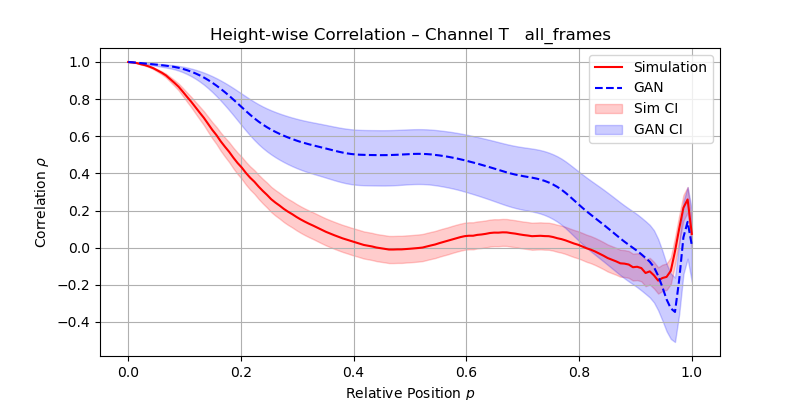}\hfill
        \includegraphics[width=0.32\textwidth]{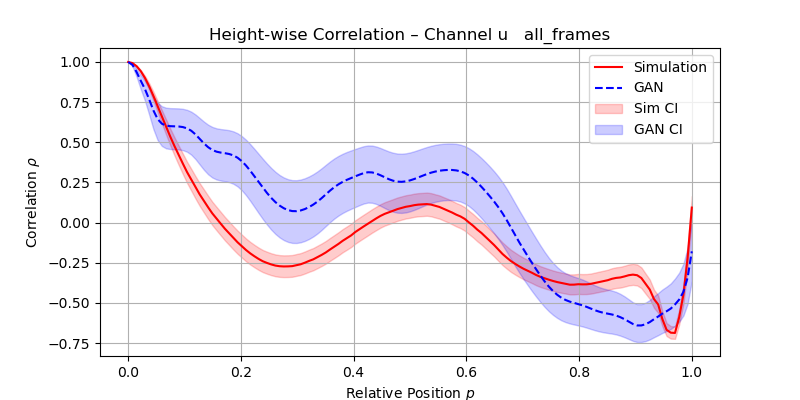}\hfill
        \includegraphics[width=0.32\textwidth]{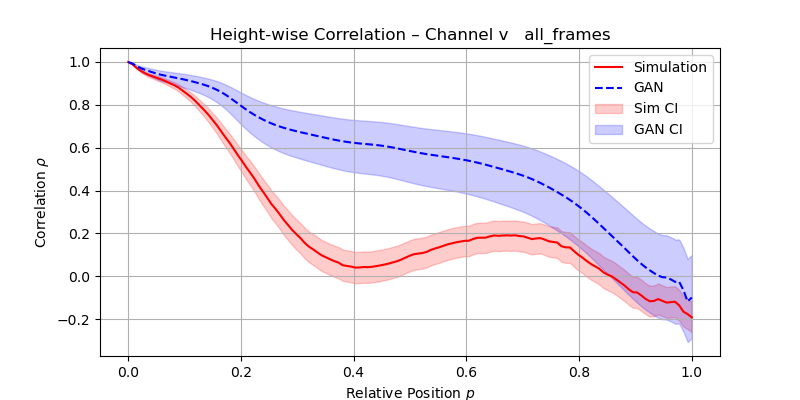}
        \caption{Row wise spatial correlations of $T$, $u$, and $v$.}
    \end{subfigure}
    \begin{subfigure}{\textwidth}
        \centering
        \includegraphics[width=0.32\textwidth]{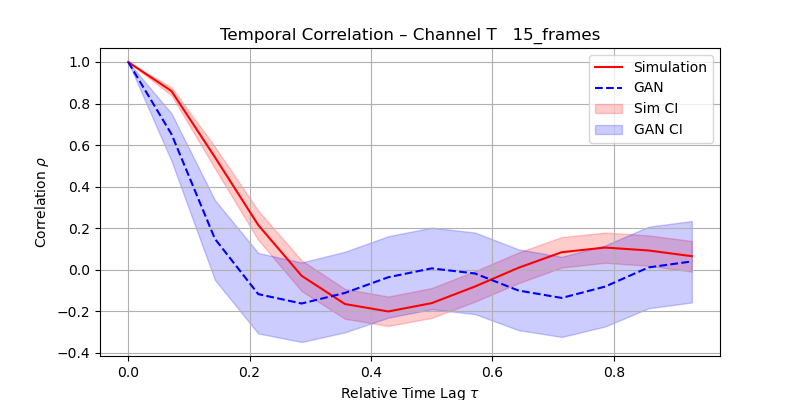}\hfill
        \includegraphics[width=0.32\textwidth]{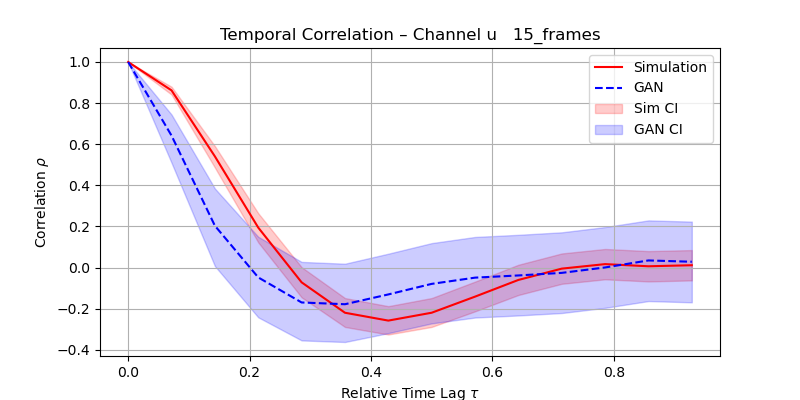}\hfill
        \includegraphics[width=0.32\textwidth]{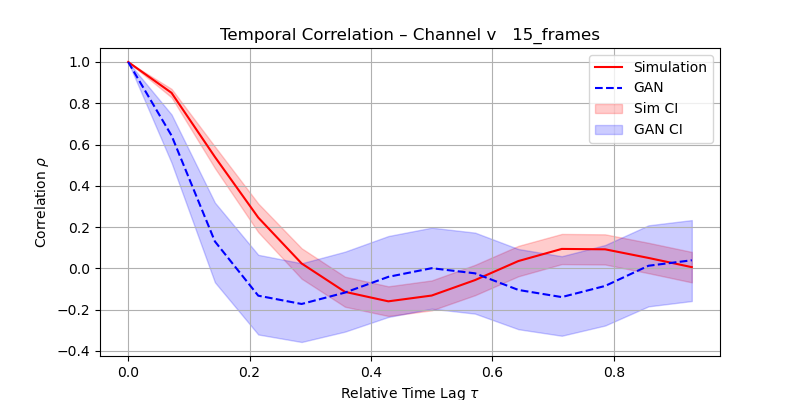}
        \caption{Temporal correlations of $T$, $u$, and $v$ in fixed spatial point 1.}
    \end{subfigure}

    \begin{subfigure}{\textwidth}
        \centering
        \includegraphics[width=0.32\textwidth]{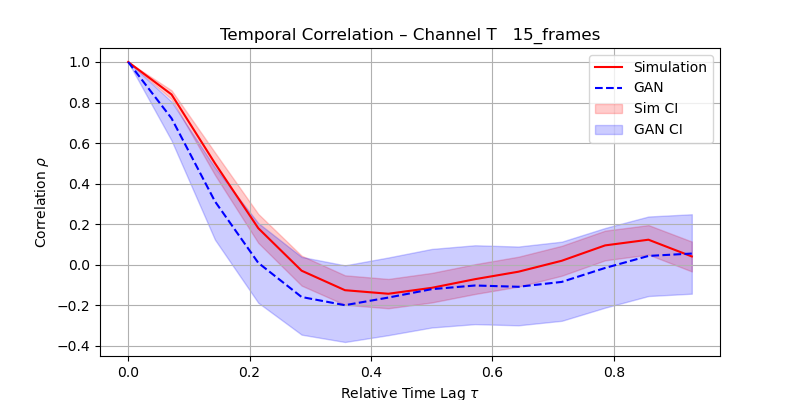}\hfill
        \includegraphics[width=0.32\textwidth]{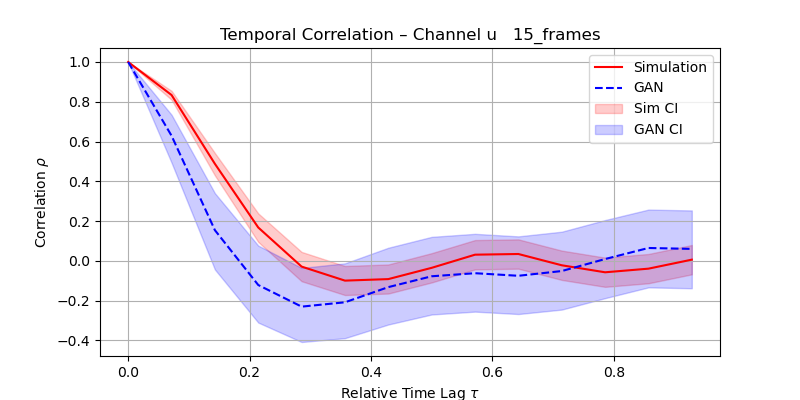}\hfill
        \includegraphics[width=0.32\textwidth]{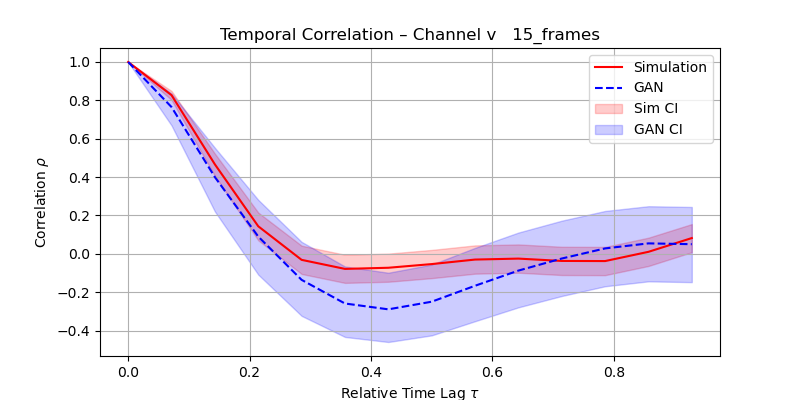}
        \caption{Temporal correlations of $T$, $u$, and $v$ in fixed spatial point 2.}
    \end{subfigure}

    \begin{subfigure}{\textwidth}
        \centering
        \includegraphics[width=0.32\textwidth]{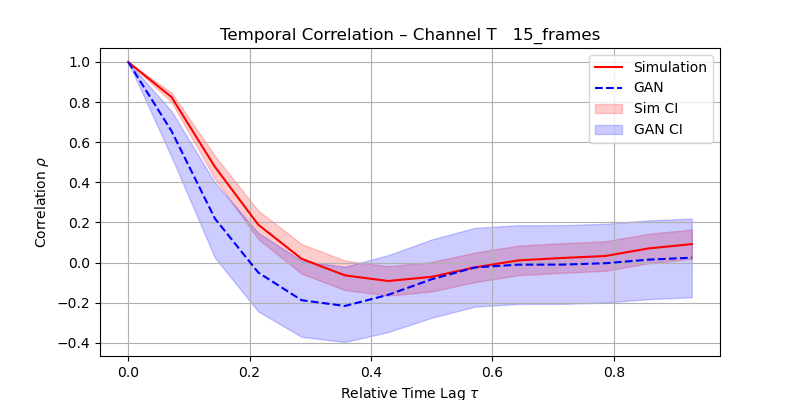}\hfill
        \includegraphics[width=0.32\textwidth]{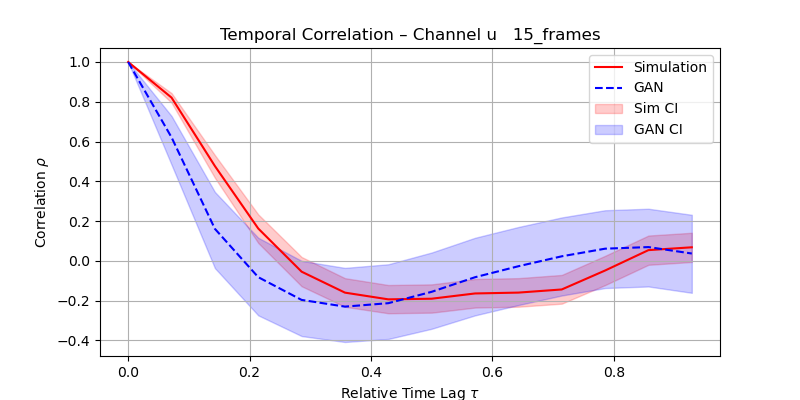}\hfill
        \includegraphics[width=0.32\textwidth]{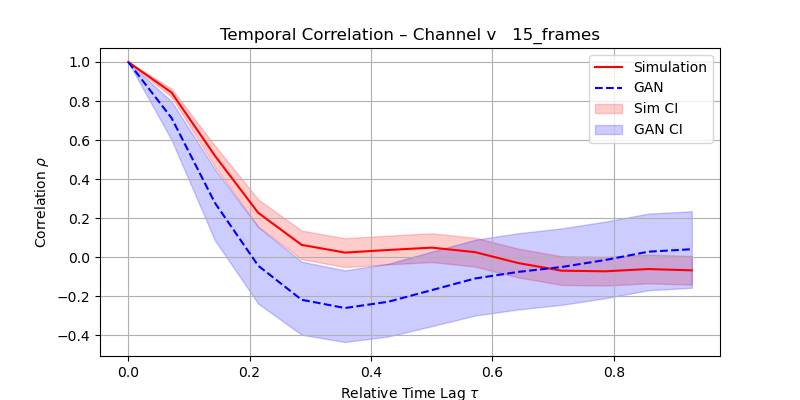}
        \caption{Temporal correlations of $T$, $u$, and $v$ in fixed spatial point 3.}
    \end{subfigure}

    \begin{subfigure}{\textwidth}
        \centering
        \includegraphics[width=0.32\textwidth]{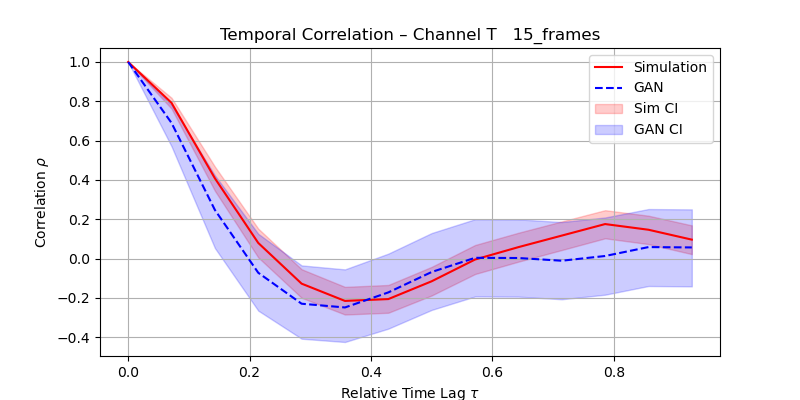}\hfill
        \includegraphics[width=0.32\textwidth]{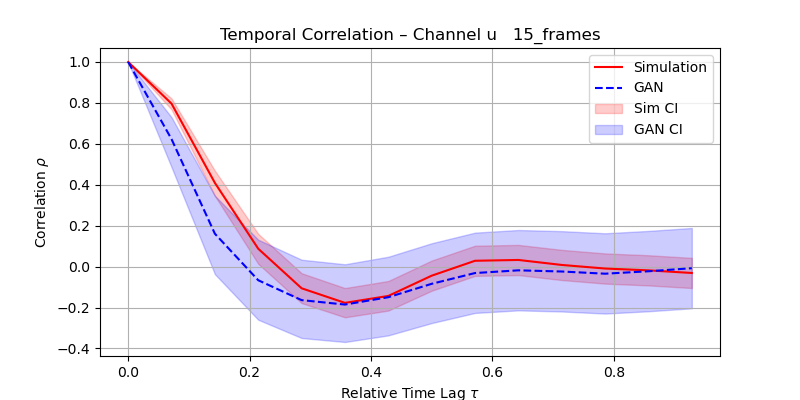}\hfill
        \includegraphics[width=0.32\textwidth]{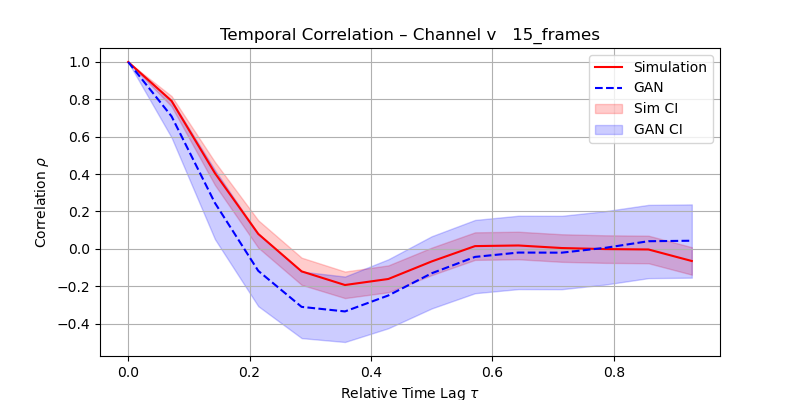}
        \caption{Temporal correlations of $T$, $u$, and $v$ in fixed spatial point 4.}
        
    \end{subfigure}

    \caption{Experiment $x_a=0.22$: Comparison of spatial and temporal correlations of simulation data (red) and GAN generated data (blue), including confidence intervals (shaded), for the fields $T$ (left), $u$ (middle), and $v$ (right).}
    \label{fig:correlations_xc022}
\end{figure}

In the experiment with $x_a=0.22$, the lower circle is near left boundary of the domain. The GAN generated frames and simulation frames are compared in Figure~\ref{fig:comp_frames022}. We observe more profound differences between the generated and simulation frames. This can be further observed in the variance plots in Figure~\ref{fig:mean_variance022} and correlation plots in Figure~\ref{fig:correlations_xc022} w.r.t.\ to the measuring points depicted in Figure~\ref{fig:corr_setup022}. The variance is still approximated reasonably well, although not as accurately as in the reproduction experiments. 
The temporal correlations are captured adequately; however, the spatial correlation curves are less accurately approximated than those reported in 
the reproduction experiment.

\section{Conclusion and Outlook}
\label{sec:conclusion_and_outlook}

 In this paper, we demonstrate for the first time that world model architectures derived from video generation neural networks are capable to learn transient physics. In particular, we consider a coupled, multi-physics heat transfer problem with internal heat flux in a solid and external buoyancy driven heat transfer in a surrounding fluid. The application case is the management of substations in volatile energy grids.  Previous operator learning methods were only shown to successfully learn instationary physics, like turbulent flow, without long term drift which is characteristic for transience.

At the same time, our model is capable of learning multi-physics and is geometry-aware. The generation times for new, previously unseen fluid flow is a few hundreds of milliseconds, making a complex physics simulation task available with almost no latency.

Despite progress being made, some limitations of our model require consideration. While the model is able to simulate transient behavior, it requires  filtering  during generation as there are instances where the model jumps to the start. Here, conditioning in time is a logical next step.  

Also, as our invesigation to generation close to birfucation points \cite{kielhofer2012bifurcation}  has shown, data generation and training operator learning models on dynamical systems which undergo bifurcation is a promising direction for future research.

Furthermore, our model is trained on and learns physics in two spatial dimensions. Despite the architecture of our world model can be extended to 3D in principle, this will come with elevated hardware requirements. We therefore believe that further architecture innovations should accompany this step. E.g., recent results using 3D graph neural networks are  promising to be efficient in learning and generation in 3D \cite{ahangarkiasari2025multistage,fortunato2022multiscale,li2023finite,pfaff2021learning,schmocker2024generalization}.

Geometries that our approach is capable to process are flexible, as they can freely be specified via 2D masks. However, we only tested on a small, parametric subset of these. A test that our architecture is capable to handle a much more generic space of geometries and generalize within it at present is unresolved. We suggest that automated simulation tool-chains that produce data for the world model to learn from in just in time pipelines might be the way to go \cite{wu2023cad,sun2025large}.  

\paragraph{Acknowledgements.}  The authors thank Claudia Drygala, Francesca di Mare and Edmund Ross for interesting discussions. This work was funded by the German Research Council (DFG) through the center of excellence Math+ under the project PaA-5 "AI Based Simulation of Transient Physical Systems – From Benchmarks to Hybrid Solutions". Hanno Gottschalk also acknowledges financial support from the German research council through SPP2403 "Carnot Batteries" project GO 833/8-1 "Inverse aerodynamic design of turbo components for Carnot batteries by means of physics
informed networks enhanced by generative learning".

\bibliographystyle{abbrv}
\bibliography{references}

\pagebreak

\appendix
\section{Additional Results}
\label{app:additional_results}
\paragraph{Reproduction.}

\begin{figure}[htbp]
\centering
    \begin{subfigure}[b]{0.48\textwidth}
        \centering
        \includegraphics[width=\textwidth]{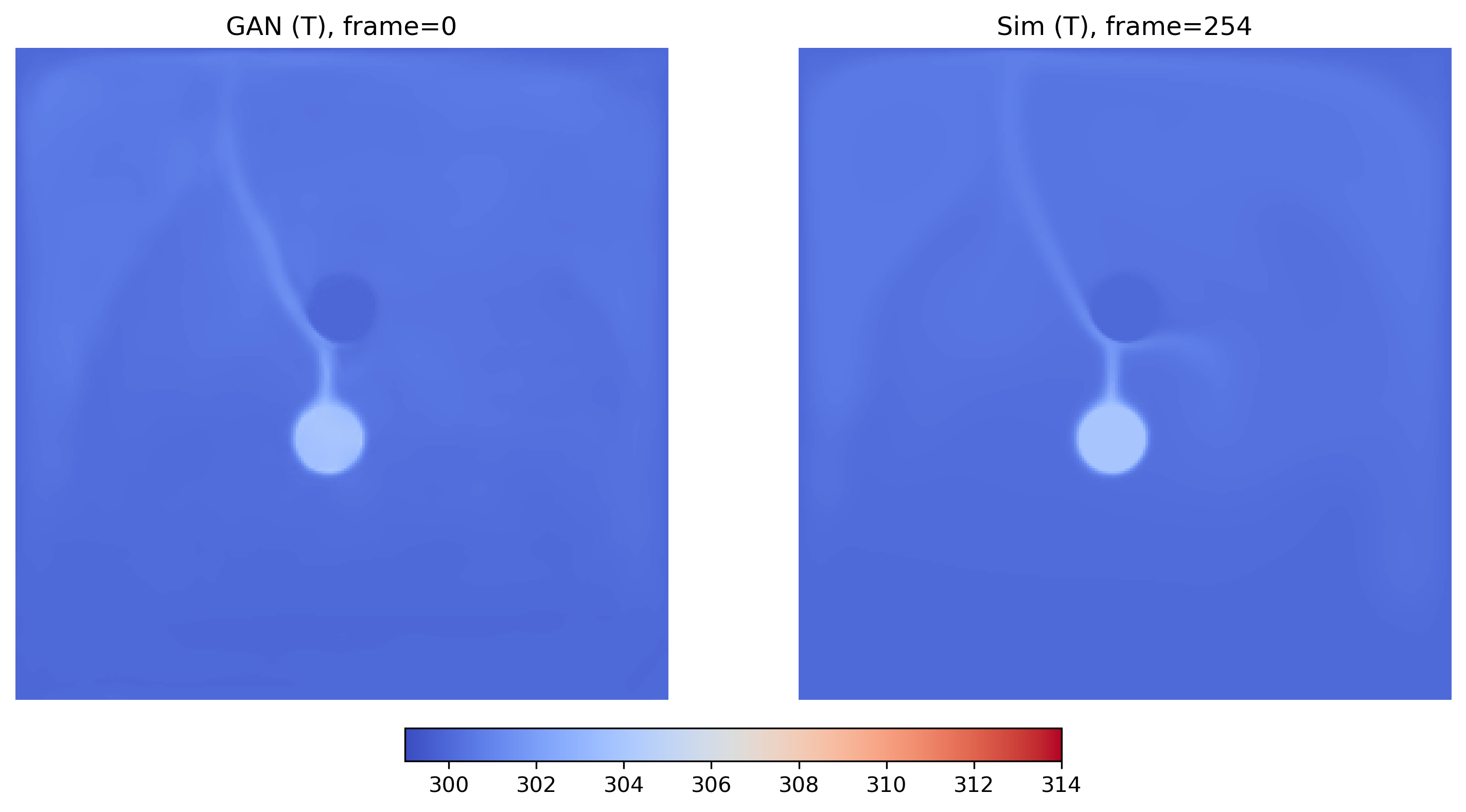}
        \caption{Comparison of $T$.}
    \end{subfigure} 
    \hfill
    \begin{subfigure}[b]{0.48\textwidth}
        \centering
        \includegraphics[width=\textwidth]{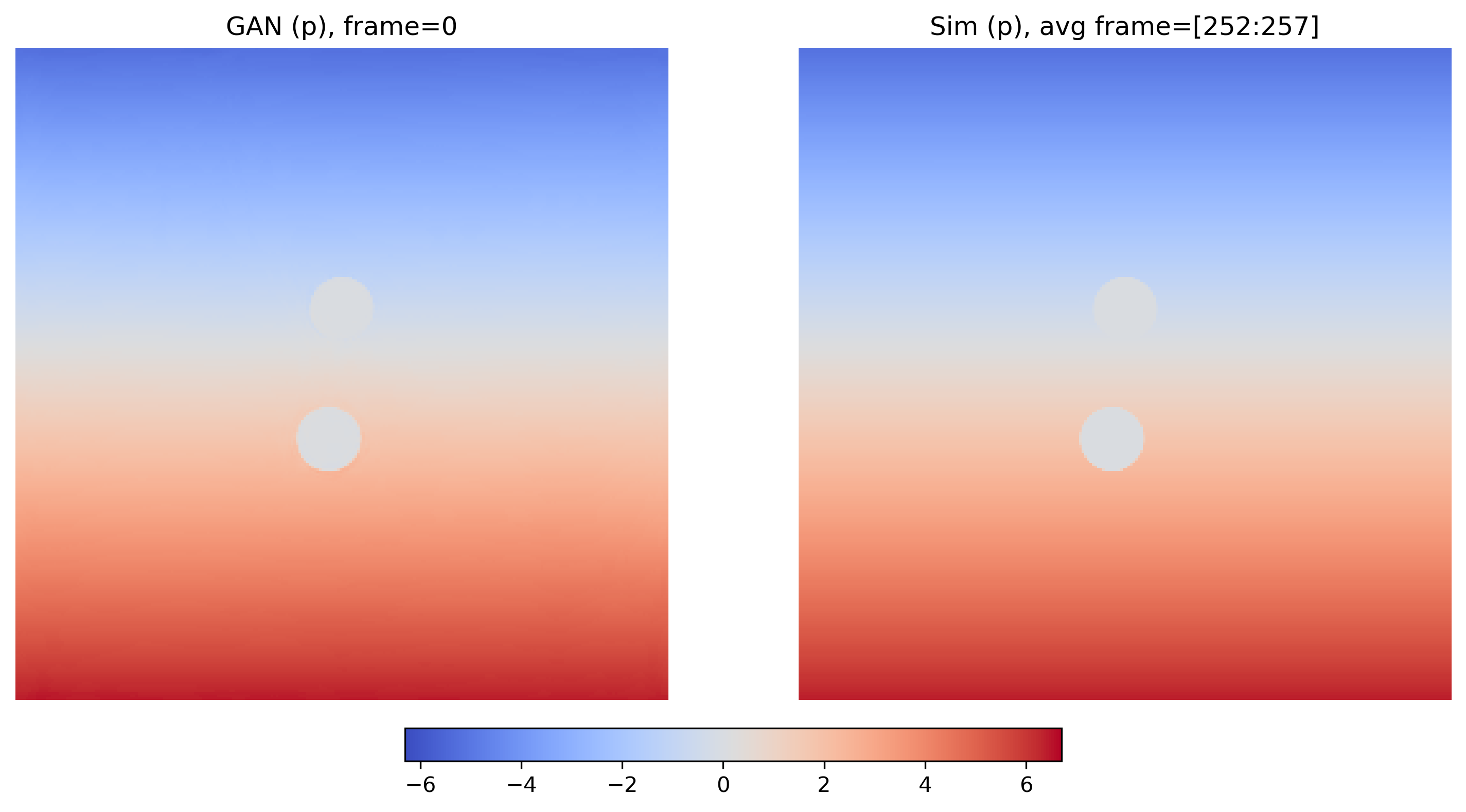}
        \caption{Comparison of $p$.}
    \end{subfigure}   
    \hfill
    \begin{subfigure}[b]{0.48\textwidth}
        \centering
        \includegraphics[width=\textwidth]{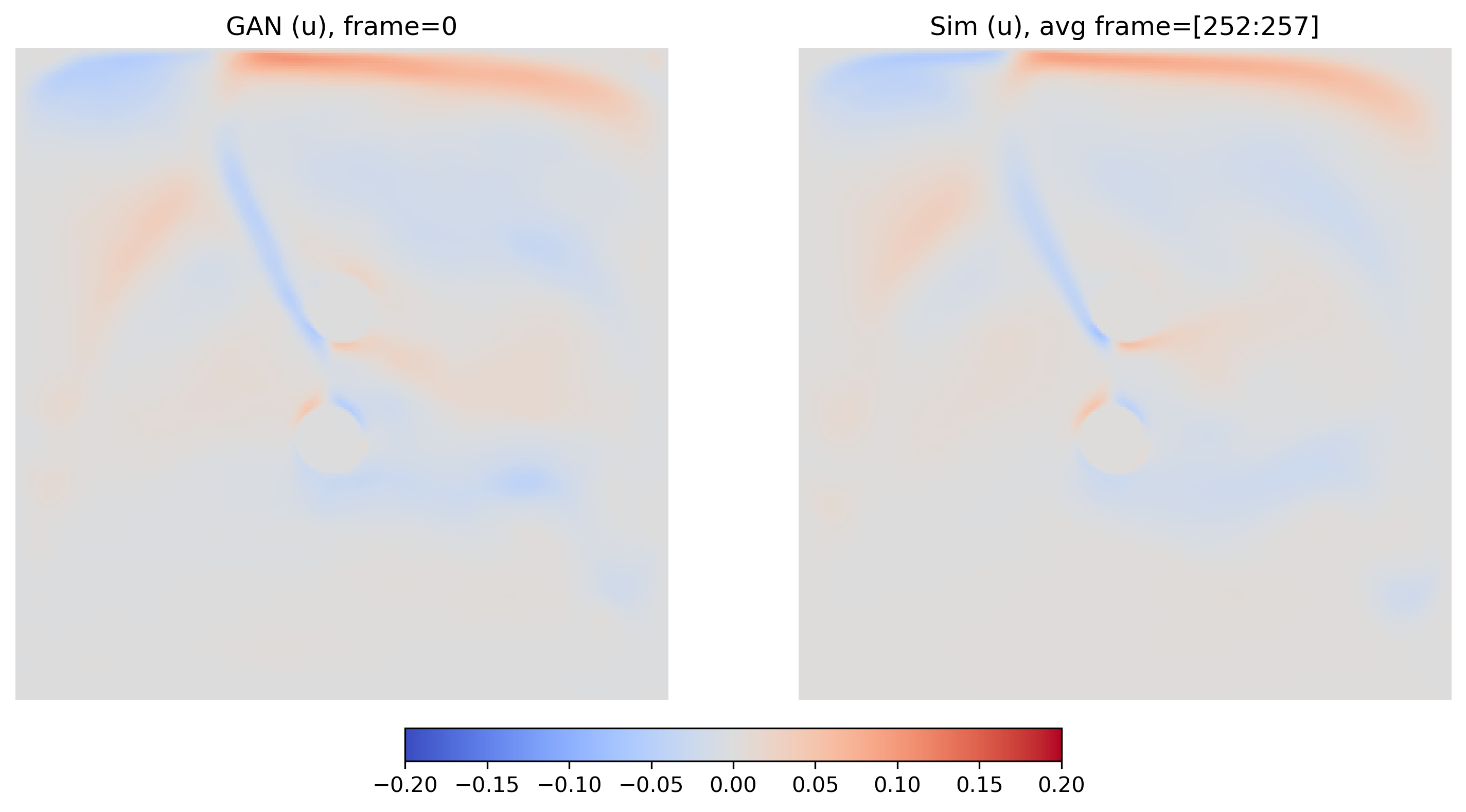}
        \caption{Comparison of $u$.}
    \end{subfigure}
    \hfill
    \begin{subfigure}[b]{0.48\textwidth}
        \centering
        \includegraphics[width=\textwidth]{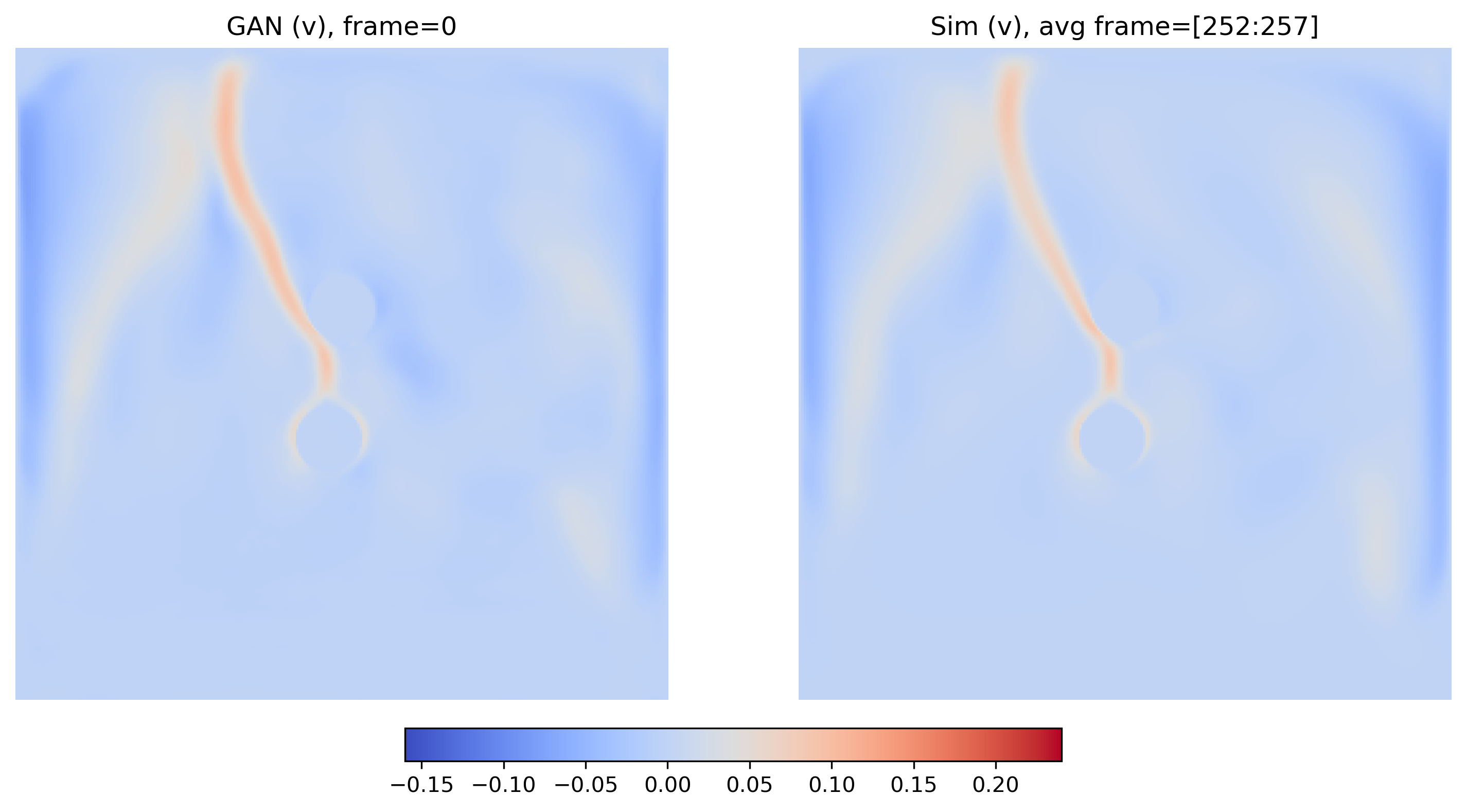}
        \caption{Comparison of $v$.}
    \end{subfigure}

    \caption{Experiment $x_a=0.48$: Comparison of GAN generated frames and simulation frames. The fields $u$ and $v$ are compared with an average over $5$ simulation frames.}
    \label{fig:comp_frames048}
\end{figure}

\begin{figure}[htbp]
    \centering
    \begin{subfigure}[b]{0.48\textwidth}
        \centering
        \includegraphics[width=\textwidth]{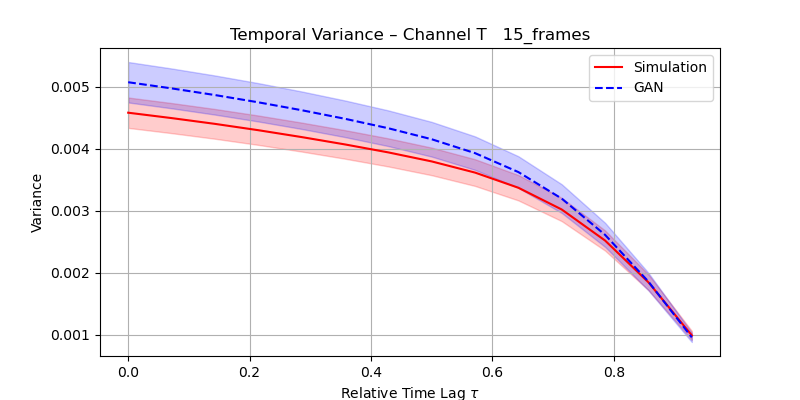}
        \caption{Comparison of $T$.}
    \end{subfigure}
    \hfill
    \begin{subfigure}[b]{0.48\textwidth}
        \centering
        \includegraphics[width=\textwidth]{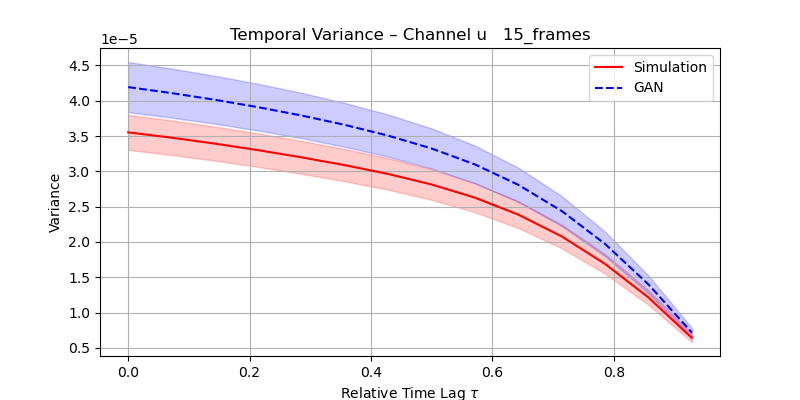}
        \caption{Comparison of $u$.}
    \end{subfigure}
    \hfill
    \begin{subfigure}[b]{0.48\textwidth}
        \centering
        \includegraphics[width=\textwidth]{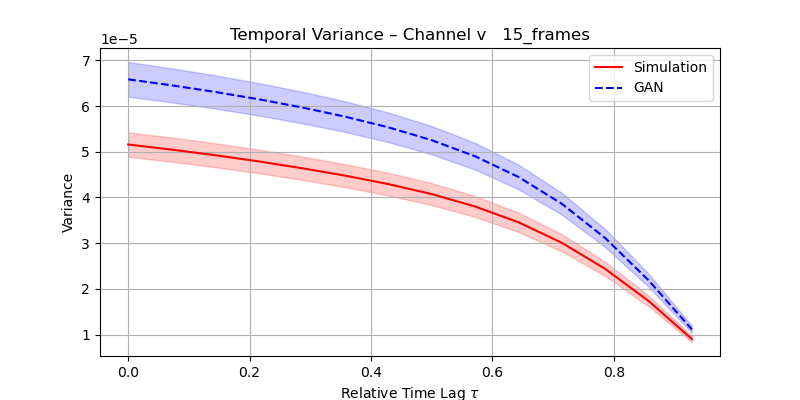}
        \caption{Comparison of $v$.}
    \end{subfigure}

    \caption{Experiment $x_a=0.48$: comparison of mean variance values computed via \eqref{eq:mean_variance} of generated (blue) and simulation frames (red). }
    \label{fig:mean_variance048}
\end{figure}

\begin{figure}[htbp]
    \centering
    \includegraphics[width=\linewidth]{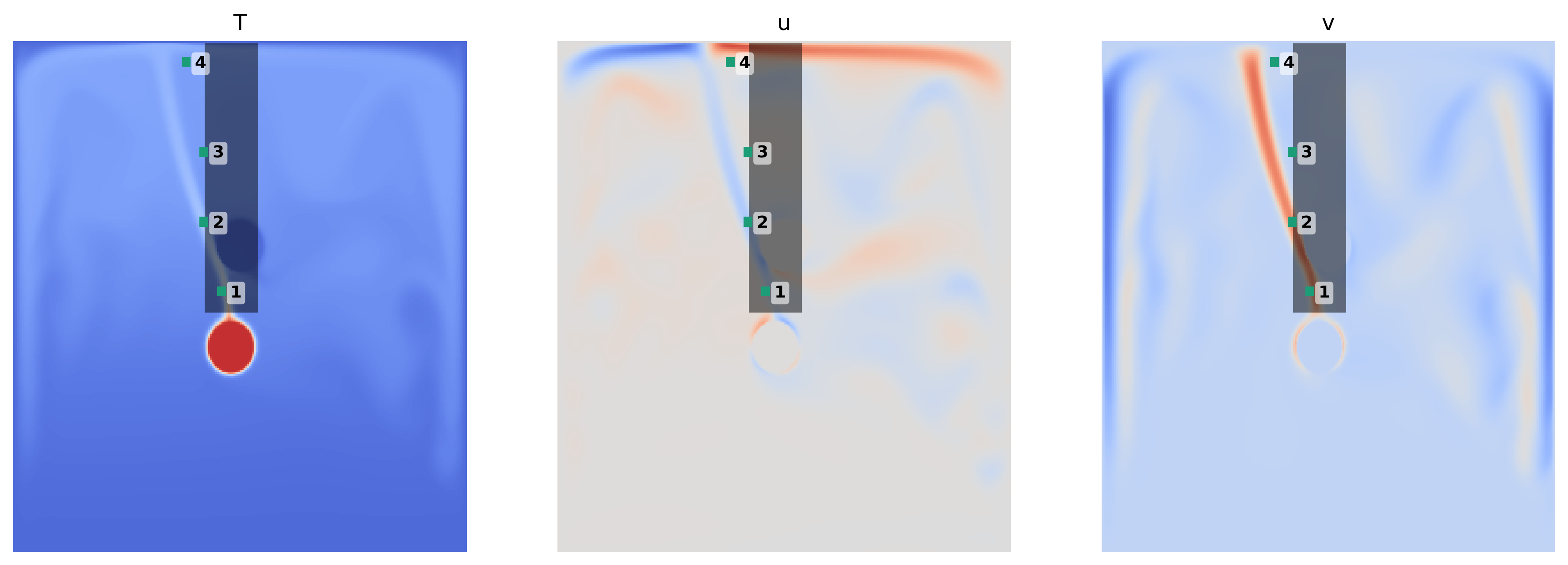}
    \caption{Experiment $x_a=0.48$: Placement of a cropped box $B$ (gray) and fixed spatial points (green) to compute spatial and temporal correlations.}
    \label{fig:corr_setup048}
\end{figure}

\begin{figure}[htbp]
    \centering
    \begin{subfigure}{\textwidth}
        \centering
        \includegraphics[width=0.32\textwidth]{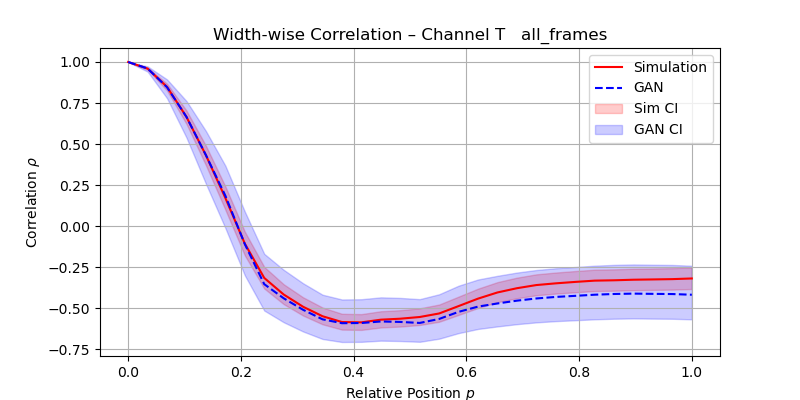}\hfill
        \includegraphics[width=0.32\textwidth]{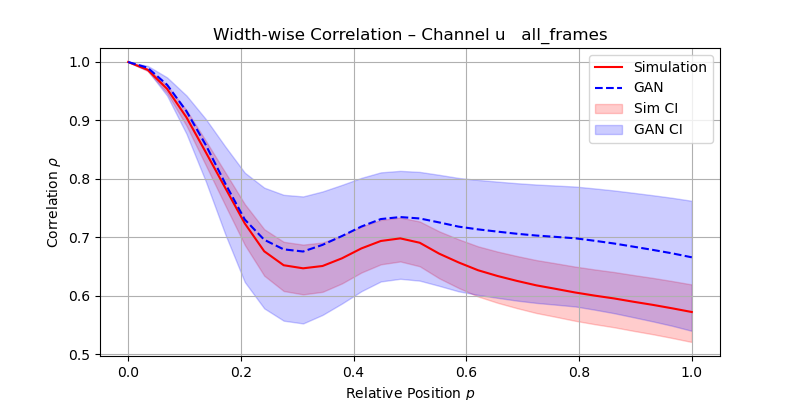}\hfill
        \includegraphics[width=0.32\textwidth]{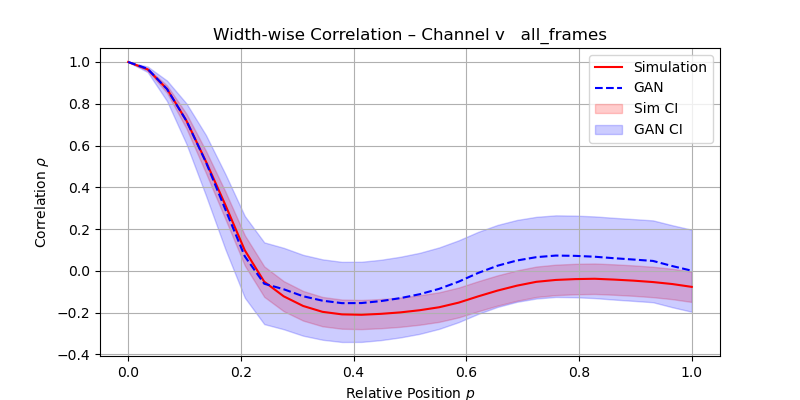}
        \caption{Column wise spatial correlations of $T$, $u$, and $v$.}
    \end{subfigure}

    \begin{subfigure}{\textwidth}
        \centering
        \includegraphics[width=0.32\textwidth]{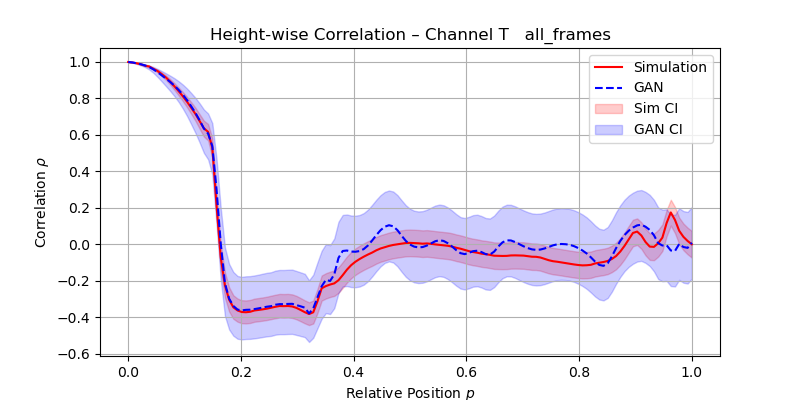}\hfill
        \includegraphics[width=0.32\textwidth]{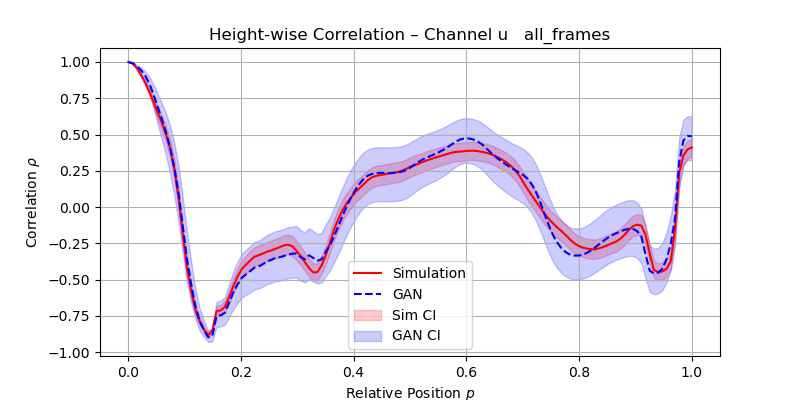}\hfill
        \includegraphics[width=0.32\textwidth]{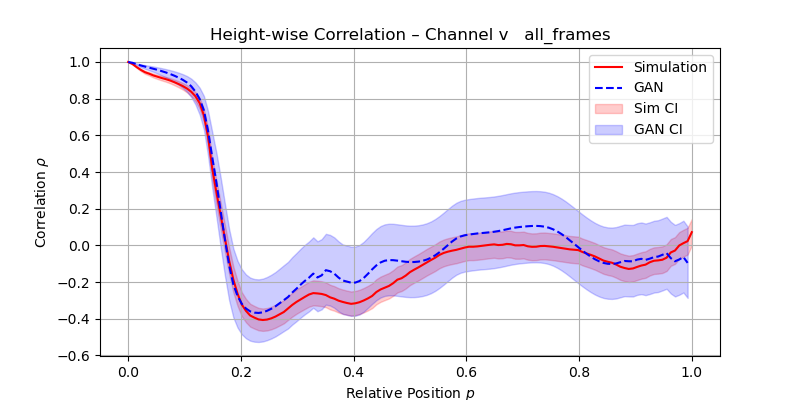}
        \caption{Row wise spatial correlations of $T$, $u$, and $v$.}
    \end{subfigure}
    \begin{subfigure}{\textwidth}
        \centering
        \includegraphics[width=0.32\textwidth]{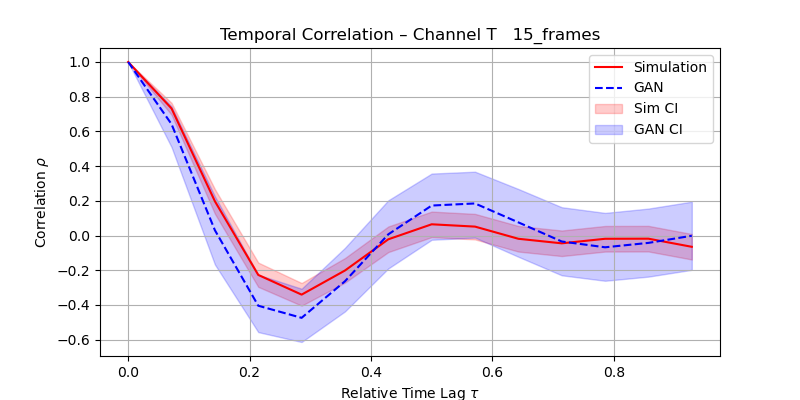}\hfill
        \includegraphics[width=0.32\textwidth]{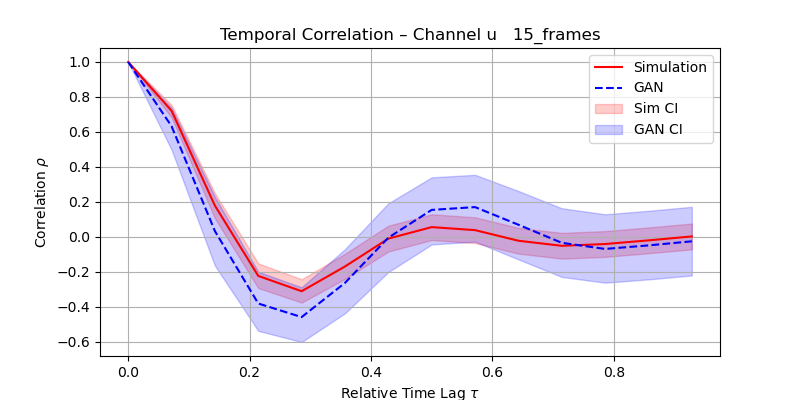}\hfill
        \includegraphics[width=0.32\textwidth]{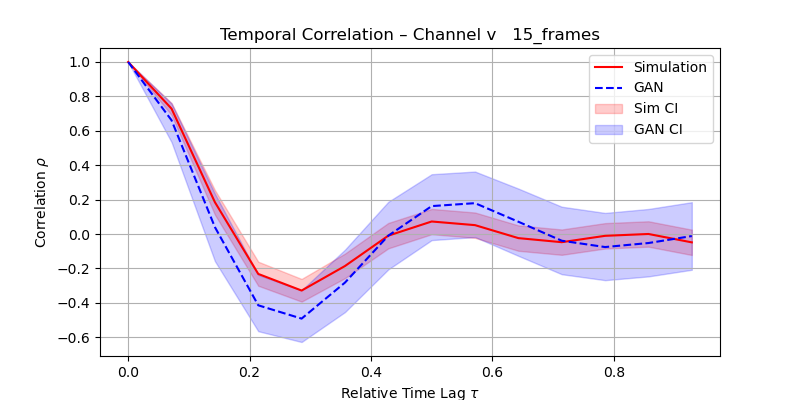}
        \caption{Temporal correlations of $T$, $u$, and $v$ in fixed spatial point 1.}
    \end{subfigure}

    \begin{subfigure}{\textwidth}
        \centering
        \includegraphics[width=0.32\textwidth]{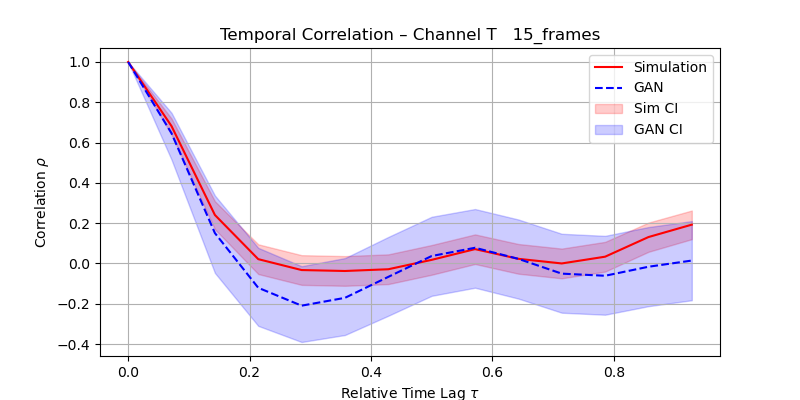}\hfill
        \includegraphics[width=0.32\textwidth]{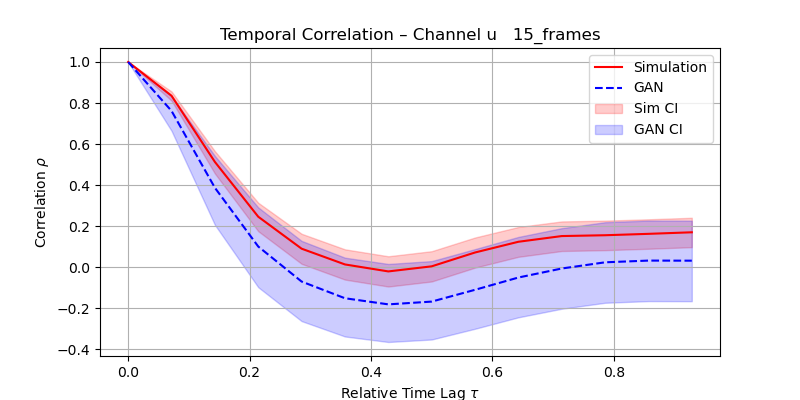}\hfill
        \includegraphics[width=0.32\textwidth]{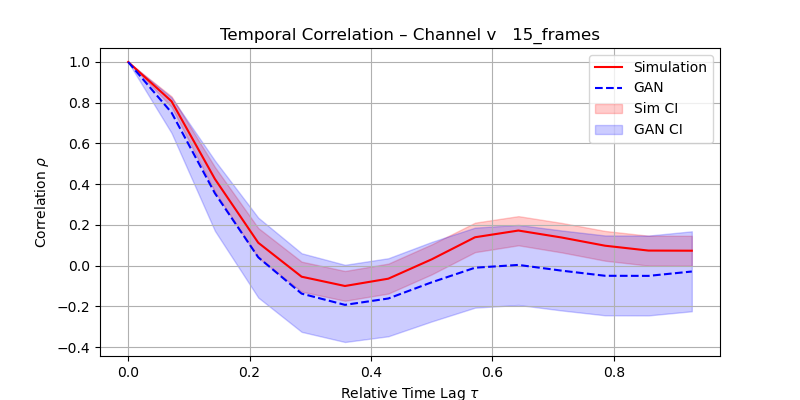}
        \caption{Temporal correlations of $T$, $u$, and $v$ in fixed spatial point 2.}
    \end{subfigure}

    \begin{subfigure}{\textwidth}
        \centering
        \includegraphics[width=0.32\textwidth]{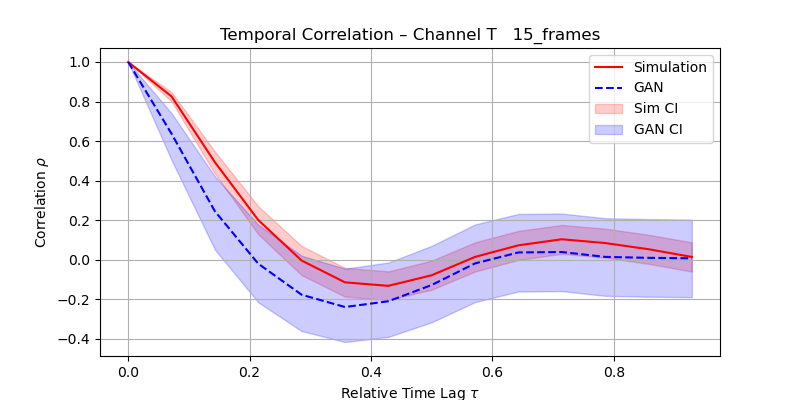}\hfill
        \includegraphics[width=0.32\textwidth]{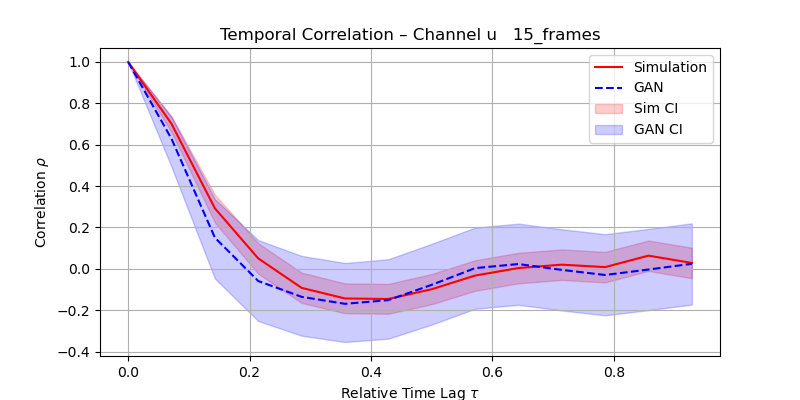}\hfill
        \includegraphics[width=0.32\textwidth]{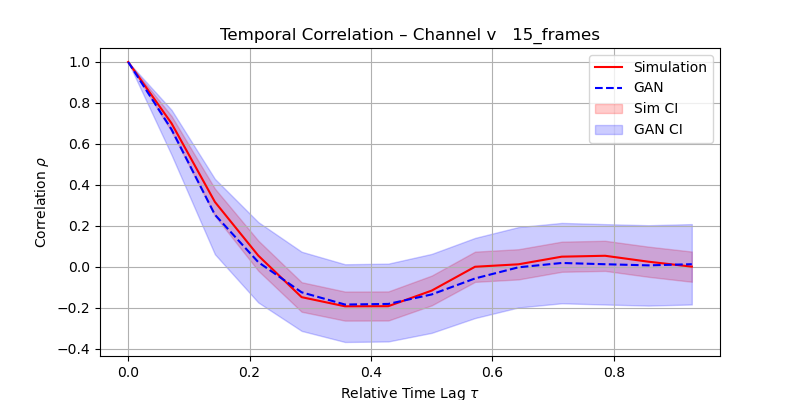}
        \caption{Temporal correlations of $T$, $u$, and $v$ in fixed spatial point 3.}
    \end{subfigure}

    \begin{subfigure}{\textwidth}
        \centering
        \includegraphics[width=0.32\textwidth]{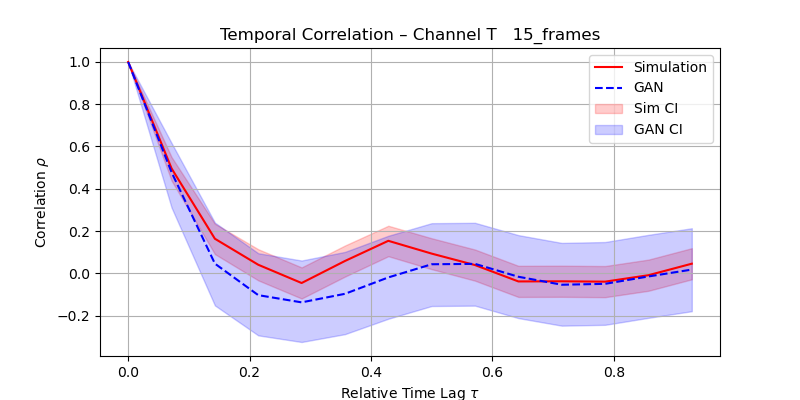}\hfill
        \includegraphics[width=0.32\textwidth]{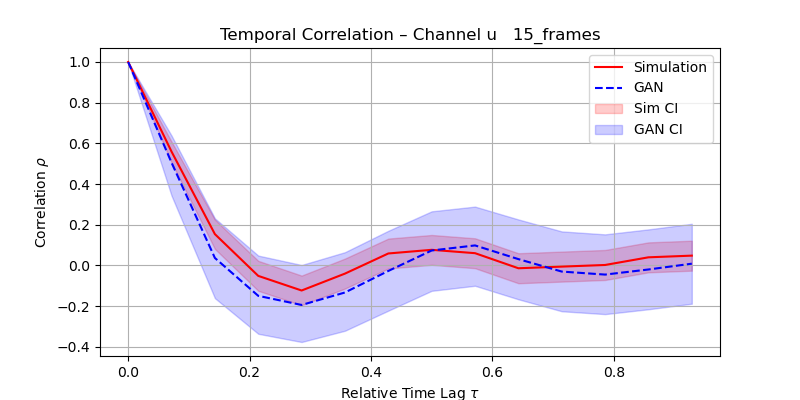}\hfill
        \includegraphics[width=0.32\textwidth]{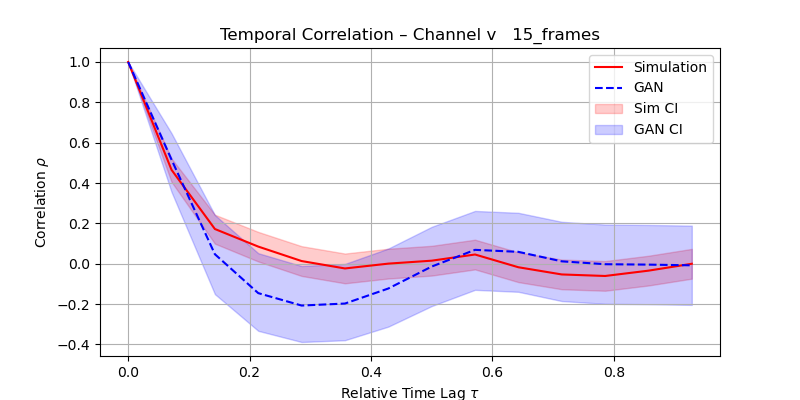}
        \caption{Temporal correlations of $T$, $u$, and $v$ in fixed spatial point 4.}
        
    \end{subfigure}

    \caption{Experiment $x_a=0.48$: Comparison of spatial and temporal correlations of simulation data (red) and GAN generated data (blue), including confidence intervals (shaded), for the fields $T$ (left), $u$ (middle), and $v$ (right).}
    \label{fig:correlations_xc048}
\end{figure}

In our experiment, consider $x_a=0.48$.
In this case, the lower circle is placed under the passive conductor (with $x_a=0.50$) and slightly moved to the left such that the upward thermal flow is split into two streams by the passive conductor. As  before we compare the GAN generated frames and simulation frames in Figure~\ref{fig:comp_frames048} and the variance in Figure~\ref{fig:mean_variance048}. Furthermore, the correlation plots and their corresponding bounding box $B$ and measuring points are depicted in Figures~\ref{fig:corr_setup048} and \ref{fig:correlations_xc048}. Also for this set of conditioning information the correlations and direct comparison of frames and averages of frames are satisfying. The model successfully learned the split of upward thermal flow into two streams, capturing the bifurcation.

\paragraph{Generalization.}

\begin{figure}[htbp]
\centering
    \begin{subfigure}[b]{0.48\textwidth}
        \centering
        \includegraphics[width=\textwidth]{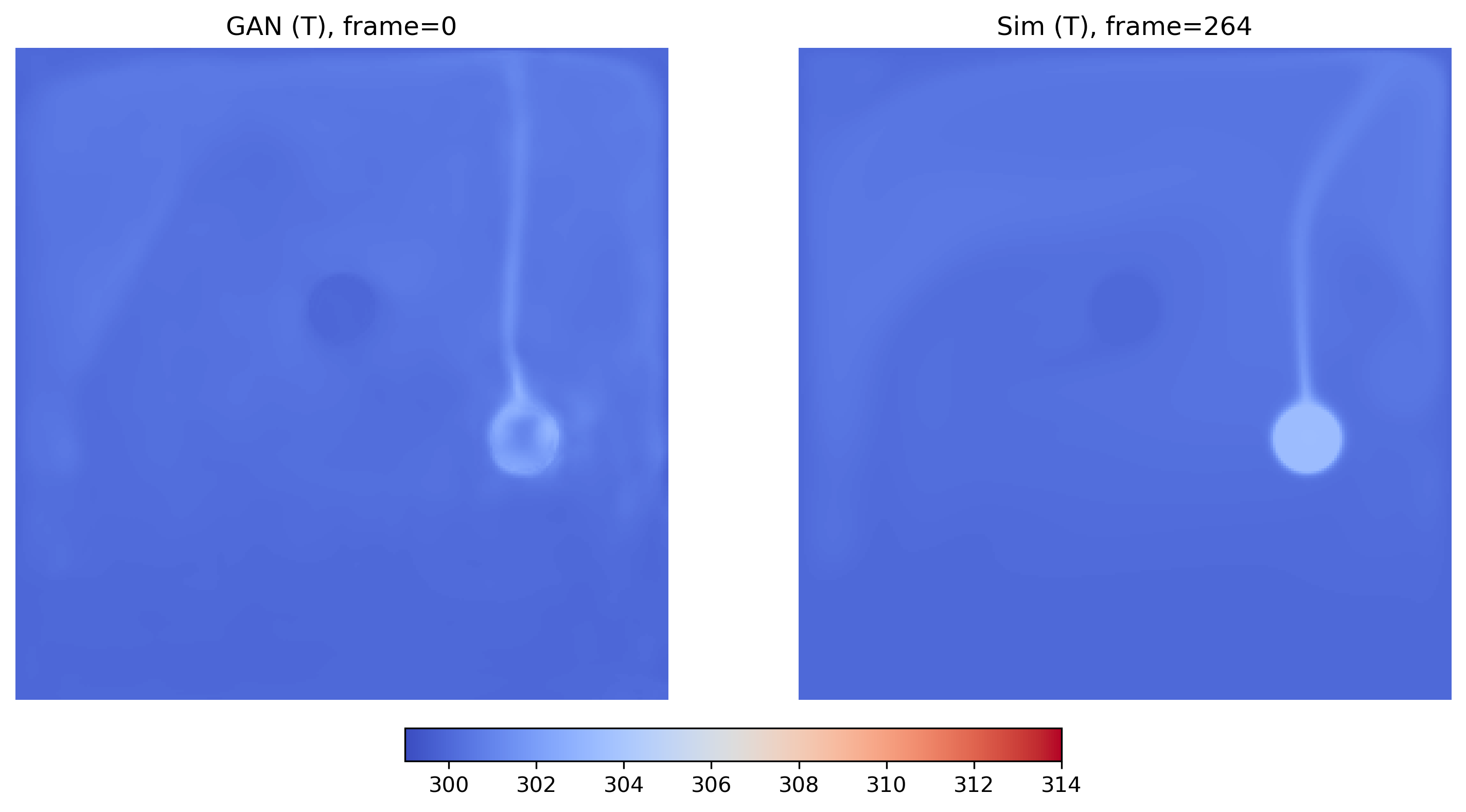}
        \caption{Comparison of $T$.}
    \end{subfigure} 
    \hfill
    \begin{subfigure}[b]{0.48\textwidth}
        \centering
        \includegraphics[width=\textwidth]{fig/frame_comparison/simdata_ref_a_xclower=0.22_tensors_ch3_s2_p.png}
        \caption{Comparison of $p$.}
    \end{subfigure}   
    \hfill
    \begin{subfigure}[b]{0.48\textwidth}
        \centering
        \includegraphics[width=\textwidth]{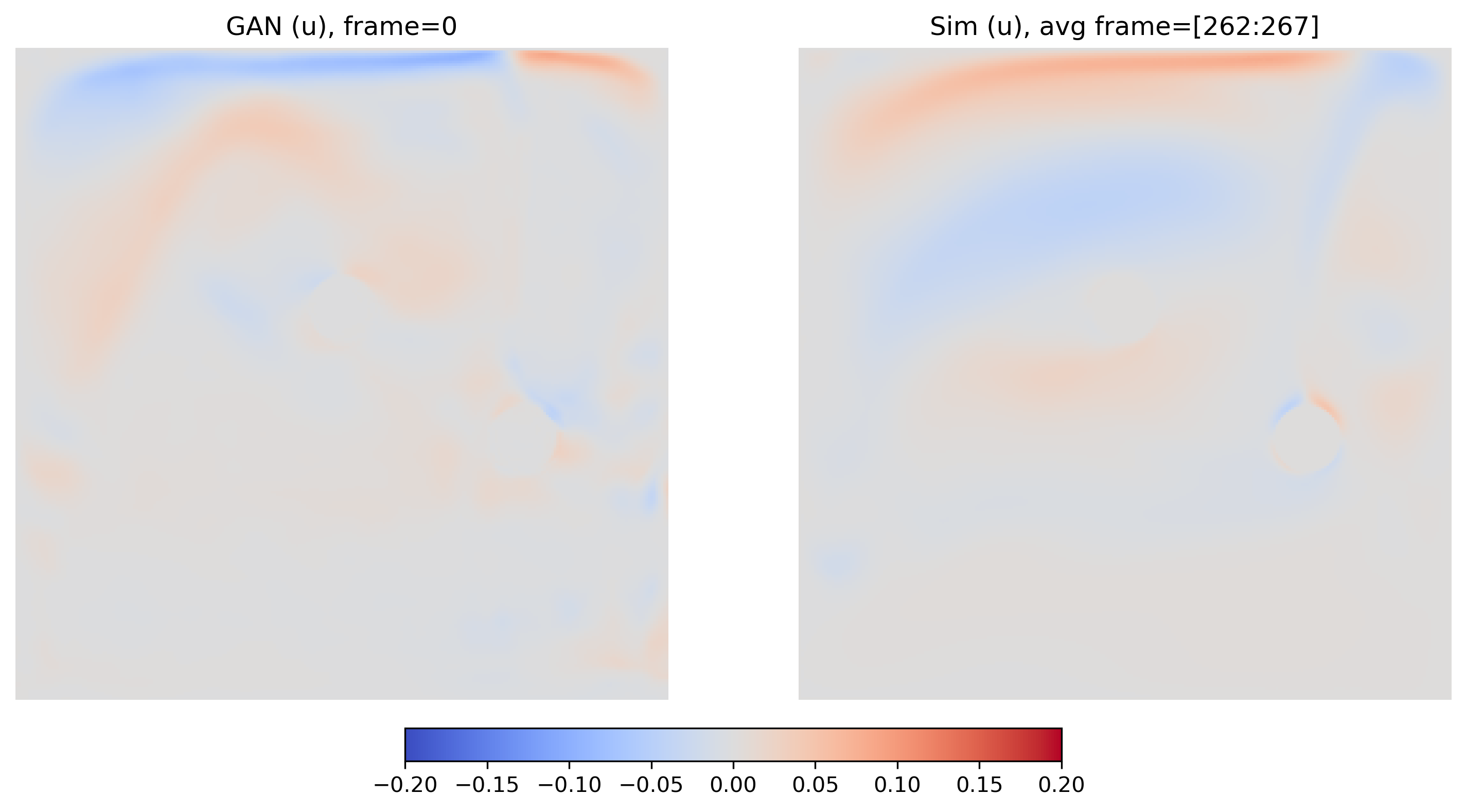}
        \caption{Comparison of $u$.}
    \end{subfigure}
    \hfill
    \begin{subfigure}[b]{0.48\textwidth}
        \centering
        \includegraphics[width=\textwidth]{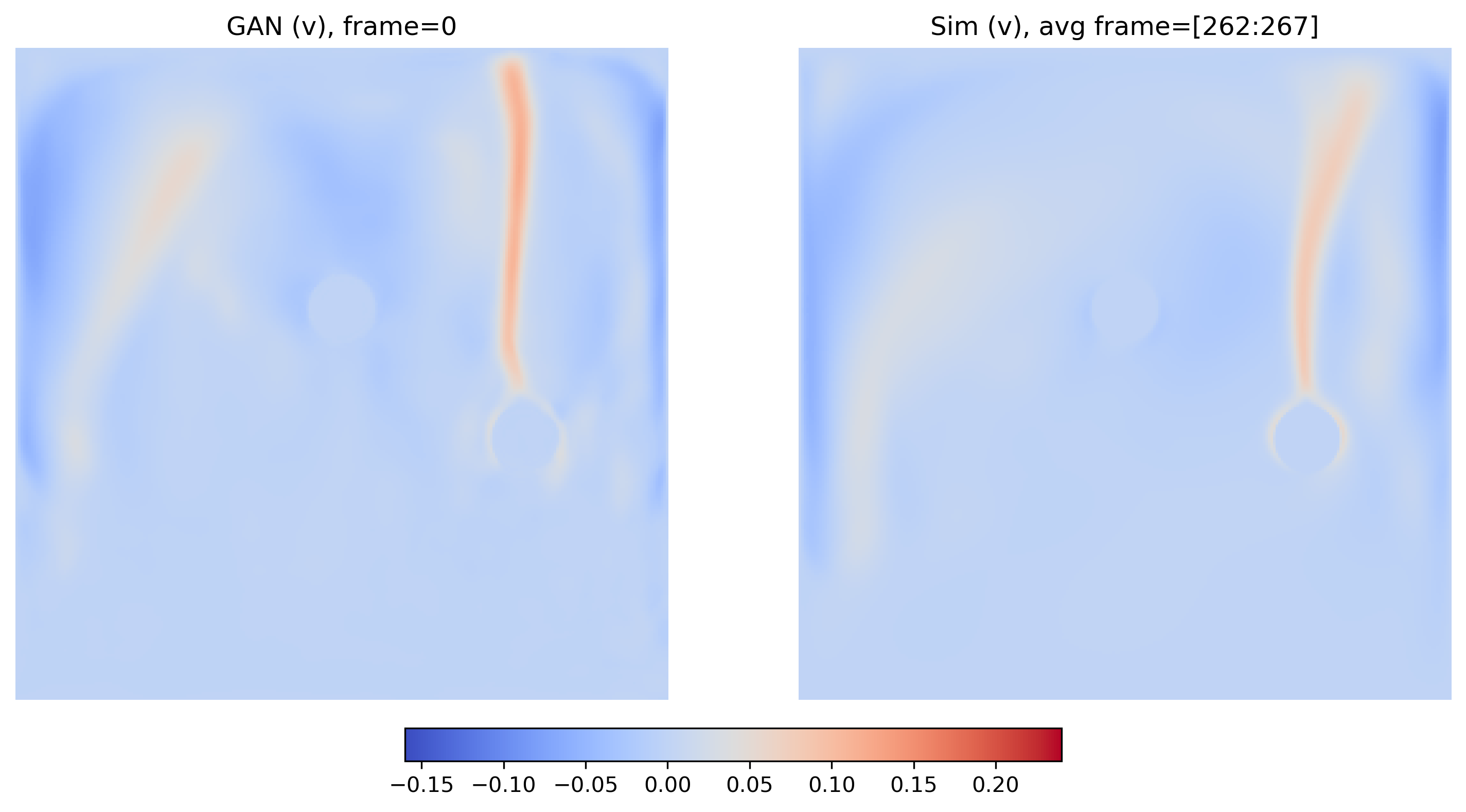}
        \caption{Comparison of $v$.}
    \end{subfigure}

    \caption{Experiment $x_a=0.78$: Comparison of GAN generated frames and simulation frames. The fields $u$ and $v$ are compared with an average over $5$ simulation frames.}
    \label{fig:comp_frames078}
\end{figure}

\begin{figure}[htbp]
    \centering
    \begin{subfigure}[b]{0.48\textwidth}
        \centering
        \includegraphics[width=\textwidth]{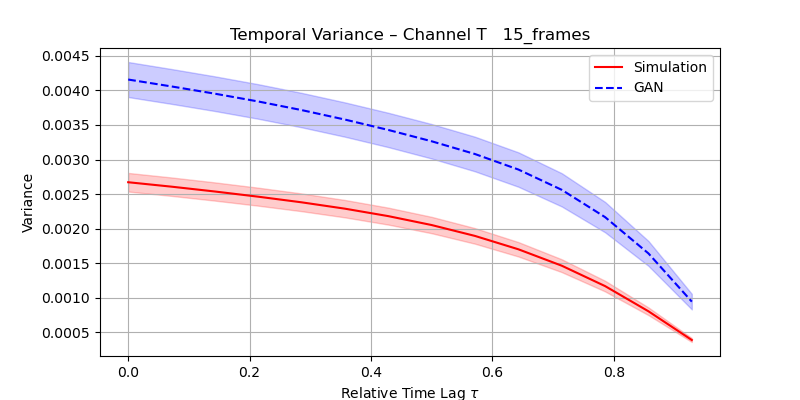}
        \caption{Comparison of $T$.}
    \end{subfigure}
    \hfill
    \begin{subfigure}[b]{0.48\textwidth}
        \centering
        \includegraphics[width=\textwidth]{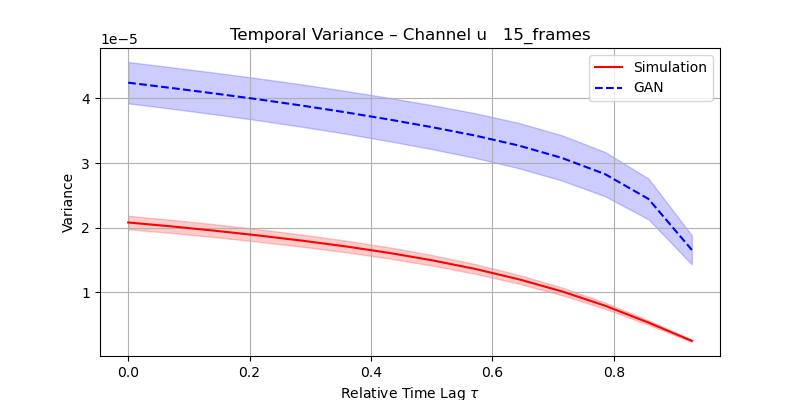}
        \caption{Comparison of $u$.}
    \end{subfigure}
    \hfill
    \begin{subfigure}[b]{0.48\textwidth}
        \centering
        \includegraphics[width=\textwidth]{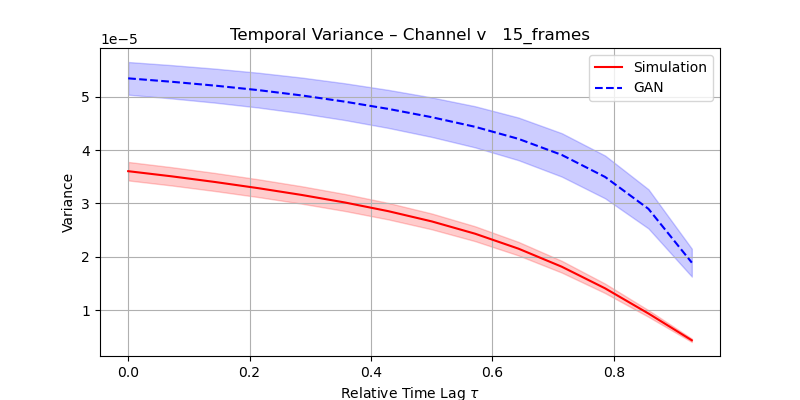}
        \caption{Comparison of $v$.}
    \end{subfigure}

    \caption{Experiment $x_a=0.78$: comparison of mean variance values computed via \eqref{eq:mean_variance} of generated (blue) and simulation frames (red). }
    \label{fig:mean_variance078}
\end{figure}

\begin{figure}[htbp]
    \centering
    \includegraphics[width=0.9\linewidth]{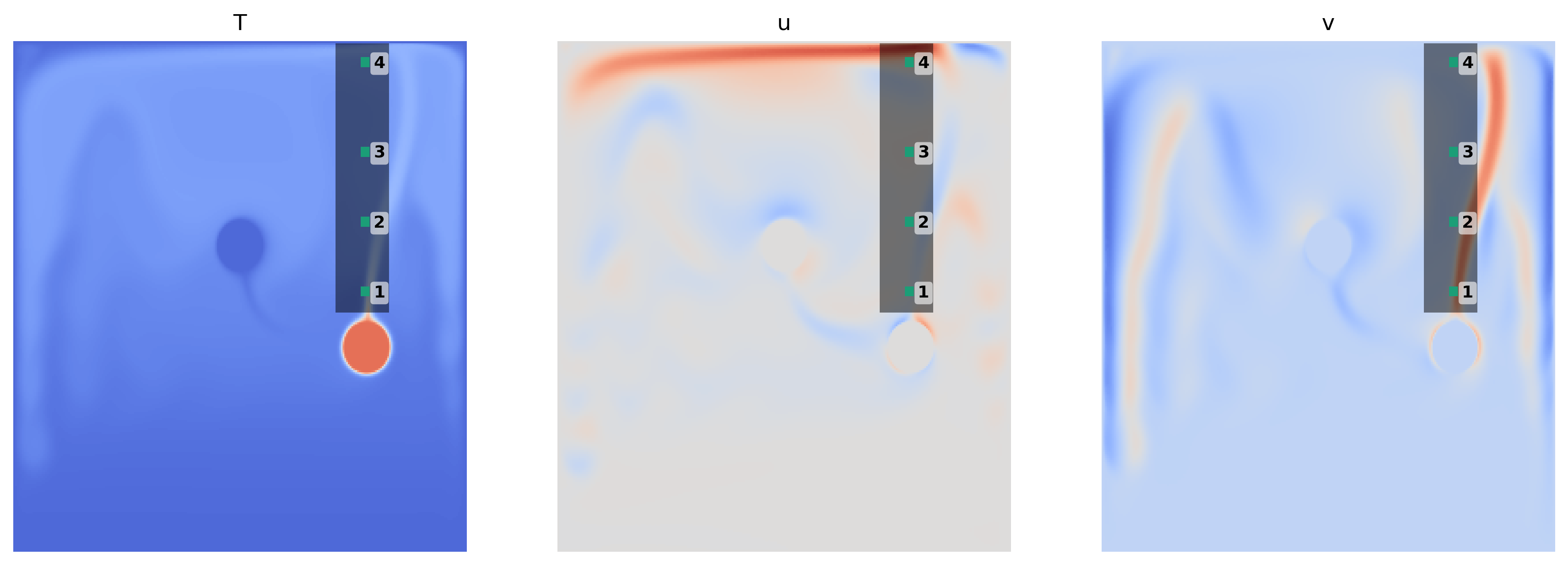}
    \caption{Experiment $x_a=0.78$: Placement of a cropped box $B$ (gray) and fixed spatial points (green) to compute spatial and temporal correlations.}
    \label{fig:corr_setup078}
\end{figure}

\begin{figure}[htbp]
    \centering
    \begin{subfigure}{\textwidth}
        \centering
        \includegraphics[width=0.32\textwidth]{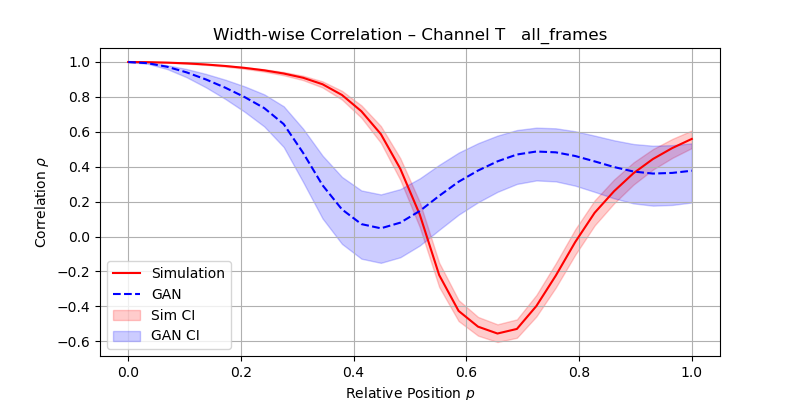}\hfill
        \includegraphics[width=0.32\textwidth]{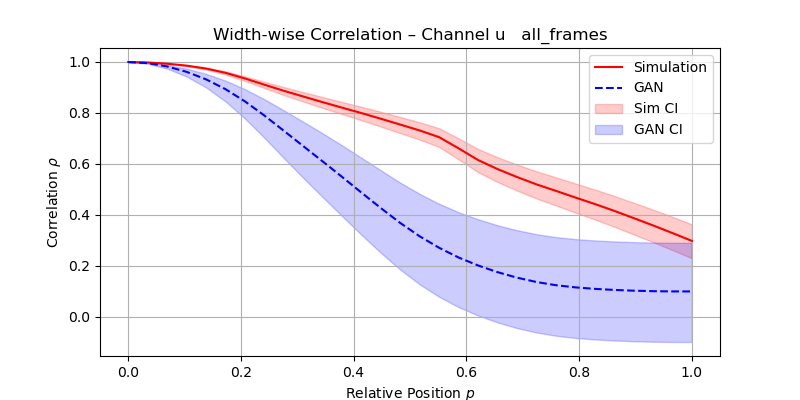}\hfill
        \includegraphics[width=0.32\textwidth]{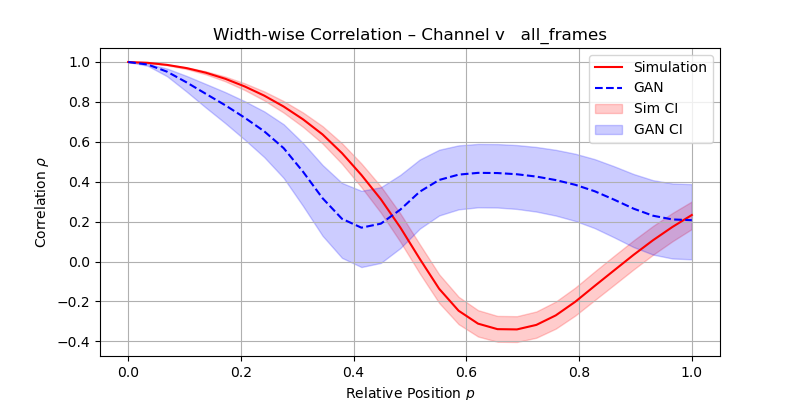}
        \caption{Column wise spatial correlations of $T$, $u$, and $v$.}
    \end{subfigure}

    \begin{subfigure}{\textwidth}
        \centering
        \includegraphics[width=0.32\textwidth]{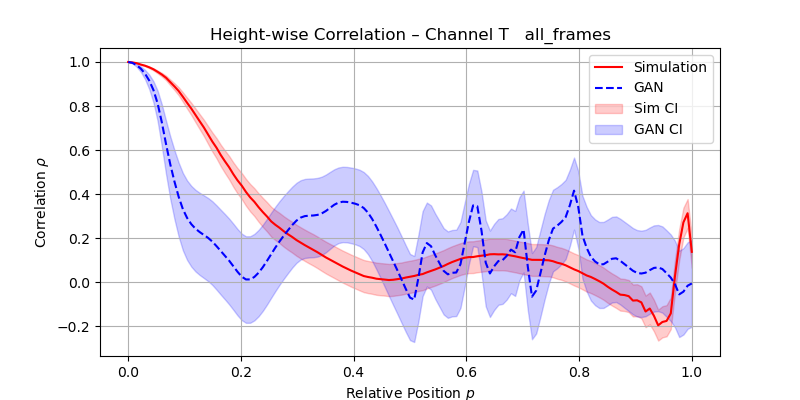}\hfill
        \includegraphics[width=0.32\textwidth]{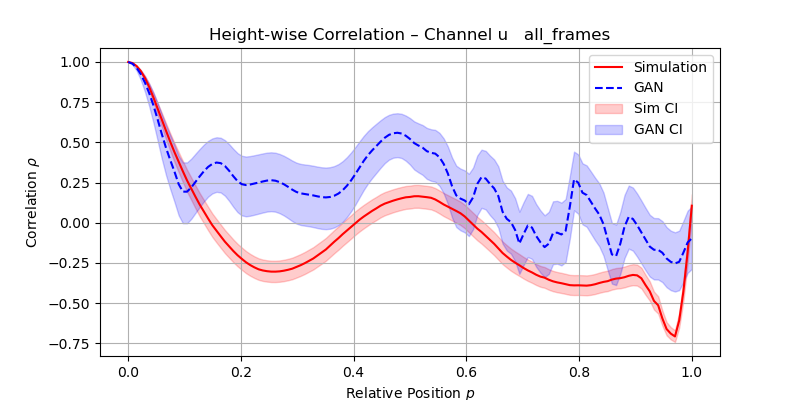}\hfill
        \includegraphics[width=0.32\textwidth]{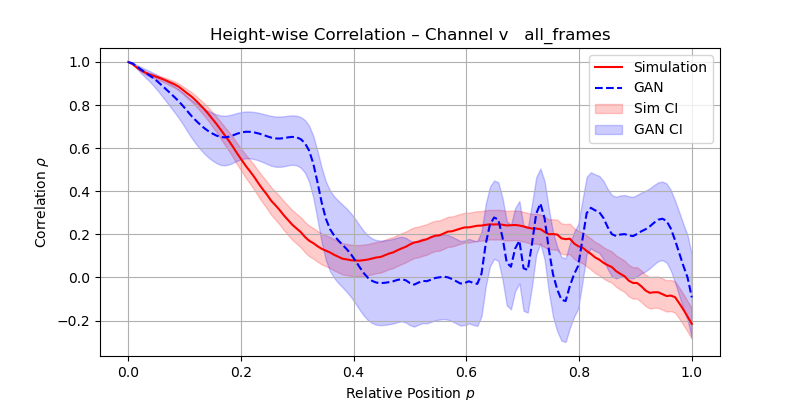}
        \caption{Row wise spatial correlations of $T$, $u$, and $v$.}
    \end{subfigure}

    \begin{subfigure}{\textwidth}
        \centering
        \includegraphics[width=0.32\textwidth]{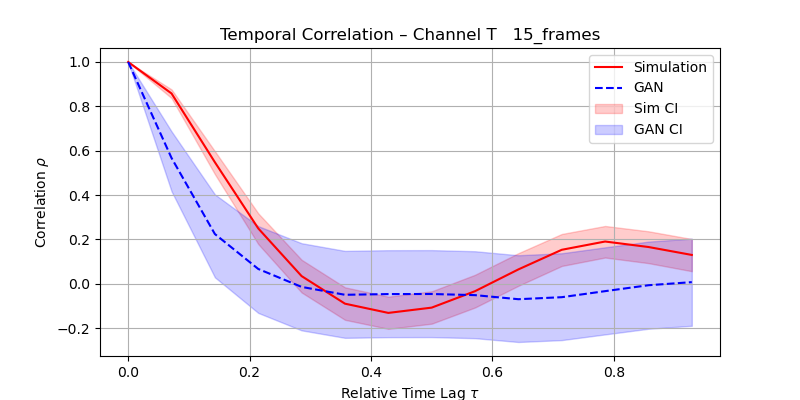}\hfill
        \includegraphics[width=0.32\textwidth]{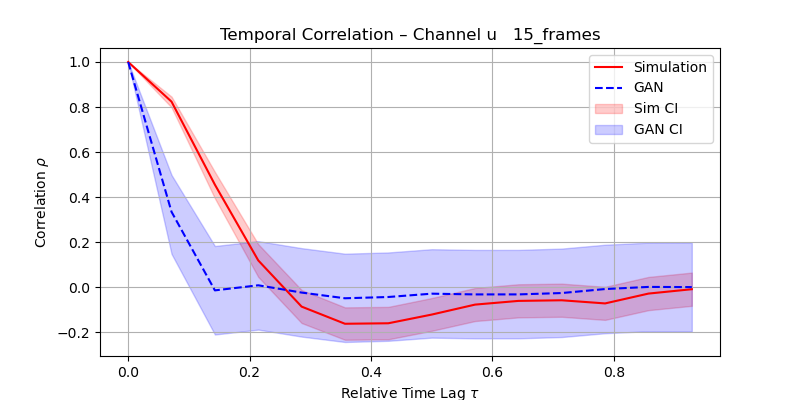}\hfill
        \includegraphics[width=0.32\textwidth]{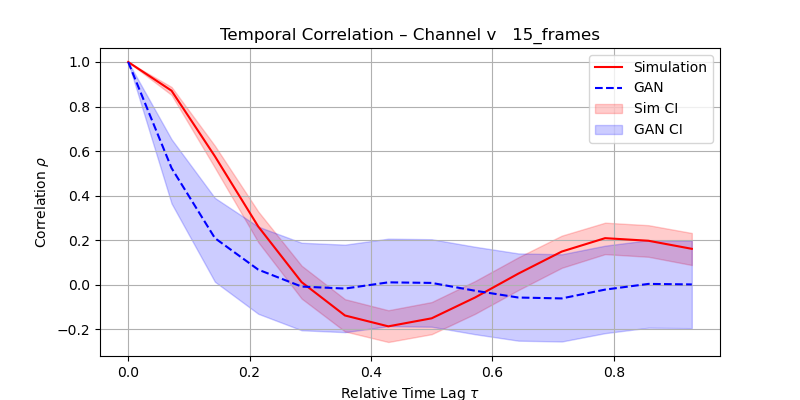}
        \caption{Temporal correlations of $T$, $u$, and $v$ in fixed spatial point 1.}
    \end{subfigure}

    \begin{subfigure}{\textwidth}
        \centering
        \includegraphics[width=0.32\textwidth]{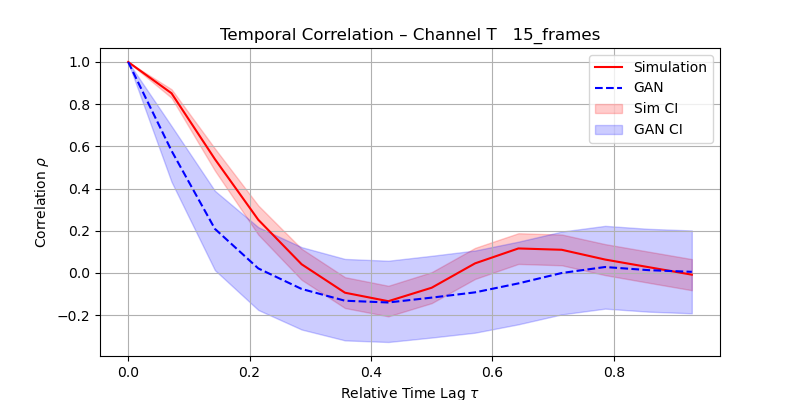}\hfill
        \includegraphics[width=0.32\textwidth]{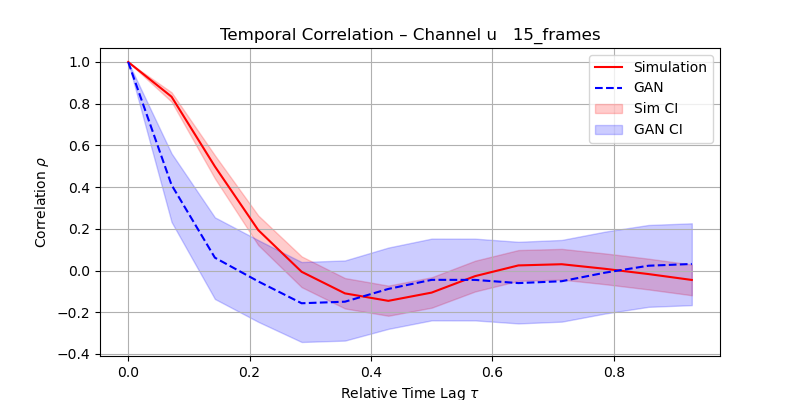}\hfill
        \includegraphics[width=0.32\textwidth]{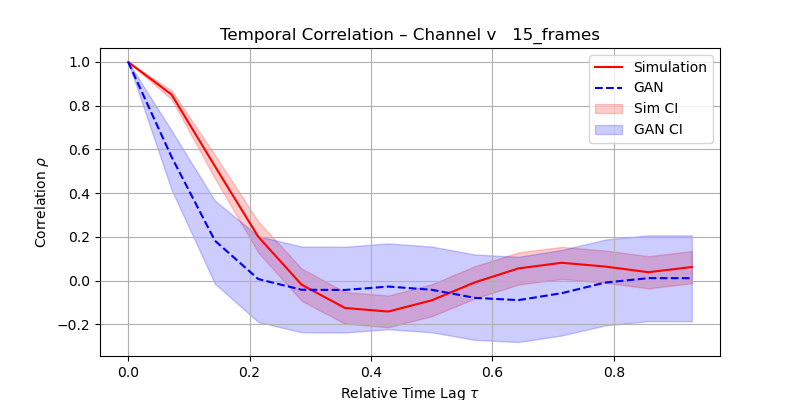}
        \caption{Temporal correlations of $T$, $u$, and $v$ in fixed spatial point 2.}
    \end{subfigure}

    \begin{subfigure}{\textwidth}
        \centering
        \includegraphics[width=0.32\textwidth]{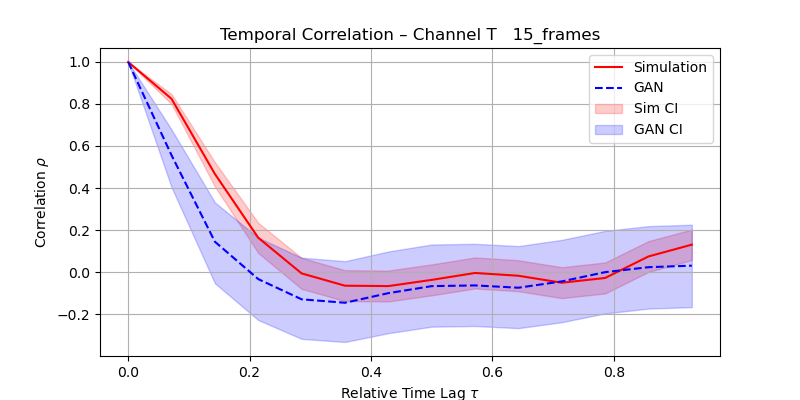}\hfill
        \includegraphics[width=0.32\textwidth]{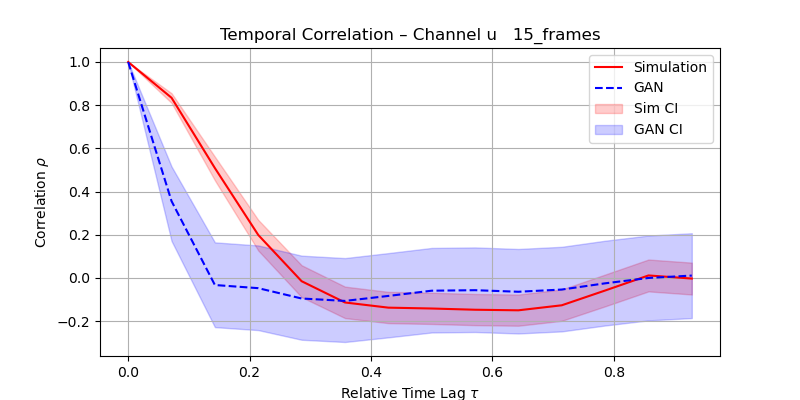}\hfill
        \includegraphics[width=0.32\textwidth]{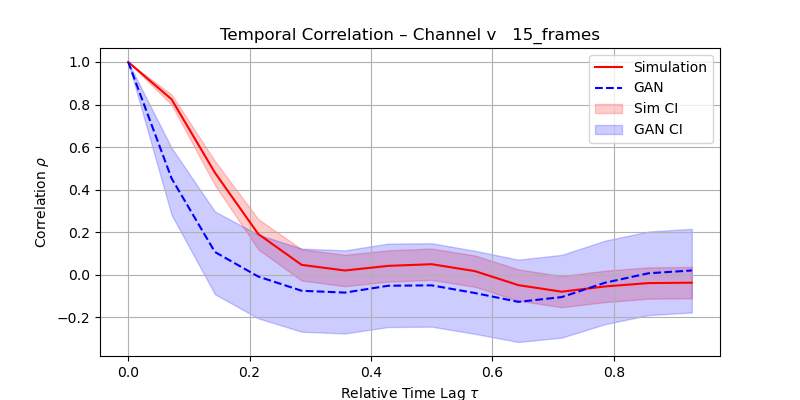}
        \caption{Temporal correlations of $T$, $u$, and $v$ in fixed spatial point 3.}
    \end{subfigure}

    \begin{subfigure}{\textwidth}
        \centering
        \includegraphics[width=0.32\textwidth]{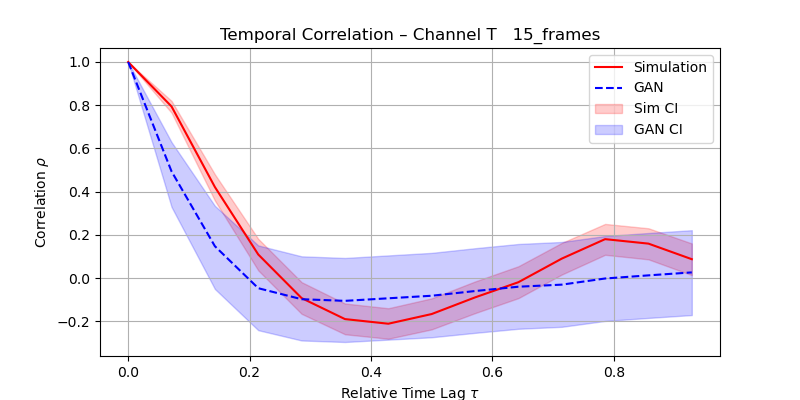}\hfill
        \includegraphics[width=0.32\textwidth]{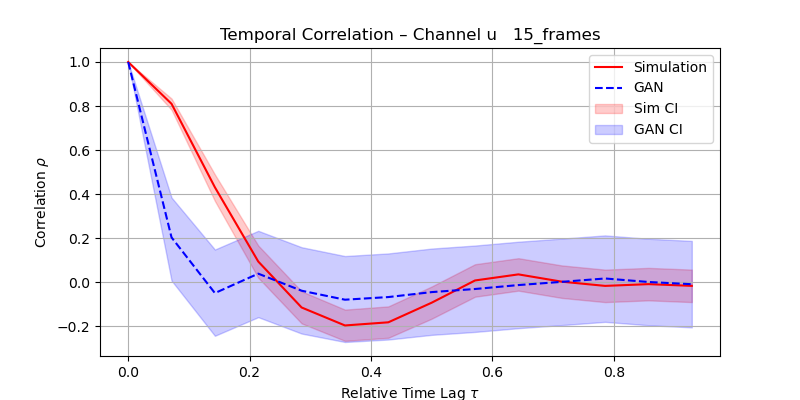}\hfill
        \includegraphics[width=0.32\textwidth]{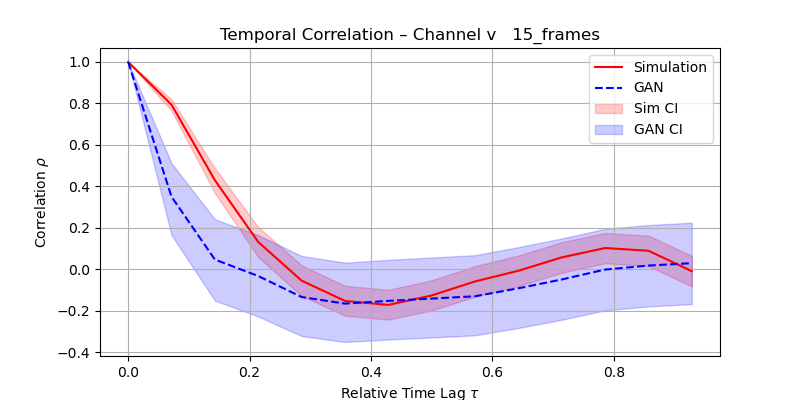}
        \caption{Temporal correlations of $T$, $u$, and $v$ in fixed spatial point 4.}
        
    \end{subfigure}

    \caption{Experiment $x_a=0.78$: Comparison of spatial and temporal correlations of simulation data (red) and GAN generated data (blue), including confidence intervals (shaded), for the fields $T$ (left), $u$ (middle), and $v$ (right).}
    \label{fig:correlations_xc078}
\end{figure}

For the experiment $x_a=0.78$, the results are weaker than for $x_a=0.22$, but could be improved with further training. In Figure~\ref{fig:comp_frames078}, the generated frames provide good, though not perfect, approximations of the simulation frames. The variance curves are also not captured perfectly, see Figure~\ref{fig:mean_variance078}. Furthermore, the positions of the measuring points are given in Figure~\ref{fig:corr_setup078} and the corresponding correlation curves in Figure~\ref{fig:correlations_xc078}. In this case, the temporal correlations are also captured, while additional training could further reduce the gap in spatial correlations.

\end{document}